\documentclass[useAMS,usenatbib]{mn2e}
\usepackage{epsfig}
\usepackage{graphicx}
\usepackage{times}
\usepackage{color}
\usepackage{rotating}

\def\cm3{cm$^{-3}$}
\def\kms{km~s$^{-1}$}

\def\lsun{L$_{\odot}$}
\def\rsun{R$_{\odot}$}

\def\msun{M$_{\odot}$}

\def\one{{\,\sc i}}
\def\two{{\,\sc ii}}

\def\beq{\begin{equation}}
\def\eeq{\end{equation}}

\def\lesssim{\mathrel{\hbox{\rlap{\hbox{\lower4pt\hbox{$\sim$}}}\hbox{$<$}}}}
\def\gtrsim{\mathrel{\hbox{\rlap{\hbox{\lower4pt\hbox{$\sim$}}}\hbox{$>$}}}}

\def\aj{AJ}
\def\pasp{PASP}
\def\apj{ApJ}
\def\apss{Ap\&SS}
\def\apjs{ApJS}
\def\apjl{ApJL}
\def\aap{A\&A}
\def\araa{ARA\&A}

\def\aaps{A\&AS}
\def\mnras{MNRAS}
\def\nat{Nature}

\def\solphys{Sol.~Phys.}

\def\gray{$\gamma$-ray}
\def\grays{$\gamma$-rays}
\def\isoni{$^{56}{\rm Ni}$}
\def\isoco{$^{56}{\rm Co}$}

\def\LC{light curve}
\def\LCS{light curves}

\newcommand{\olevel}[4]{#1\,#2\,$^{#3}$#4$^{\scriptsize \rm  o}$}
\newcommand{\elevel}[4]{#1\,#2\,$^{#3}$#4}

\title[Radiative transfer of SN Ib/Ic ejecta]{
On the nature of supernovae Ib and Ic}
\author
[Luc Dessart et al.]{\vspace{0.3cm} Luc Dessart,$^{1,2}$\thanks{email: Luc.Dessart@oamp.fr} D. John Hillier,$^{3}$,
Chengdong Li,$^3$, and Stan Woosley,$^4$ \\
 $^1$: Laboratoire d'Astrophysique de Marseille, Universit\'e Aix-Marseille \& CNRS,
UMR\,7326, 38 rue Fr\'ed\'eric Joliot-Curie, 13388 Marseille, France \\
 $^2$: TAPIR, Mailcode 350-17, California Institute of Technology, Pasadena, CA 91125, USA \\
  $^3$: Department of Physics and Astronomy, University of Pittsburgh, 3941 O'Hara Street, Pittsburgh,
        PA, 15260, USA \\
  $^4$: Department of Astronomy and Astrophysics, University of California, Santa Cruz, CA 95064, USA}

\date{Accepted . Received }

\pagerange{\pageref{firstpage}--\pageref{lastpage}} \pubyear{2010}

\begin{document}

\maketitle

\label{firstpage}

\begin{abstract}
Utilizing non-LTE time-dependent radiative-transfer calculations we investigate the 
impact of mixing and non-thermal processes associated with radioactive decay on SN IIb/Ib/Ic 
light curves and spectra. Starting with short-period binary models of $\lesssim$\,5-\msun\  
helium-rich stars, originally 18 and 25\,\msun\ on the main-sequence, we produce 1.2\,B 
ejecta which we artificially mix to alter the chemical stratification. While the 
total \isoni\ mass influences the light-curve peak, the spatial distribution of  \isoni, 
controlled by mixing processes, impacts both the multi-band light curves and spectra. 
With enhanced mixing, our synthetic \LCS\ start their post-breakout re-brightening 
phase earlier, follow a more gradual rise to peak, appear redder, and fade faster 
after peak due to enhanced \gray\ escape. Non-thermal electrons, crucial for  the 
production of He\one\ lines, deposit a large fraction of their energy as heat, 
and this fraction approaches 100\% under fully ionized conditions. Because energy 
deposition is generally local well after the light-curve peak, the broad He\one\ lines 
characteristic of maximum-light SN IIb/Ib spectra require mixing that places \isoni\ 
and helium nuclei to within a \gray-mean-free-path. This requirement indicates that SNe IIb 
and Ib most likely arise from the explosion of  stripped-envelope massive stars (main-sequence 
masses $\lesssim$\,25\,\msun) that have evolved through mass-transfer in a binary system, rather 
than from more massive {\it single} Wolf-Rayet stars. In contrast, the lack of He\one\ lines 
in SNe Ic may result from a variety of causes:  A genuine helium deficiency; strongly-asymmetric 
mixing; weak mixing; or a more massive, perhaps single, progenitor characterized by a larger 
oxygen-rich core. Helium deficiency is not a prerequisite for SNe Ic. Our models, subject to 
different mixing magnitudes, can produce a variety of SN types, including IIb, IIc, Ib, and Ic. 
As it is poorly constrained by explosion models, mixing challenges our ability to infer 
the progenitor and explosion properties of SNe IIb/Ib/Ic.
 \end{abstract}


\section{Introduction}

While the number of peculiar transients has grown in the last decade, much debate still surrounds the generic
properties of stellar explosions, including core-collapse supernovae (CCSNe) and their progenitors.
Ultimately, we aim to make an accurate prediction of the explosion energy, explosion morphology, ejecta mass, remnant mass,
nucleosynthetic yields etc. of a massive star given its mass, rotation rate, metallicity on the main sequence, and binarity.
At present, the exercise is done the opposite way -- for any new SN a model is crafted to match observations
with various levels of success. To reach a higher level of consistency, considerable work is still needed in stellar evolution,
stellar explosion, and radiative-transfer modeling of CCSNe. The present work is devoted to this later domain.

There is strong evidence that SNe II-P are linked to red-supergiant star explosions, either
through direct progenitor identification on pre-explosion images or through \LC\ modelling
\citep{falk_arnett_77,litvinova_nadezhin_85,eastman_etal_94,utrobin_07,KW09,DLW10b,DH11,bersten_etal_11}.
The failure to detect the progenitors of nearby SNe Ib/c on pre-explosion images \citep{smartt_09} prevents
such a robust identification. However, \LC\ \citep{EW88,shigeyama_etal_90,dessart_etal_11}
as well as spectral \citep{swartz_etal_93a} modelling  favors low-mass ejecta
likely resulting from the explosion of essentially hydrogen-less cores of moderate mass evolved in binary systems
\citep{podsiadlowski_etal_92,yoon_etal_10}.
The surprisingly high occurrence rates of SNe Ib and Ic versus II-P in existing surveys \citep{smith_etal_11}, together with the
non-detection of SNe Ib/c progenitors so far, gives further support to such binary-star progenitors \citep{eldridge_etal_08}.

\citet{georgy_etal_09} have studied the evolution of single massive stars and find a range of
metallicities and initial masses for the production of helium-rich and helium-deficient Wolf-Rayet (WR) stars at death.
Assuming a successful explosion follows the collapse of the core, they infer Ib to Ic SN rates that
corroborate the observations as well as the expectations from binary-star evolution.\footnote{In all these stellar-evolution studies,
the definition of a SN IIb, Ib, and Ic is based on the progenitor surface composition which does not directly translate into a SN type
due to the subtleties of atomic physics and radiative transfer. Much ambiguity thus surrounds
the definition of a SN Ib or Ic, which is distinct for SN observers, SN modelers, and modelers of stellar-evolution.
However, with the exception of those evolved at twice-solar metallicity,
single-star models have final masses $\gtrsim$10\,\msun, and thus are much larger than those of pre-SN ``hydrogen-less" cores produced
from binary-star evolution.}
However, this low-mass binary-star population is distinct from the recorded WR star population, which is dominated
by higher-mass objects\footnote{The reason is
simply that WR stars are searched for using a luminosity/mass-loss criterion, reflected by the strength of their optical
emission lines relative to the overlapping continuum. Only highly luminous WRs with dense stellar winds, which are primarily single
massive stars, are detected in such surveys (see \citealt{crowther_07} for a review).} that shed their envelope through radiatively-driven winds
rather than Roche-lobe overflow, and endowed with a higher mass at collapse.
Such properties at death  \citep{crowther_07} do not seem compatible with what makes the bulk of SNe Ib/c \citep{dessart_etal_11}.

Unstable \isoni\ nuclei, created by explosive nucleosynthesis,  are ultimately responsible for the long-term brightness of CCSNe.
In Type I CCSNe, the radioactive-decay energy from \isoni\ and its daughter nucleus \isoco\ also powers the peak of the \LC,
whose evolution is controlled by the dynamic radiative diffusion of the heat released at depth by decay \citep{colgate_mckee_69}.

As the densities decline with age, the assumption that radioactive decay energy is deposited only as heat becomes
untenable. Decaying nuclei produce \grays\ that Compton-scatter with free and bound electrons, producing
a high-energy tail to the otherwise Maxwellian electron distribution \citep{spencer_fano_54}.
These non-thermal electrons are a source of heat, critical for the \LC, and a source of non-thermal excitation and ionization.
The latter affect SN spectroscopic signatures, and are critical for the production of the optical He\one\ lines
that characterize the Type Ib SN class \citep{lucy_91,swartz_91}. They also play a considerable role in determining the
excitation/ionization of the gas during the nebular phase \citep{KF92,KF98a,KF98b}. Because of such complicated physics,
in addition  to non-LTE and time-dependent issues \citep{DH08_time,DH10}, estimating the exact composition of these ejecta,
and in particular the helium abundance, is a challenge.

In the last decade, the observational diversity of SNe Ib/c has been documented \citep{richardson_etal_06,modjaz_07,drout_etal_11}.
Several nearby events received special attention, such as the SN Ib 2008D \citep{soderberg_etal_08,modjaz_etal_09} or the
SN Ic 1994I \citep{filippenko_etal_95,richmond_etal_96, clocchiatti_etal_96}. While helium is undoubtedly
present in SNe Ib, it is unclear whether the lack of obvious He\one\ lines in SNe Ic spectra is evidence
for helium deficiency \citep{filippenko_etal_95,clocchiatti_etal_96}. Line overlap and blanketing prevents
a definite resolution of this problem based on a single-epoch optical or near-IR spectrum. The task instead requires
a full modelling of the \LC\ and spectra, from photospheric to nebular phase, since the successful model must
support observations at  all times and wavelength ranges. Such results must also be confronted with the
predictions from single- or binary-star evolution.

So far, CCSN radiation modeling has been split into two distinct efforts, one devoted to the \LC\ and the other,
usually using entirely different assumptions, to the spectra.
Light-curve modelling of SNe Ib/c has been performed based on theoretical models of the hydrogen-less cores produced through
binary-star evolution  \citep{EW88,shigeyama_etal_90}.
Because it is so critical to the problem, time-dependent radiation transport is treated but the gas is assumed in LTE and
opacities are approximated.
Using this approach, \citet{iwamoto_etal_94} modelled the SN 1994I \LC\ to infer a record-low ejecta mass of 0.9\,\msun\ for a CCSN.

 In contrast, spectroscopic modelling typically allows for departures from LTE but assumes steady-state radiation transport
 and is limited to the photospheric layers exclusively. In this spirit, \citet{sauer_etal_06} and \citet{tanaka_etal_09} used a Monte Carlo
 technique to model multi-epoch spectra of SNe 1994I and 2008D, respectively.
A similar photospheric steady-state approach assuming an homogeneous composition and a diffusive inner boundary
(but with allowance for non-thermal effects) was performed by \citet{baron_etal_99} on SN 1994I.
The assumption of steady-state makes the modelling more practical since one can adjust model parameters any number
of ways in order to fit the spectrum but this limits both the accuracy and the consistency of the simulation. The neglect of time-dependent effects impacts
the ionization and temperature structure; the prescribed diffusive inner-boundary flux, which becomes inadequate as the ejecta
thins out (i.e., as early as the peak of the \LC), compromises the accuracy of synthetic spectra longward of the visual band.
In addition, the piecemeal approach of these studies, which compute separately \LCS\ and spectra, is not globally consistent: the best-fit
model to the observed \LC\ may conflict with multi-epoch spectra, and vice-versa.

To remedy such shortcomings, we have developed a new non-LTE line-blanketed time-dependent radiative-transfer modelling approach
that simultaneously computes the multi-band \LCS\ and spectra with the capability to cover {\it continuously} from early times when the ejecta
is optically thick to late times when it is thin \citep{DH10,HD12}. We have
applied this method to study the early-time evolution of SN 1987A, and extended this work to SNe II-Plateau up to 1000\,d after explosion \citep{DH11}.

More recently, to gauge the conditions under which one may see H\one\ and He\one\ lines in the optical and near-IR ranges, we studied the early-time evolution of SNe IIb/Ib/Ic \citep{dessart_etal_11}.
These full-ejecta simulations were based on realistic binary-star evolution models \citep[Woosley et al. in prep]{yoon_etal_10},
evolved until iron-core collapse, and exploded by means of a piston with allowance for explosive burning.
These simulations indicate that as little as 0.001\,\msun\ of hydrogen with surface mass fraction as low as 0.01
can produce a strong H$\alpha$ line and a likely classification as IIb (or IIc as we argue in this paper). Contrary to myth, we also find
He\one\ lines can be produced in the absence of \isoni, but their presence is short lived and hinges on a very high helium mass
fraction in the ejecta. From the general agreement of our synthetic \LCS\ with observations, we argued that SNe IIb/Ib/Ic are generally low-mass
ejecta, likely resulting from low-mass massive stars evolved in interacting binaries, as generally proposed.

In this work, we repeat a sample of such SN IIb/Ib/Ic  simulations but this time we include the treatment of non-thermal effects
associated with \gray-energy deposition. As shown by \citet{lucy_91}, non-thermal effects are crucial for understanding
the production of He\,{\sc i} lines in SN Ib spectra. However, with advances in computational power and  the availability of better atomic data,
we relax many of the assumptions made in his treatment, such as steady-state (applied to single-epoch spectra), not fully non-LTE,
neglect of line-blanketing, and the use of fully mixed (i.e., homogeneous) ejecta.
Our paper is also a continuation of the earlier explorations of \citet{woosley_eastman_97} on the role of mixing
and non-thermal processes in SN Ib and Ic ejecta.

To investigate the effect of mixing on CCSN \LCS\ and spectra originating from our binary-star progenitors,
we use a ``boxcar'' algorithm to modulate the species distribution in the ejecta and hence mimic multi-dimensional effects
associated with massive-star explosions \citep{kifonidis_etal_03,kifonidis_etal_06,joggerst_etal_09,hammer_etal_10}.
Our crude mixing prescription only affects the distribution of the elements --- the density structure remains unchanged.
While some mixing seems to occur in all SNe Ib/c, we find that it must be very efficient in helium-rich ejecta
to yield a SN Ib -- moderate mixing, even in helium-rich ejecta, will still produce a SN Ic spectrum. This sets strong
constraints on the nature of these events, both concerning the pre-SN star evolution and the explosion physics.

This paper is structured as follows. In the next section, we discuss the various issues that surround
mixing in CCSN ejecta (origin, nature, effects). In Section \ref{sect_model}, we briefly summarize our
modelling approach, focusing primarily on the treatment of non-thermal processes since this aspect is
an addition to the method described in \citet{DH10,DH11,dessart_etal_11,HD12}.
We then describe in Section~\ref{sect_nonth} the thermal and non-thermal effects associated with \gray\ energy deposition and
the production of Compton-scattered high-energy electrons, and how these affect the formation of He\one\ lines
(Section~\ref{sect_rates}). Section \ref{sect_results_LC} describes the ejecta properties, in particular the photospheric
properties, and how these are reflected in the \LCS.
We then discuss in Section \ref{sect_results_spec} the spectral properties of our grid of SN Ib/c models
and their sensitivity to the efficiency of ejecta mixing.
In Section \ref{sect_comp}, we compare these synthetic spectra to observations in order to gauge the adequacy of our models.
We finally discuss the relevance of these results and present our conclusions in Section \ref{sect_conc}.

\begin{figure*}
\epsfig{file=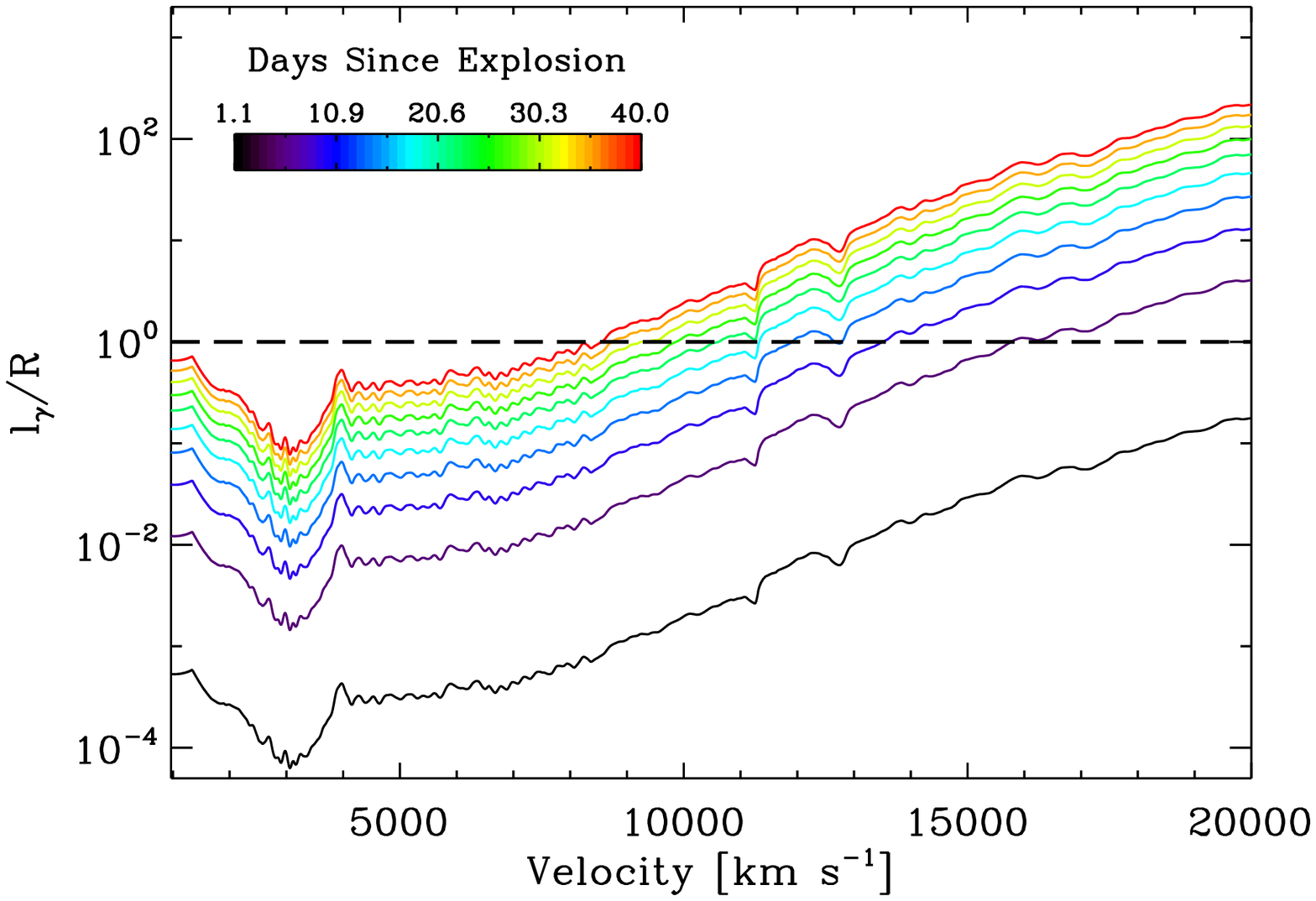,width=8.5cm}
\epsfig{file=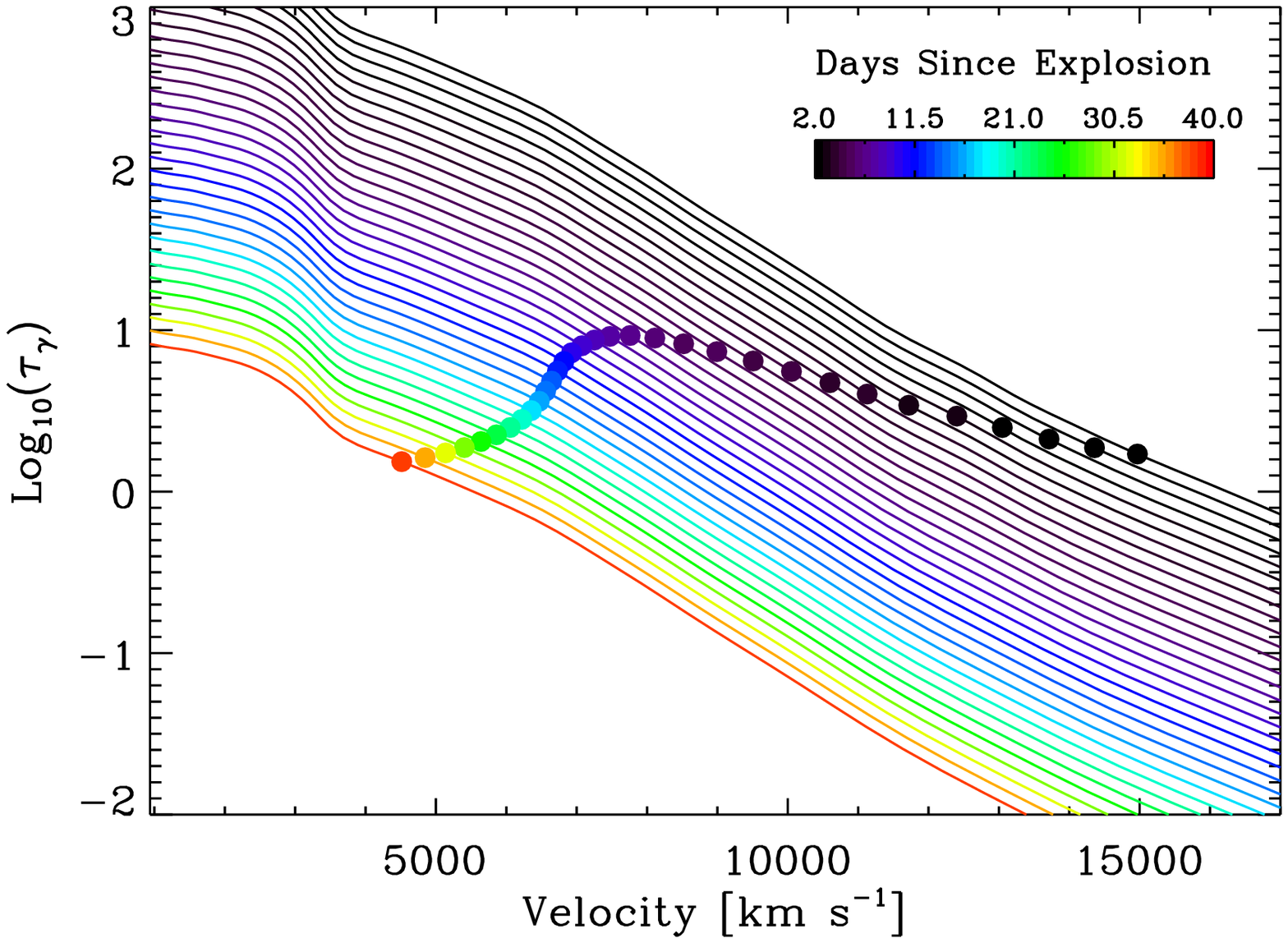,width=8.5cm}
\caption{{\it Left:} Evolution of the radial variation of the \gray\ mean free path, normalized to the local radius
in the SN IIb model Bmi18mf4p41z1 of \citet{dessart_etal_11}.
A color coding is used to distinguish different epochs, selected at fixed intervals between 1.1 and 40\,d
after explosion. The dashed line refers to $l_{\gamma} = R$, which is a representative threshold for
the non-local influence of \grays.
{\it Right:} Same as left, but now showing the \gray\ optical depth integrated inwards from the outer boundary
of the ejecta. The dots indicate the location of the electron-scattering photosphere (i.e., $\tau_{es}=2/3$).
\label{fig_lgray}
}
\end{figure*}

\begin{figure*}
\epsfig{file=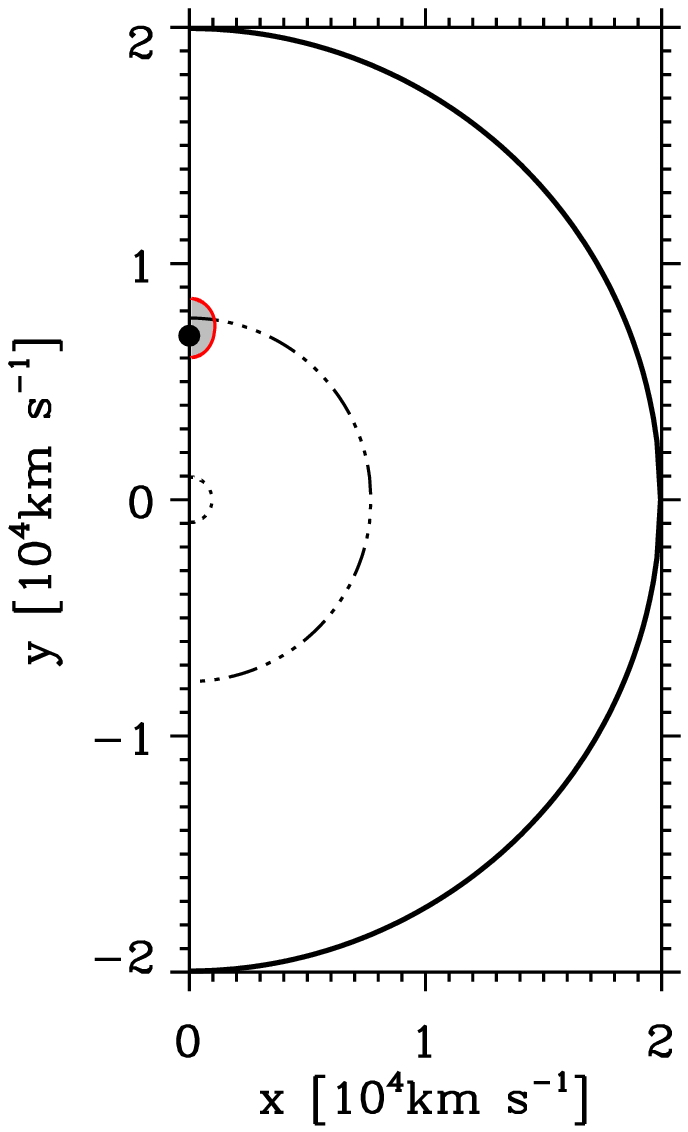,width=3.6cm}
\hspace{0.4cm}
\epsfig{file=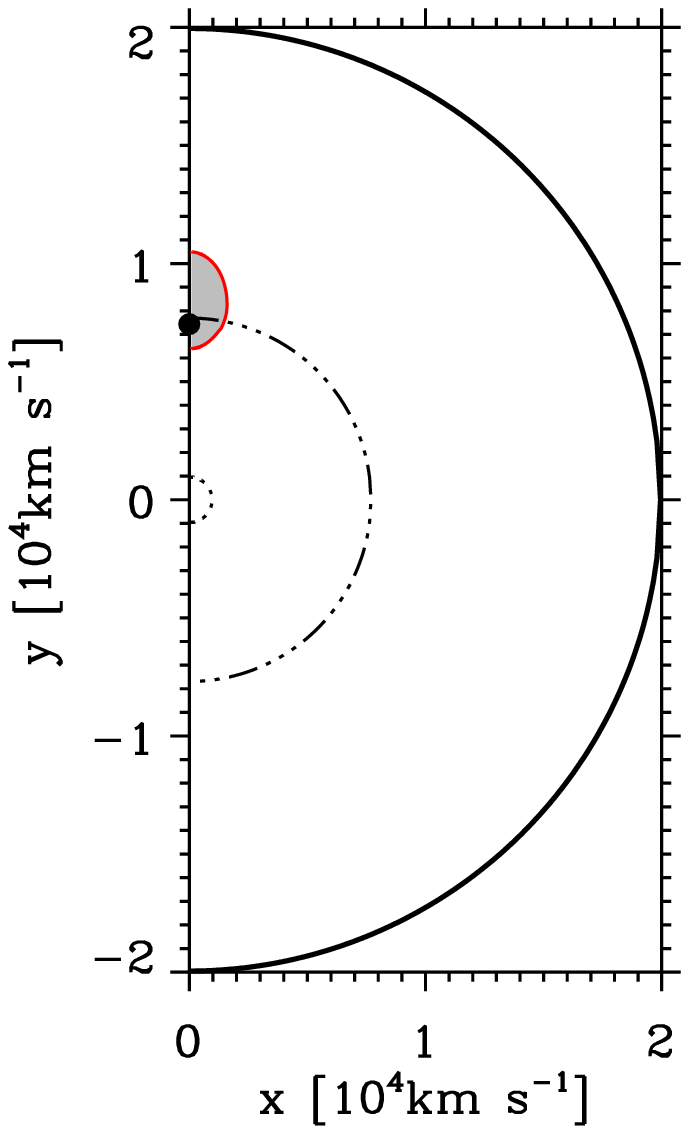,width=3.6cm}
\hspace{0.4cm}
\epsfig{file=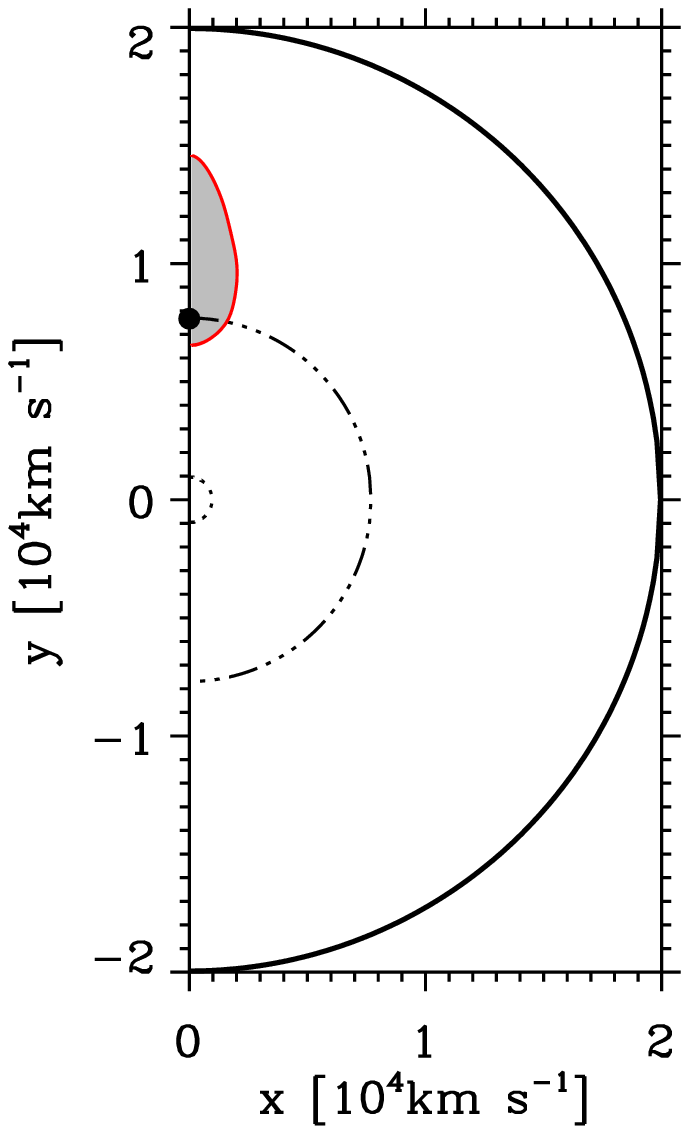,width=3.6cm}
\hspace{0.4cm}
\epsfig{file=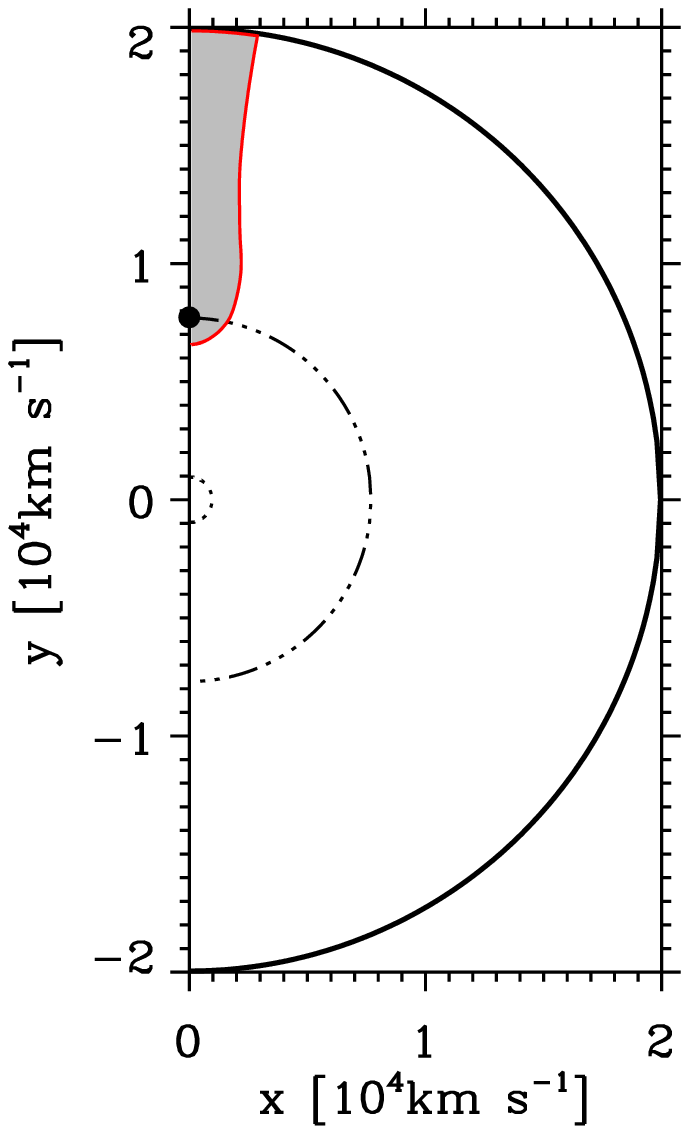,width=3.6cm}
\epsfig{file=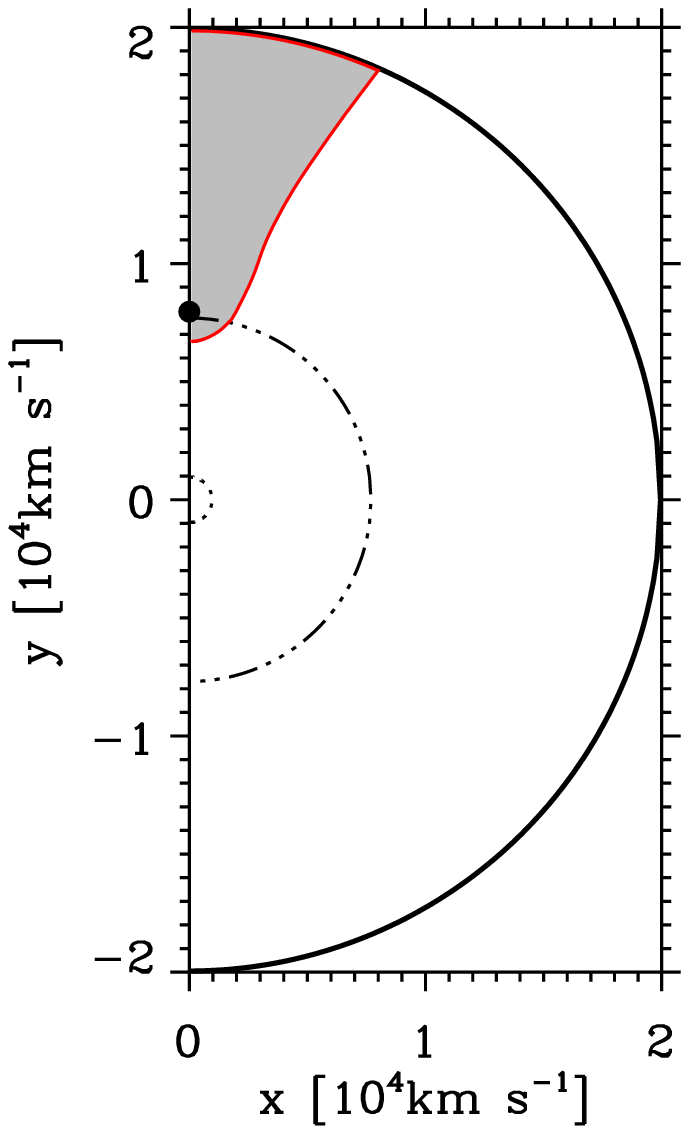,width=3.6cm}
\hspace{0.4cm}
\epsfig{file=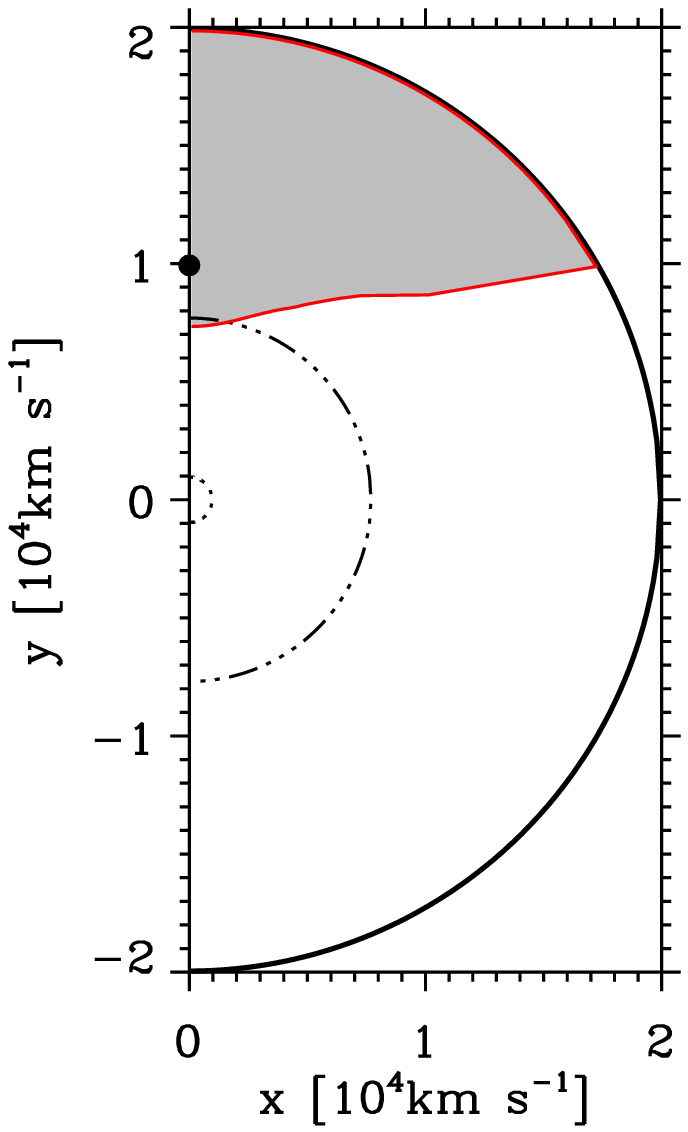,width=3.6cm}
\hspace{0.4cm}
\epsfig{file=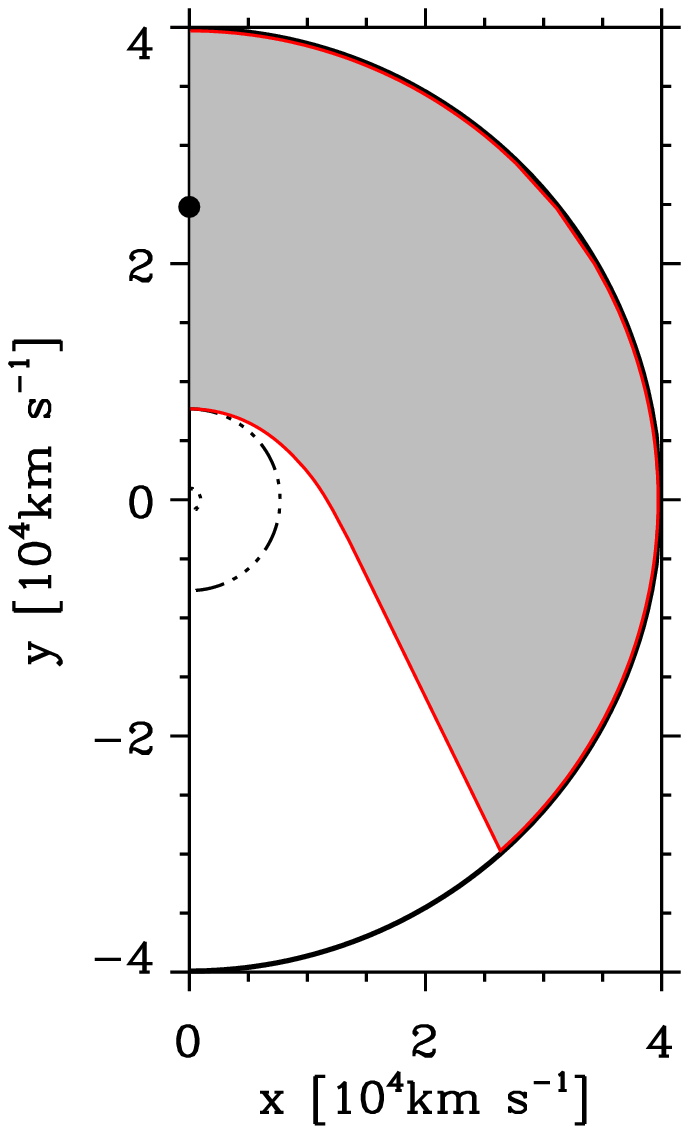,width=3.6cm}
\hspace{0.4cm}
\epsfig{file=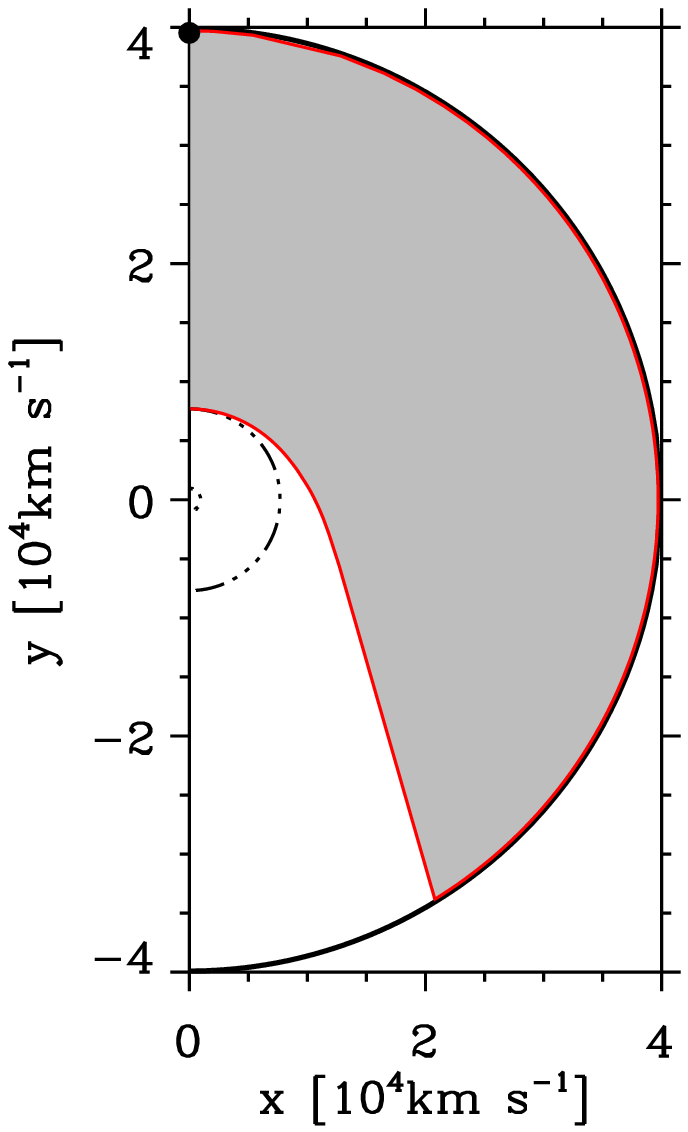,width=3.6cm}
\caption{\label{fig_gray}
Slices through a SN ejecta (containing its center of mass; we use model Bmi18mf4p41z1 of \citet{dessart_etal_11} at 26.8\,d
after explosion) and illustrating the area (shown in gray) that is within a \gray\
optical depth of 2/3, assuming various heights along the vertical axis for the \gray\ emission site (shown as a black dot).
We assume the SN ejecta has a spherically-symmetric mass distribution, so that this area corresponds to a volume of revolution
about the vertical axis. Its morphology changes drastically as
the \gray-emission site crosses the \gray\ photosphere (dash-dotted line) located at 7700\,\kms\ at this time
(where the optical photosphere is at that time depends on the model; see Section~\ref{sect_results_LC}).
For emission sites near the \gray\ photosphere, \grays\ are incapable of reaching regions of
similar radius/velocity but at different latitudes/longitudes: the larger \gray\ mean free path in the radial direction makes it
the preferred direction of escape.}
\end{figure*}

\section{The need for mixing}
\label{sect_gray}

Before presenting detailed radiative-transfer models that include a treatment of non-thermal effects,
we start by motivating the need for mixing, since it modulates the magnitude of these processes
through the local variation of the \isoni\ mass fraction.

The  \gray\ mean free path $l_{\gamma}$ is defined by $l_{\gamma} = 1 / \kappa_\gamma \rho$, where $\kappa_\gamma$ is
the mass-absorption coefficient to \grays\ and $\rho$ is the mass density.
For a uniform ejecta composition, $\kappa_\gamma$ is a constant,\footnote{At the energies of interest, gamma-rays lose most of
their energy through Compton scattering. While the scattering is forward peaked, the gamma-rays loose little energy until they undergo
a large angle scattering. As shown by  \cite{swartz_etal_95} accurate answers for the gamma-ray transport can be obtained
by treating gamma-ray energy deposition as a purely absorptive process.} which we set to 0.03\,cm$^2$\,g$^{-1}$ \citep{KF92},
so that $l_{\gamma}$ just scales inversely with $\rho$. The density distribution in WR-star explosions is well described above
$\sim$3000\,\kms\ by a power law with exponent $n \sim 8$ \citep{EW88,dessart_etal_11} and is thus a steep declining
function of radius (or equivalently velocity). Even with the corresponding strong increase in $l_{\gamma}$ with radius,
the ejecta density is very large at early times so that $l_{\gamma}/R$ remains generally small.
Hence, \grays\ cannot  travel the distance that separates the inner ejecta layers, where the \isoni\ was explosively produced,
and the optical photosphere where the bulk of the SN radiation escapes.\footnote{The lower the main sequence
mass the smaller the (oxygen-rich) mass buffer between these two regions. Intuitively, we
anticipate such \gray-transport to occur sooner in lower-mass ejecta from lower-mass main-sequence stars in binary systems.}
Such deeply seated \grays\ will completely ``thermalize'' and trigger the formation of the heat wave that will eventually affect the
\LC. However, they will induce no visible non-thermal effect at early times.

In homologously-expanding ejecta, the density in any mass shell varies with the inverse cube of the time since explosion,
and thus the optical depth to the surface of the ejecta varies as the inverse square of the time. Hence, deeply emitted  \grays\
will eventually be able to directly influence the bulk of the ejecta mass.
However, this will happen no earlier than about 2 months after explosion in {\it unmixed} SN IIb/Ib/Ic ejecta \citep{dessart_etal_11},
and thus after the medium is thin to optical photons.
Using as a reference the spherically-symmetric ejecta produced from the explosion of the Bmi18mf4p41z1 model
of \citet{dessart_etal_11}, we illustrate in Fig.~\ref{fig_lgray} (left plot) the depth and time variation of $l_\gamma/R$.
This quantity is indeed less than unity below 10000\,\kms\ for up to $\sim$30\,d
after explosion, so that the \gray\ influence is essentially local (there is no ``transport'') for up to a month.
An alternative method of examining gamma-ray transport is to examine the radial optical depth, integrated
inward from the outermost grid point down to $\tau_\gamma=2/3$ (Fig.~\ref{fig_lgray}, right plot). This gives the location above
which gamma-ray photons can ``freely'' escape.

The critical observation that requires mixing and the early influence of non-thermal processes on the observed radiation
in SNe IIb/Ib  is the persistence of He\one\ lines throughout the photospheric phase \citep{modjaz_07}.
To produce non-thermal excitation/ionization early-on, \isoni\ nuclei have to be mixed  into regions where helium atoms
are present, although some helium will be mixed inwards too.
Quantitatively, the width of the He\one\,5875\AA\ line at early times in the standard SN Ib 2008D implies \isoni\ has to
be present in ejecta shells traveling at $\sim$10000\,\kms\ (what this implies for the SN explosion energy depends on
the ejecta mass, which is debated; see, e.g., \citealt{tanaka_etal_09}). When we discuss mixing to large velocities, we
mean the large velocities commensurate with the observed line profile (half) widths.
What clearly emerges from our modeling is that helium richness alone is not sufficient to ensure a SN Ib classification; the ``right'' mixing has to occur to allow efficient non-thermal excitation \citep{lucy_91}. In this respect, SNe Ib set a very stiff constraint on
the explosion physics through this mixing efficiency, as well as on the composition and mass of the ejecta in which this mixing operates.

Mixing of \isoni\ out to large radii/velocities in CCSN ejecta may take various forms but it is fundamentally associated with
multi-dimensional effects.
Mixing may arise from a large scale asymmetry of the explosion associated with a jet/bipolar/unipolar
morphology, as in the collapsar model \citep{woosley_93}, an intrinsic large-scale asymmetry of the shock in
an explosion through the neutrino mechanism \citep{scheck_etal_06,marek_janka_09} or the magneto-rotational
mechanism \citep{leblanc_wilson_70,bisnovatyi_etal_76,khokhlov_etal_99,maeda_etal_02,burrows_etal_07a,burrows_etal_07b,dessart_etal_08}.
Mixing may also stem from the small-scale structure of the SN shock, associated with convection and slowly-developing
Rayleigh-Taylor and Kelvin-Helmholtz instabilities \citep{kifonidis_etal_03,kifonidis_etal_06,joggerst_etal_09,hammer_etal_10,nordhaus_etal_10}.
Note that even with mixing, the bulk of the \isoni\ mass remains deeply seated because mixing only affects the species
distribution while the mass density is essentially unaffected and steeply rises towards the inner ejecta.

Because the \gray\ mean free path tends to be much smaller than the scale of the system at early times (Fig~\ref{fig_lgray}),
these different aspherical morphologies behave very differently for the purpose of non-local \gray\ transport and
non-thermal excitation/ionization of the gas.
To illustrate this, we use the model shown in Fig.~\ref{fig_lgray} at an age of 26.8\,d (which typically corresponds to the
\LC\ peak in a SN Ib) and study the geometrical properties imposed on mixing to generate non-thermal effects at the
photosphere at that time. For this model, $\tau_\gamma=2/3$, is located at  7700\,\kms.

The definition of the \gray\ mean free path $l_{\gamma}$ contains no information on the radial or angular dependence of
the escape probability of \grays\  at a given time from a given location in the ejecta.
A better quantity is the angle-dependent \gray\ optical depth, whose variation $d\tau_\gamma$ over a step size $ds << R_{\rm phot}$
is equal to $\kappa_\gamma \rho ds$ (the photospheric radius  $R_{\rm phot}$ is used as a representative scale for the system).

\begin{figure*}
\epsfig{file=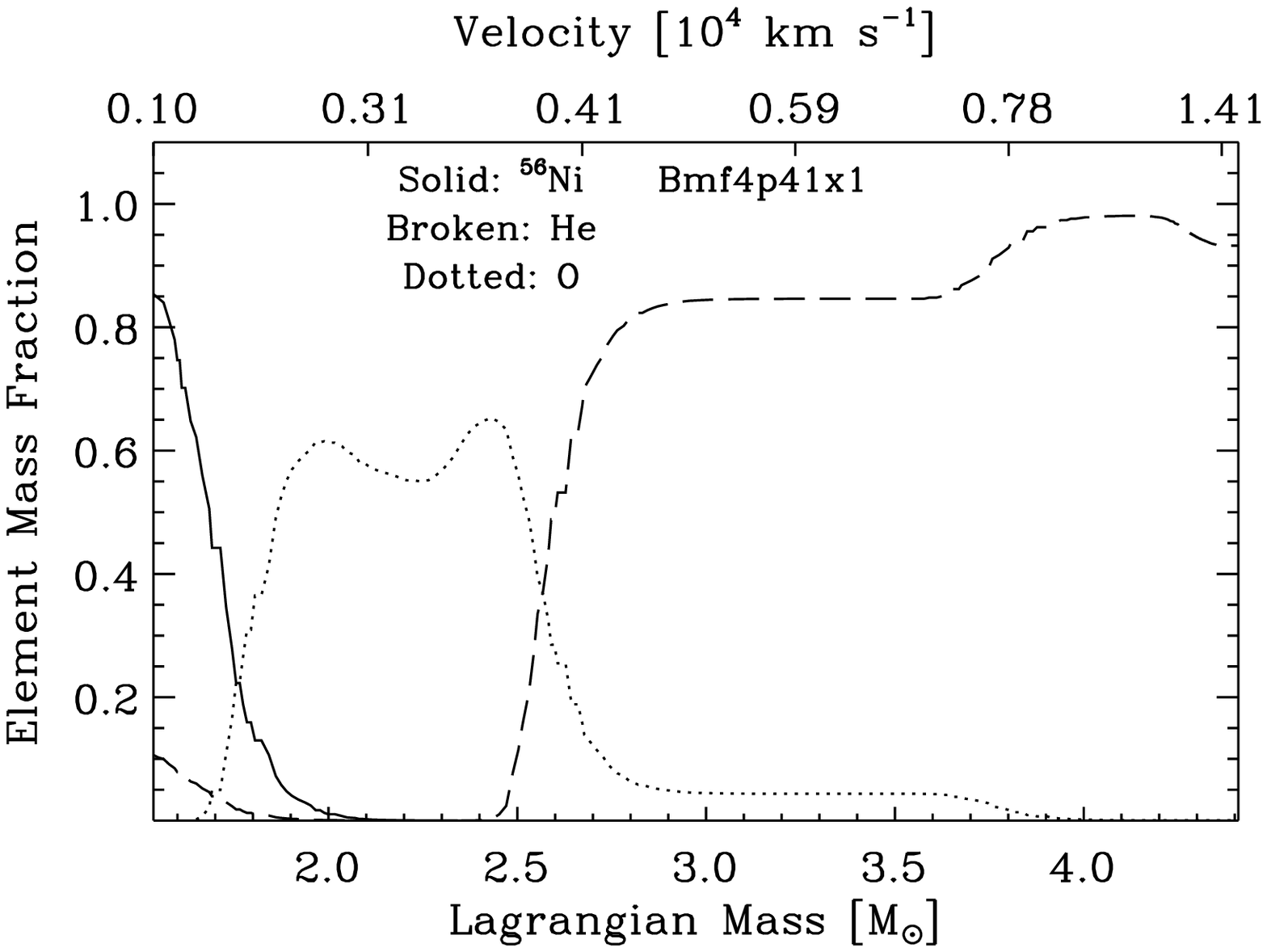,width=8.cm}
\epsfig{file=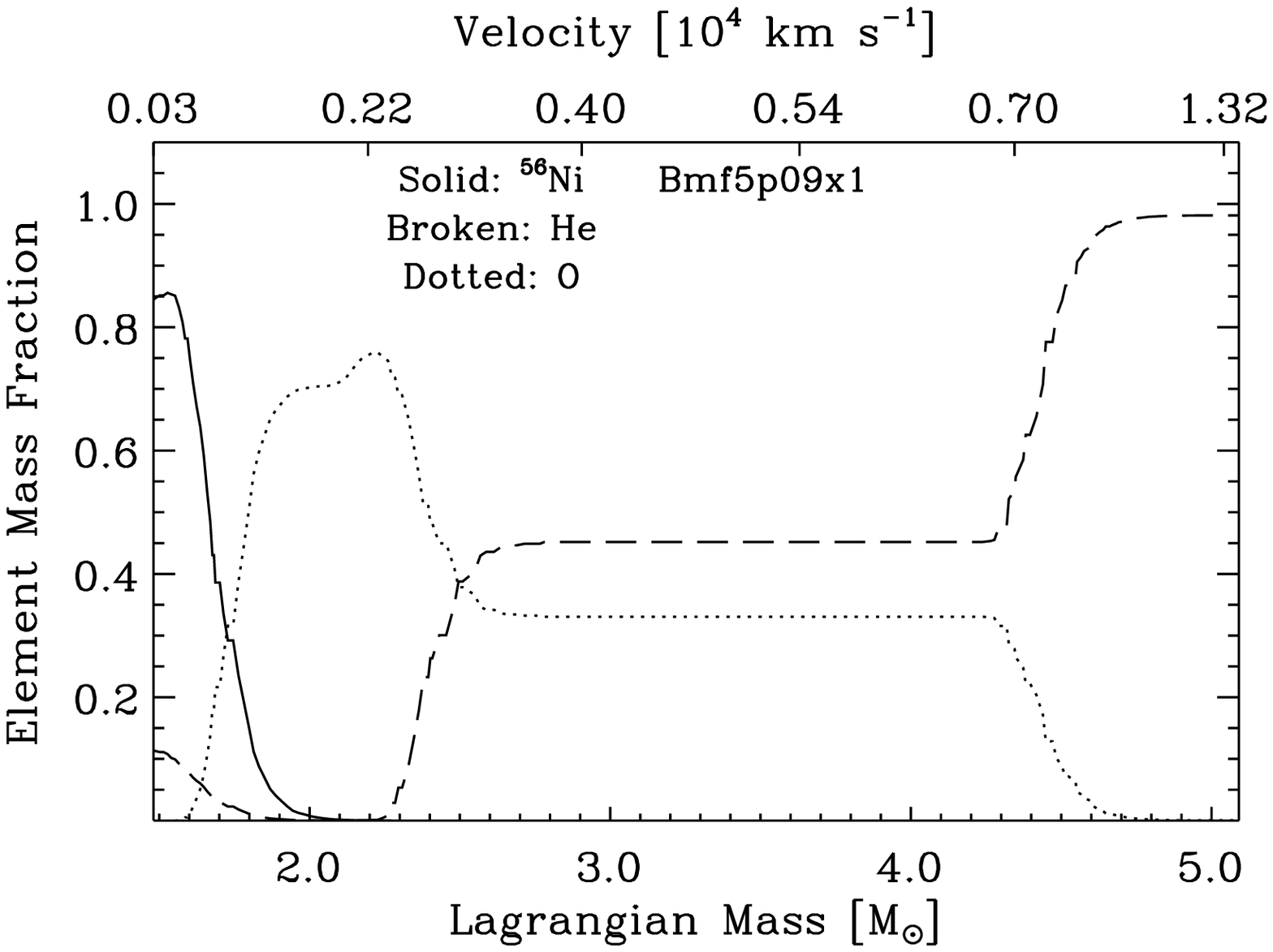,width=8.cm}
\vspace{-0.5cm}
\epsfig{file=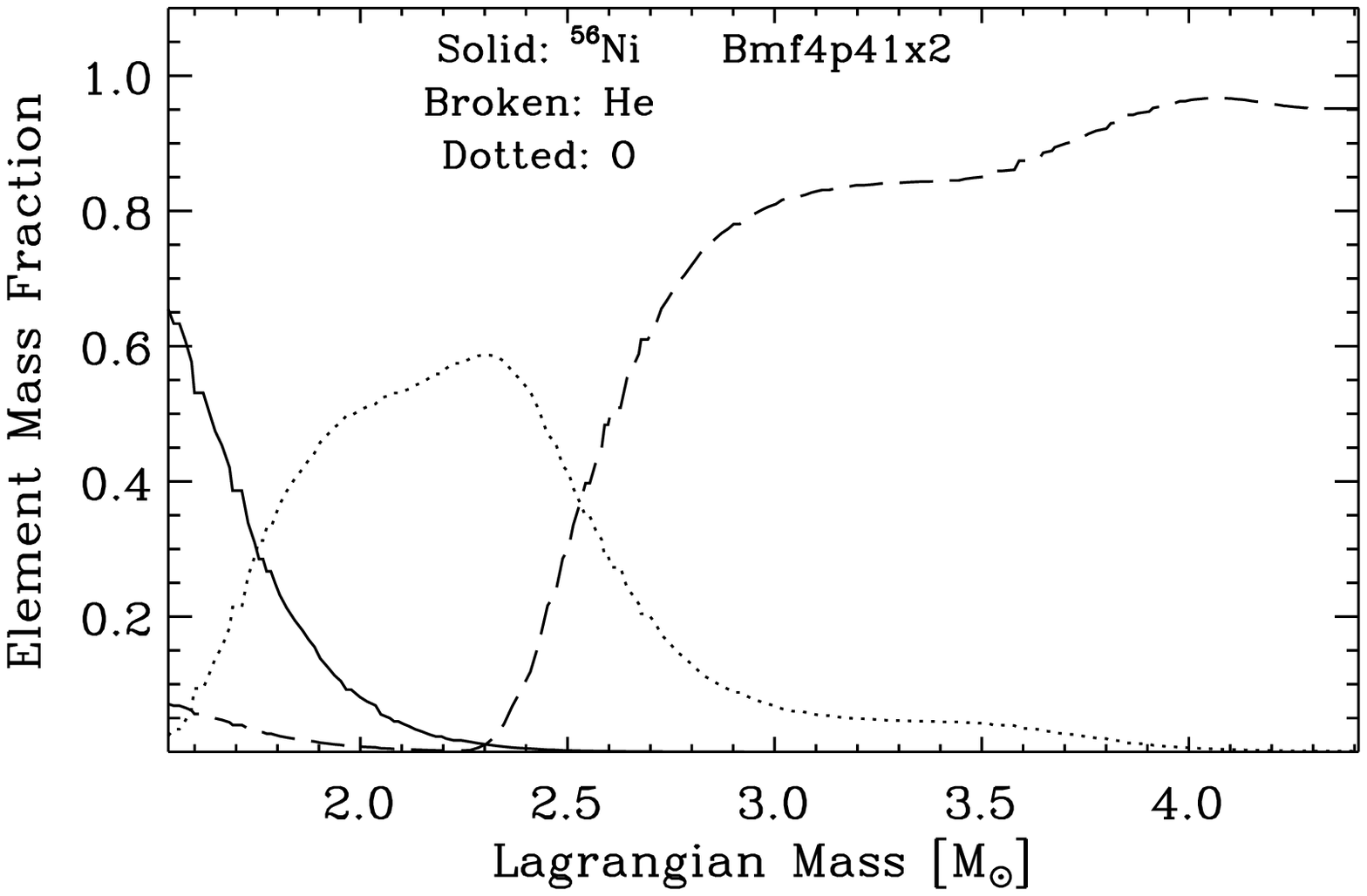,width=8.cm}
\epsfig{file=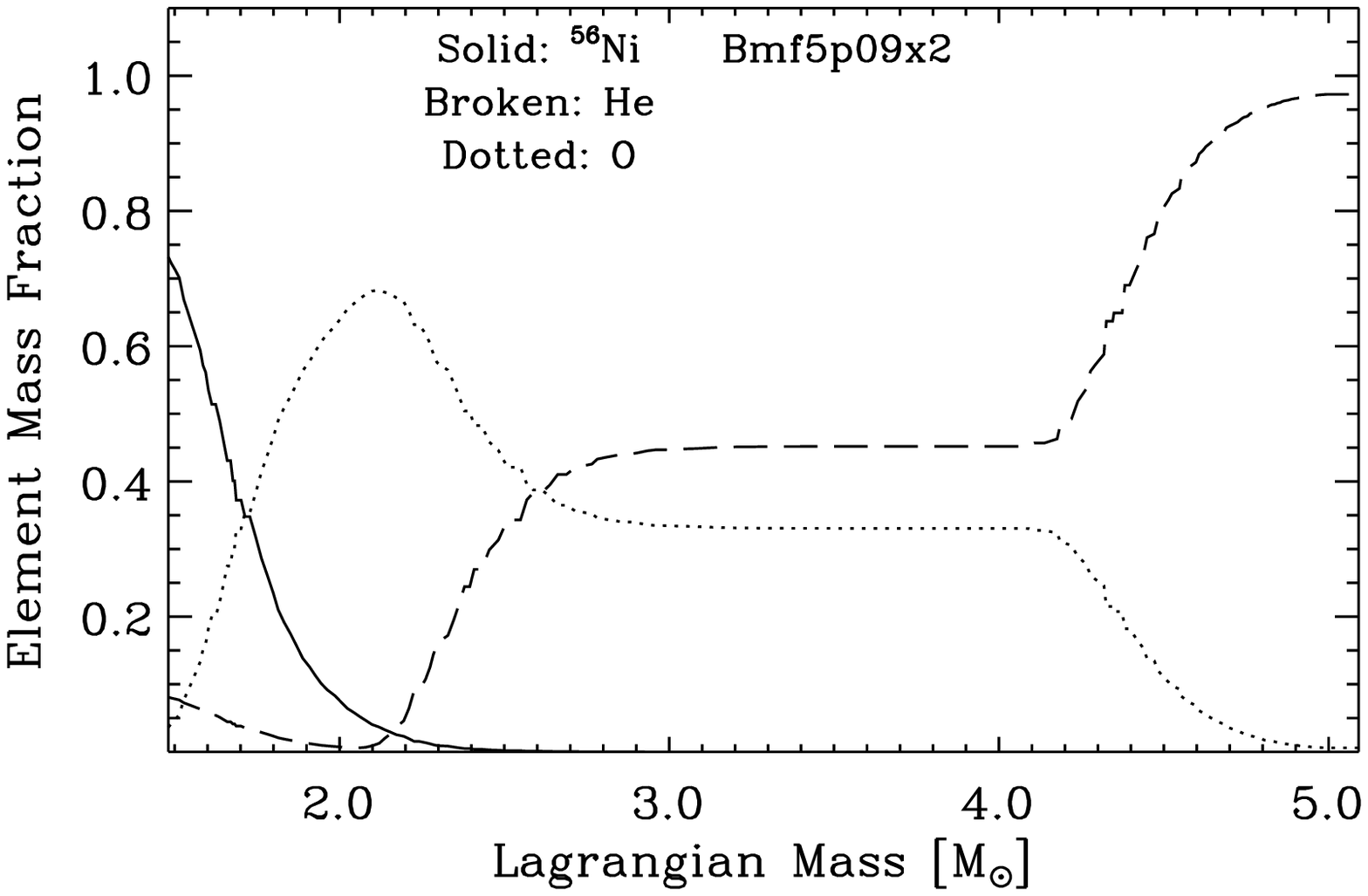,width=8.cm}
\vspace{-0.5cm}
\epsfig{file=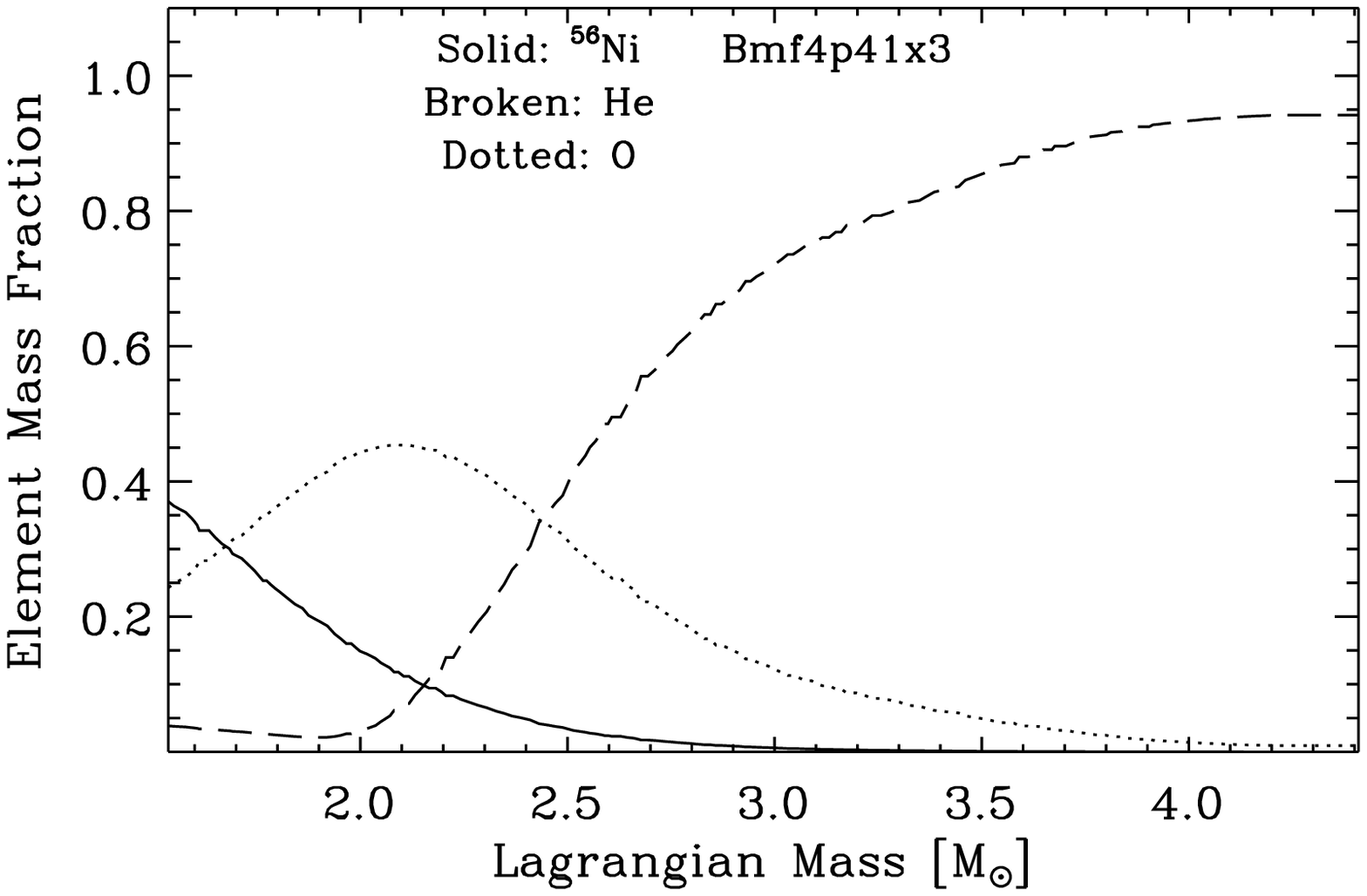,width=8.cm}
\epsfig{file=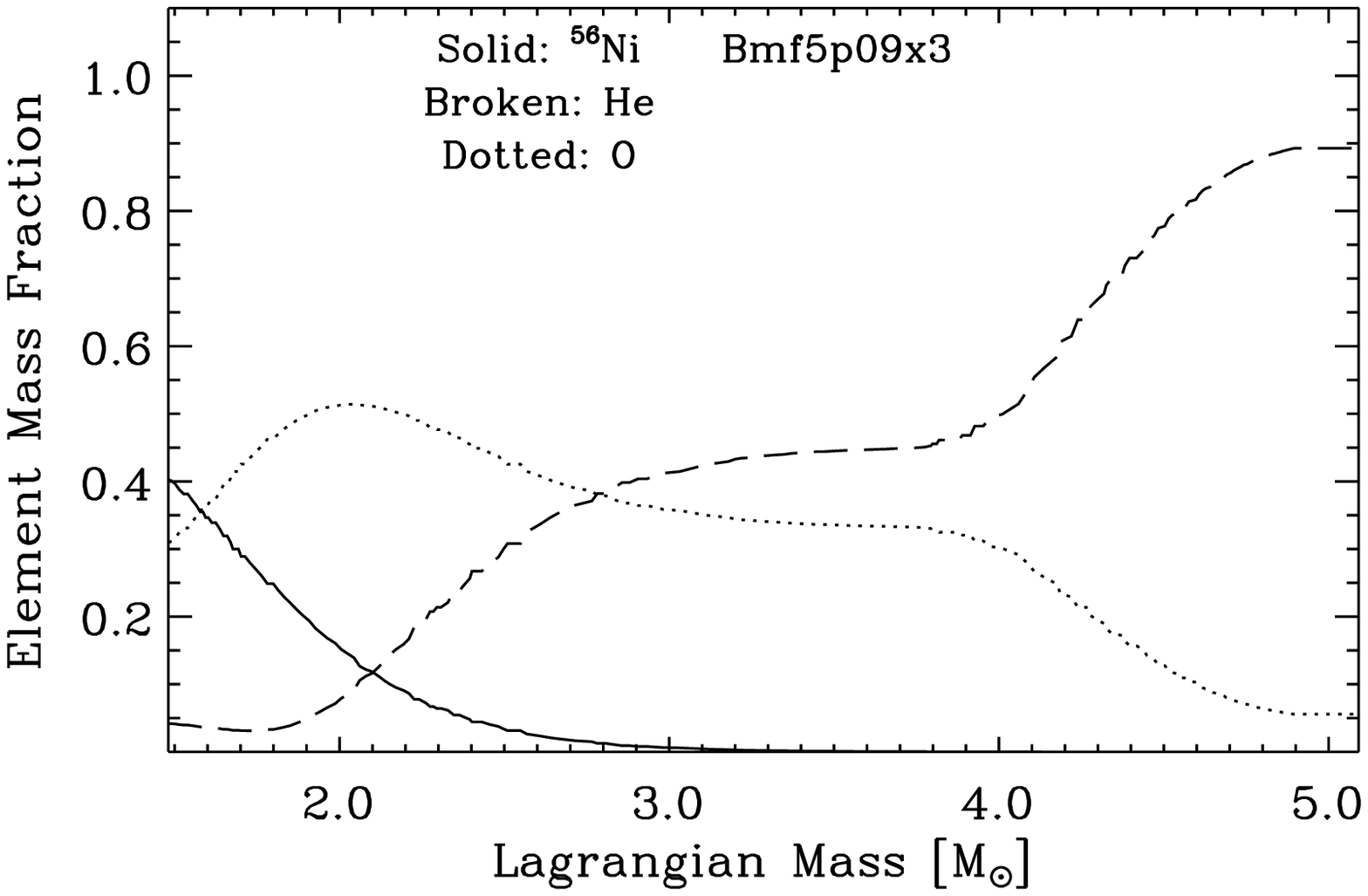,width=8.cm}
\vspace{-0.25cm}
\epsfig{file=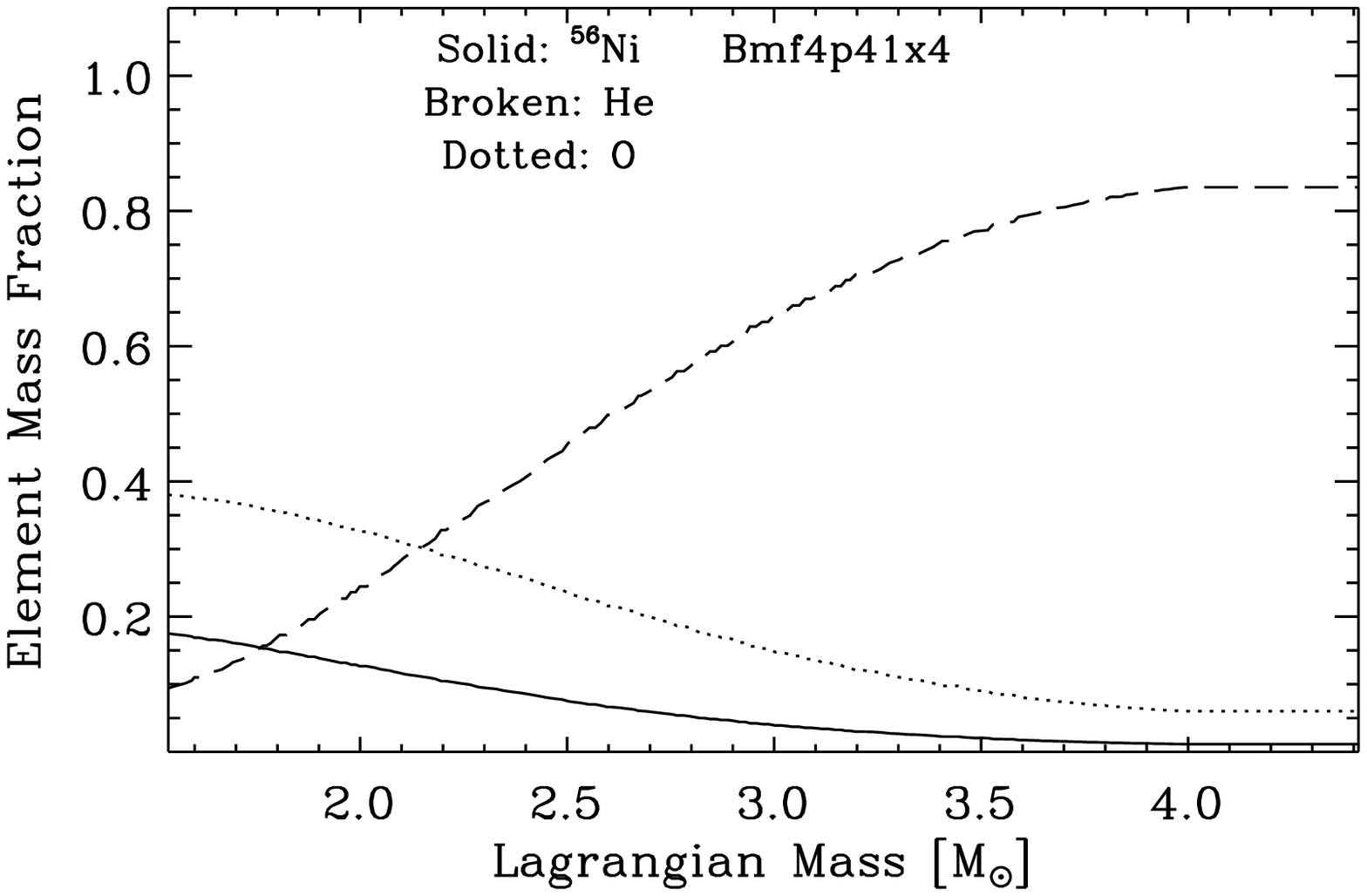,width=8.cm}
\epsfig{file=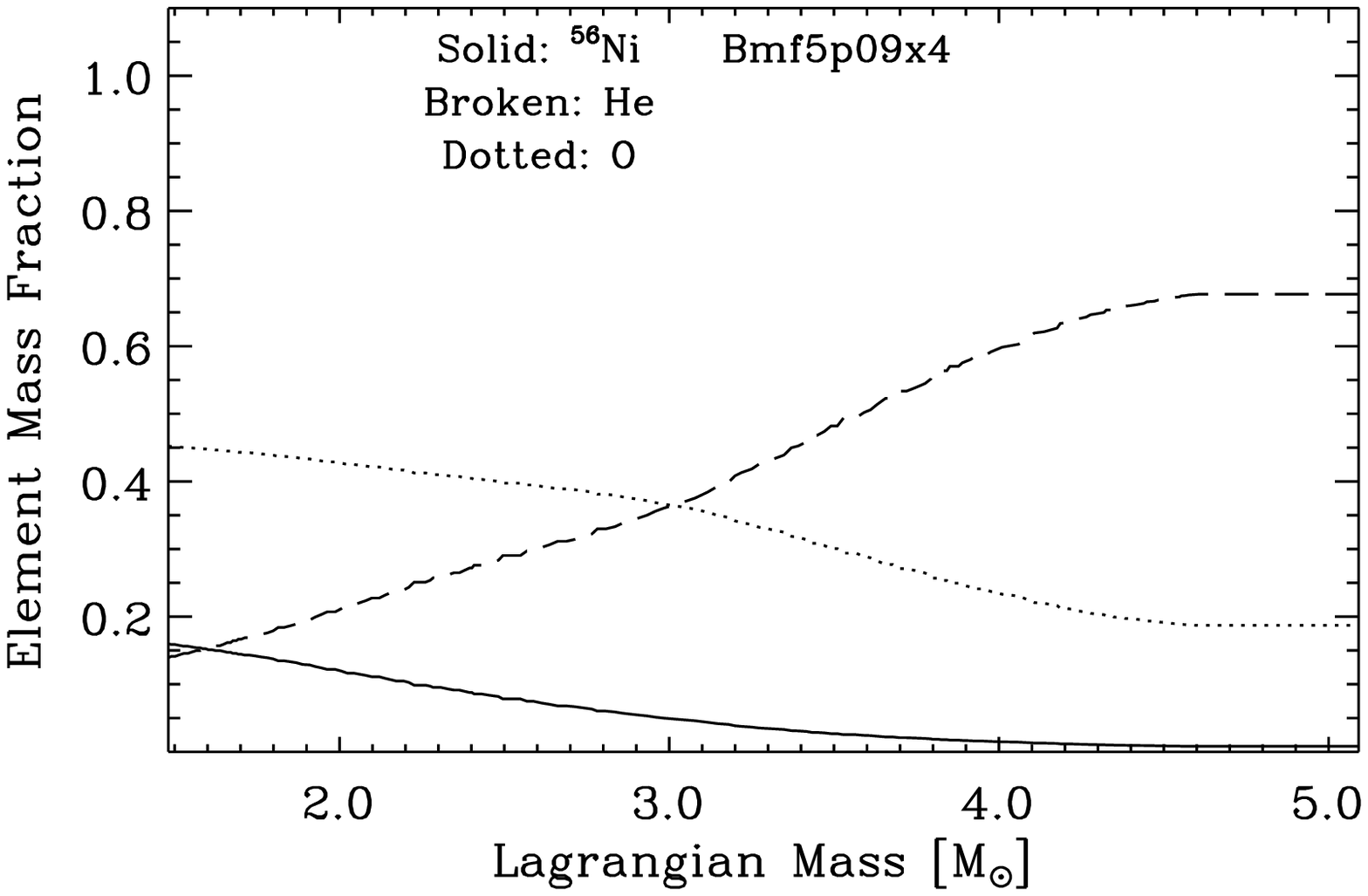,width=8.cm}
\caption{Helium (dashed line), Oxygen (dotted line), and \isoni\ (solid line; we show here the initial distribution unaffected
by decay) mass distribution versus ejecta mass for  Bmf4p41 (left) and Bmf5p09 (right)
model sets. The suffix x1--x4 refers to increasing levels of mixing (see Tables~\ref{tab_comp}--\ref{tab_xfrac}).
In the top row panels, we also add the velocity information.
\label{fig_ejecta_prop}
}
\end{figure*}

\begin{table*}
\begin{center}
\caption[]{Progenitor and ejecta properties for models Bmf4p41x[1--4] and Bmf5p09x[1--4].
In our nomenclature, ``B" stands for binary (evolution), ``mfXpYZ" for the final star mass X.YZ in \msun\ 
(i.e., at the time of explosion). The suffix x1--x4 refers to increasing levels of mixing.
Since mixing affects the ejecta distribution but not the cumulative
ejecta masses of species, the yields are the same within a mixing sequence. The remnant mass is 1.48\,\msun\ (1.5\,\msun) for model
sequence Bmf4p41x[0--4] (Bmf5p09x[0--4]), and all model ejecta have the same kinetic energy of 1.2\,B.
For both series, the pre-SN evolution is computed at solar metallicity.
In model sequence Bmf4p41x[0--4], the inner 4 zones of the {\sc kepler} ejecta are infalling. We trim these regions, and because of the different levels
of mixing, this causes slightly different \isoni\ cumulative masses, in the range 0.157 to 0.177\,\msun.
Note that in the subsequent evolution with {\sc cmfgen}, the steep density variation in the innermost ejecta layers of model Bmf5p09xn
causes numerical diffusion issues, which are worse in cases of weaker mixing.
\label{tab_comp}
}
\begin{tabular}{l@{\hspace{1.6mm}}c@{\hspace{1.6mm}}c@{\hspace{1.6mm}}c@{\hspace{1.6mm}}
c@{\hspace{1.6mm}}c@{\hspace{1.6mm}}c@{\hspace{1.6mm}}c@{\hspace{1.6mm}}
c@{\hspace{1.6mm}}c@{\hspace{1.6mm}}c@{\hspace{1.6mm}}c@{\hspace{1.6mm}}
c@{\hspace{1.6mm}}c@{\hspace{1.6mm}}c@{}}
\hline
Model   &  $M_{\rm i}$  & $M_{\rm f}$ &       $M_{\rm ejecta}$ & $M_{\rm H}$ &$M_{\rm He}$ & $M_{\rm C}$  & $M_{\rm N}$  & $M_{\rm O}$
& $M_{\rm Ne}$ & $M_{\rm Mg}$& $M_{\rm Si}$& $M_{\rm Fe}$ & $M_{^{56}{\rm Ni}}$     \\
        &   [\msun]    & [\msun]    &    [\msun]           &   [\msun]  &   [\msun]  &    [\msun]  &   [\msun]   &   [\msun]
        &   [\msun]   &    [\msun] &[\msun]     & [\msun]     & [\msun]  \\
\hline
Bmf4p41x[0--4]    &     18.0     &   4.41    &  2.91 & 6.88(-3) &     1.63(0) &     1.63(-1) &     8.83(-3) &     5.36(-1) &     1.47(-1) &     6.99(-2) &     1.05(-1) &     6.87(-3) &     1.57(-1) ... 1.77(-1) \\
Bmf5p09x[0--4]    &     25.0     &   5.09    &  3.61 & 1.39(-6) &     1.58(0) &     4.31(-1) &     8.56(-3) &     1.15(0)  &     7.60(-2) &     7.30(-2) &     7.01(-2) &     8.85(-3) &     1.84(-1) ... 2.20(-1) \\
\hline
\end{tabular}
\end{center}
\end{table*}

\begin{table*}
\begin{center}
\caption[]{Same as for Table~\ref{tab_comp}, but now showing the surface composition for each model in each series.
The models with suffix x0 correspond to the unmixed ejecta studied in \citet{dessart_etal_11} and are repeated
here for completeness and reference.
We also quote the ejecta velocity that encompasses 99\% of the $^{56}$Ni mass initially, i.e., $V_{^{56}{\rm Ni}}$, the
peak bolometric luminosity $L_{\rm peak}$ and the time to peak luminosity $t_{\rm peak}$.
\label{tab_xfrac}}
\begin{tabular}{
l@{\hspace{1.6mm}}c@{\hspace{1.6mm}}c@{\hspace{1.6mm}}c@{\hspace{1.6mm}}c@{\hspace{1.6mm}}c@{\hspace{1.6mm}}
c@{\hspace{1.6mm}}c@{\hspace{1.6mm}}c@{\hspace{1.6mm}}c@{\hspace{1.6mm}}c@{\hspace{1.6mm}}c@{\hspace{1.6mm}}
c@{\hspace{1.6mm}}c@{\hspace{1.6mm}}
}
\hline
 Model    & $X_{\rm H}$    &$X_{\rm He}$ & $ X_{\rm C}$ &$ X_{\rm N}$ & $X_{\rm O}$& $X_{\rm Ne}$& $X_{\rm Mg}$ & $X_{\rm Si}$
  & $X_{\rm Fe}$ & $X_{\rm Ni}$  & $V_{^{56}{\rm Ni}}$ & $L_{\rm peak}$ & $t_{\rm peak}$ \\
  & & & & & & & & & & & \kms & $10^8$\lsun & [d] \\
\hline
Bmf4p41x0                &  1.25(-1) &     8.55(-1) &     1.48(-4) &     1.33(-2) &  4.53(-4) & 1.85(-3) &    7.25(-4) &     7.34(-4) &     1.36(-3) & 0.0 & 2541      & 9.02 & 31.65 \\
Bmf4p41x1                &  4.91(-2) &     9.32(-1) &     2.63(-4) &     1.32(-2) &  4.09(-4) & 1.85(-3) &    7.24(-4) &     7.35(-4) &     9.08(-4) & 0.0 & 2755      & 8.37 & 33.27 \\
Bmf4p41x2                &  2.76(-2) &     9.51(-1) &     1.99(-3) &     1.30(-2) &  1.19(-3) & 2.05(-3) &    7.27(-4) &     7.35(-4) &     9.08(-4) & 3.0(-10) & 3040 & 8.27 & 34.37 \\
Bmf4p41x3                &  1.44(-2) &     9.42(-1) &     1.65(-2) &     1.10(-2) &  9.11(-3) & 4.13(-3) &    8.63(-4) &     7.85(-4) &     9.10(-4) & 5.23(-5) & 3680 & 7.95 & 35.93 \\
Bmf4p41x4                &  6.83(-3) &     8.35(-1) &     4.35(-2) &     6.95(-3) &  6.01(-2) & 1.79(-2) &    6.00(-3) &     7.28(-3) &     1.24(-3) & 1.14(-2) & 7500 & 7.61 & 34.20 \\
\hline
Bmf5p09x0                &  6.33(-5) &     9.81(-1) &     2.92(-4) &     1.33(-2) &  3.21(-4) & 1.85(-3)   &    7.23(-4)  &     7.34(-4) &     1.36(-3) & 0.0 & 1230      & 7.88 & 50.46 \\
Bmf5p09x1                &  1.12(-5) &     9.81(-1) &     3.54(-4) &     1.32(-2) &  3.42(-4) & 1.85(-3)   &    7.24(-4)  &     7.35(-4) &     1.36(-3) & 0.0 & 1510      & 7.54 & 52.33 \\
Bmf5p09x2                &  5.81(-6) &     9.72(-1) &     3.68(-3) &     1.30(-2) &  5.98(-3) & 2.11(-3)   &    7.80(-4)  &     7.35(-4) &     1.36(-3) & 0.0 & 1930      & 7.43 & 52.23 \\
Bmf5p09x3                &  2.94(-6) &     8.93(-1) &     3.29(-2) &     1.10(-2) &  5.57(-2) & 4.41(-3)   &    1.28(-3)  &     7.37(-4) &     1.36(-3) & 2.92(-6) & 3100 & 6.91 & 47.20 \\
Bmf5p09x4                &  1.16(-6) &     6.77(-1) &     9.88(-2) &     6.06(-3) &  1.87(-1) & 1.14(-2)   &    5.21(-3)  &     3.62(-3) &     1.36(-3) & 8.24(-3) & 6430 & 7.15 & 36.56 \\
\hline
\end{tabular}
\end{center}
\end{table*}

Using the tangent-ray method to compute optical-depth
integrals along multiple directions \citep{mihalas_78}, we compute the location of the $\tau_\gamma=2/3$ surface as seen from
a selection of heights/radii/velocities in the SN ejecta.\footnote{This calculation uses the model density distribution, interpolated
on a logarithmic radial grid, with 400 radial points to ensure a good resolution in particular for lateral directions.
The tangent-ray method is fast since it does not require any interpolation; all ray quantities are calculated at grid points.
Depending on the location of the \gray\ source, we adjust the location of the outermost grid point to be at 20000 or 40000\,\kms.}
This surface bounds the ejecta volume over which a \gray\ source may cause non-thermal effects.
In Fig.~\ref{fig_gray}, we show results for a  \gray-emission site that is below, at, or above the
\gray\ photosphere.\footnote{This exercise is illustrative and thus the message conveyed is not altered by the somewhat arbitrary choice for $\tau_\gamma$.
In practice, using 2/3 yields the volume over which $\sim$50\% of the \grays\ are absorbed, while non-thermal processes
may be sufficiently strong with only 10\% of such \grays. In that case, the edge of that volume would correspond to $\tau_\gamma = 2.3$
from the emission site and the contours shown in Fig.~\ref{fig_gray} would correspond to 14.5\,d after explosion rather than 26.8\,d.}

For locations below the \gray\ photosphere, the volume of influence is essentially a sphere centered on the emission site.
As the emission site is brought outward nearer the photosphere, this sphere gets stretched on its outer side, bulges somewhat laterally,
while its inner side stays very close to the emission site. As the emission site crosses the photosphere, the $\tau=2/3$ surface breaks open
along the radial direction, reaching all the way to infinity (at this time, \grays\ would be observable along this beam).
In comparison, the opening angle of that surface remains rather
small, and wraps tightly to the emission site at depth. As the emission site is brought further out, the opening
angle broadens, eventually reaching the opposite side of the ejecta for the outermost locations. Even in this situation,
\grays\ do not penetrate significantly the ejecta layers interior to the emission site.

We see that the influence of \grays\ is highly angle dependent, with a strong bias for escape in the radial direction
for locations at and above the photosphere. This results from the contrast in scale height between the radial direction
(i.e., $\sim R/n$ for a power law density distribution with exponent $n$) and the lateral direction (i.e., $\sim 2R/\sqrt{n}$).
At early times, $l_\gamma$ is so small that \grays\ will only influence their immediate vicinity.
Thus to allow non-thermal ionization/excitation of the bulk of the He atoms (preferentially located in the outer ejecta),
the \isoni\ and \isoco\  \gray\ sources  need to be thoroughly mixed with the material in the helium-rich shell.
By thorough mixing, we mean that fingers or clumps of \isoni\ will need to be sufficiently
densely distributed so that the inter-clump distance is on the order of $\l_\gamma$.
At a radial distance $R$,  this translates into an inter-clump angular distance $\Delta \theta \sim \l_\gamma / R$.
For the example shown in Fig.~\ref{fig_gray} at 26.8\,d after explosion, clumps with a thickness $\Delta R$ and
spaced  every $\sim$10$^\circ$ would allow \grays\ to influence the entire shell at $R$.
Strongly asymmetric explosions propelling \isoni\ nuclei to large velocities in a ``jet" or a strong bipolar/unipolar explosion
are not attractive alternatives since the non-thermal processes would only affect a restricted volume of the SN ejecta --
primarily the high-velocity material associated with the ``jet'' that does not contribute much to the bulk of the SN radiation.

\section{Numerical approach and setup}
\label{sect_model}

 The numerical procedure is the same as that used in our recent simulations on SN 1987A \citep{DH10},
 SNe II-Plateau \citep{DH11}, and SNe IIb/Ib/Ic \citep{dessart_etal_11}, and will thus be briefly summarized below.
 A full description of the code is presented in \citet{HD12}.
 The first distinction with those works is that instead of adopting a unique ejecta composition structure, we explore different
 stratifications produced by different levels of mixing.
 Secondly, we now treat explicitly the effects of non-thermal excitation and ionization --- in
 our former studies the entire decay energy was treated as heat \citep{li_etal_12}.

\subsection{Pre-SN evolution and explosion models}

In this study, we employ a subset of the models discussed in \citet{dessart_etal_11}, whose pre-SN evolution was described
by \citet[see also Woosley et al., in prep]{yoon_etal_10}.\footnote{In future work, we will investigate a broader range of
progenitors, including single rotating and non-rotating WR stars \citep{georgy_etal_12}, and wider interacting binaries
\citep{claeys_etal_11}.}
These two models corresponds to their models 21 and 28. Namely,  we focus on the binary-star models Bmi18mf4p41z1
and Bmf25mi5p09z1, which result from the evolution of an 18 and an 25\,\msun\ star in a binary system at solar metallicity,
with, at the time of explosion, a mass of 4.41 and 5.09\,\msun, and a stellar radius of 12.3 and 4.35\,\rsun, respectively. 
These two models were exploded with {\sc kepler}
\citep{weaver_etal_78} by driving a piston at the iron-core edge to yield an asymptotic ejecta kinetic energy of 1.2\,B.
Explosive burning was computed from first principles with an approximate 13-isotope network that includes \isoni\
(for simplicity, \isoni\ is also the only unstable isotope that we consider in this work).
In \citet{dessart_etal_11}, the ejecta were unmixed, i.e., the composition was taken directly from the results of stellar evolution
as modified by the explosion.
The radiative-transfer simulations were started with {\sc cmfgen} from such hydrodynamical inputs evolved to 1\,d after explosion.
To be more concise, we adjust our former model nomenclature and assign to these two ejecta the root-names Bmf4p41 and Bmf5p09.

In this work, we wish to explore the effects of mixing and non-thermal processes on SN radiation. 
We selected these two models from the study of \citet{dessart_etal_11} to investigate the role of 
non-thermal processes in helium-rich low-mass cores with and without hydrogen at their surface, thus covering 
potential SN types of IIb/IIc and Ib/Ic.
All numerical codes used
here are 1D, i.e., the stellar evolution is done in 1D, as is the radiation-hydrodynamics of the explosion, and the subsequent
detailed radiative-transfer modelling of the ejecta gas and radiation. So, to mimic the multi-dimensional effects that may be at the origin
of mixing in CCSN ejecta (see Section~\ref{sect_gray}), we run a ``boxcar'' through the unmixed ejecta models Bmf4p41x0 and
Bmf5p09x0, and increase the efficiency of mixing by increasing the mass range over which the material is mixed.
This mixing is applied to the original model shortly after explosive burning ceases so that the evolution to 1\,d
properly accounts for the heating associated with this ejecta stratification (in particular for the \isoni\ heat source).
We produce Bmf4p41 and Bmf5p09 models that cover unmixed (suffix x0), weakly mixed (x1 and x2), moderately mixed (x3)
and strongly mixed (x4) configurations. Mixing does not change the cumulative yields of models Bmf4p41x0 and Bmf5p09x0, shown
in Table~\ref{tab_comp}, but it gradually modifies the chemical stratification, shown for each model in Fig.~\ref{fig_ejecta_prop}.
Table~\ref{tab_xfrac} gives additional information on the surface composition for each model.\footnote{Since the mixing
is performed over a specified mass range, it can affect all the way to the surface so a more desirable mixing approach
would be to use a velocity criterion when setting the range of mass shells that are homogenized.}

Mixing softens composition gradients. In the most mixed models, the ejecta is nearly homogeneous --- any species
abundant at one depth is present at all depths.
For stars of increasing mass, a greater mass buffer exists between the ejecta base and the helium-rich outer shells.
Progressing outwards from the innermost ejecta layers, one encounters an oxygen-rich helium-deficient shell (here bounded between
1.8 and 2.5\,\msun), followed by a shell with a composition shared between He, C and O. In our models, 
only the outermost layers ($\lesssim$0.5\,\msun\ below the surface) are nearly pure helium (with a mass fraction that depends on the level of mixing).
In model Bmf4p41x0-3, hydrogen is also present with a mass fraction
in excess of 0.01, which is sufficient to produce H$\alpha$ emission at early times \citep{dessart_etal_11}.

The velocity of the ejecta shell that bounds 99\% of the total \isoni\ mass is a good indicator of the magnitude of mixing in our models.
This velocity varies from $\sim$1500\,\kms\ up to $\lesssim$8000\,\kms\ (Table~\ref{tab_xfrac}). Note that the bulk of the mass of such unstable
isotopes is still located at small velocities because the mixing, which affects the profile of species' mass fractions, leaves the steeply-rising
density distribution at depth unaffected.

\subsection{non-LTE time-dependent radiative-transfer including non-thermal processes}

The critical improvement in this work over all past studies of SNe Ibc explosions
is the simultaneous computation of the SN multi-color \LC\ and spectra with allowance for line-blanketing,
non-LTE, time-dependence in the
statistical, energy, and radiative-transfer equations, and non-thermal processes associated with high-energy electrons Compton
scattered by the \grays\ arising from radioactive decay. All ingredients but the last have been described in detail in
\citet{DH10,DH11,dessart_etal_11} and \cite{HD12}. 
In these former works and in the present study, time dependence is considered both for the treatment of the 
radiation field (i.e., $DJ/Dt$ and $DH/Dt$ terms in the 0th and 1st moment of the radiative-transfer equation) 
and of the gas ($De/Dt$ and $Dn/Dt$ terms appearing in the 
energy  and the statistical equilibrium equations; \citealt{DH08_time} ; we thus naturally 
account for time-dependent ionization).
The new addition is the treatment of non-thermal processes through a solution of the Spencer-Fano equation 
\citep{spencer_fano_54} using the approach presented in \citet{KF92}.
Assuming a \gray-energy deposition profile,
we inject the corresponding energy in the form of electrons at 500\,eV and compute the {\it local} degradation of such
electrons as they collide with electrons and ions/atoms in the plasma.
Once this degradation function is known, we can determine how this deposited energy is channelled into
the various forms of heat, ionization and excitation. The non-thermal terms associated with ionization and excitation
are entered as rates in the statistical-equilibrium equations and are thus fully integrated in the {\sc cmfgen} solution
to the non-LTE time-dependent ionization and level populations.
A full description of the technique and an in-depth analysis of non-thermal effects in core-collapse SNe 
is presented in a companion paper \citep{li_etal_12}. Here, we will limit the discussion of non-thermal
effects relevant to understanding the formation of He\one\ lines (Section~\ref{sect_nonth}) in SNe Ibc.

In practice, the non-thermal solver in {\sc cmfgen} is only used if the mass fraction of unstable isotopes at the photosphere
is indeed sizeable, typically above 0.0001. This was decided for each model by switching on the non-thermal processes
at selected times and checking if this produced any effect on the gas properties and/or on the SN radiation. As discussed
in Section~\ref{sect_gray}, $l_\gamma$ in the \isoni-rich regions is small at early times in all but the most mixed
models (suffix x4), so we switch on non-thermal processes at the start of the time sequence only in the x4 series.
To save space and since the unmixed models have already been discussed in \citet{dessart_etal_11}, we do not show
results for models Bmf4p41x0 and Bmf5p09x0;  models Bmf4p41x1 and Bmf5p09x1 behave very similarly to them and are thus
used as substitutes.

For the \gray-energy deposition, we can either assume it is local or we can compute its distribution in the ejecta
using a \gray\ Monte-Carlo transport code we developed \citep{HD12}.
We find that the \gray-energy deposition is essentially local up to about 30\,d after the \LC\ peak in all models.
Because we focus on the photospheric-phase evolution all simulations shown here assume a local energy deposition.
Non-local energy deposition is not crucial for the non-thermal effects discussed here since it merely stretches the deposition profile
outward, with an enhanced escape at large velocity. It does, however, change the amount of energy trapped in the ejecta and
thus the luminosity decline rate during the nebular phase.

The model atoms adopted for all simulations in this work are identical to that used in \citet{DH11}.
The completeness of the model atoms influences the spectral appearance but it
does not alter the He\one-line strength or the non-thermal effects we observe. The main impact of increasing the size of
the model atoms in our simulations, in particular for metals, is to enhance the magnitude of line blanketing,
which tends to make the spectral-energy distribution (SED) redder. This effect is generally weaker than obtained here
through variations in composition caused by mixing (i.e., \isoni, C/O versus He etc.), although it noticeably alters the colors
after the light-curve peak. Since atomic data represents an important part component of the work we do, we wish to
provide the sources and references for the data we use.
The sources of atomic data are varied, and in many cases multiple data sets for a given ion are available.
In some cases these multiple data sets represent an evolution in data quality and/or quantity, while in
other cases they represent different sources and/or computational methods. Comparisons of models
calculated with different data sets and atomic models potentially provide insights into the sensitivity
of model results to the adopted atomic atoms and models (although such calculations have yet to be
undertaken for SN). Oscillator strengths for CNO elements were originally taken from
\citet{NS83_LTDR, NS84_CNO_LTDR}. These authors also provide transition probabilities to states in the ion
continuum. The largest source of oscillator data is from \citet{Kur09_ATD,Kur_web}; its principal advantage
over many other sources (e.g., Opacity Project) is that LS coupling is not assumed. More recently, non-LS
oscillator strengths have become available through the Iron Project \citep{HBE93_IP}, and work done by
the atomic-data group at Ohio State University \citep{Nahar_OSU}. Other important
sources of radiative data for Fe include \citet{BB92_FeV,  BB95_FeVI, BB95_FeIV}, \cite{Nahar95_FeII}.
Energy levels have generally been obtained from NIST. Collisional data is sparse, particularly for
states far from the ground state. The principal source for collisional data among low lying states
for a variety of species is the tabulation by \citet{Men83_col}; other sources include
\citet{BBD85_col}, \citet{LDH85_CII_col}, \citet{LB94_N2}, \citet{SL74},
\citet{T97_SII_col,T97_SIII_col}, Zhang \& Pradhan (\citeyear{ZP95_FeII_col,ZP95_FeIII_col,ZP97_FeIV_col}).
Photoionization data is taken from the Opacity Project \citep{Sea87_OP} and the  Iron Project
\citep{HBE93_IP}. Unfortunately Ni and Co photoionization data is generally unavailable,
and we have utilized crude approximations. Charge exchange cross-sections are from the tabulation
by \citet{KF96_chg}. 

Finally, we have identified a numerical-diffusion problem which arises from the repeated remappings performed at each step
in a time sequence. These remappings are necessary to adjust the grid onto the optical depth scale, which changes due to
expansion and recombination/re-ionization. Unfortunately, the 1D hydrodynamical inputs we employ have strong composition
and density gradients which are smoothed by such remappings. Since the model at a new time step uses the
model at the previous time step, these remappings produce systematic shifts which become continually worse as we evolve the
models to later time. This alters the total mass of some species by 10\%, but by as much as 30\% for \isoni\ in unmixed models
because \isoni\ is most abundant in a narrow shell at the ejecta base where the density steeply rises. While this does not impact
the discussion on non-thermal effects, it induces a slight shift between model \LCS\ (most visible at \LC\ peak). In future modeling
this problem, at least for models where the velocity of a given mass shell is constant, will be rectified by interpolating the density
structures from the original hydrodynamic model grid.
Radiation-hydrodynamics codes do not suffer from this shortcoming because they use a high-resolution Lagrangian mass grid,
but it is then the poor resolution of the optical-depth scale that affects their results.

\begin{figure}
\epsfig{file=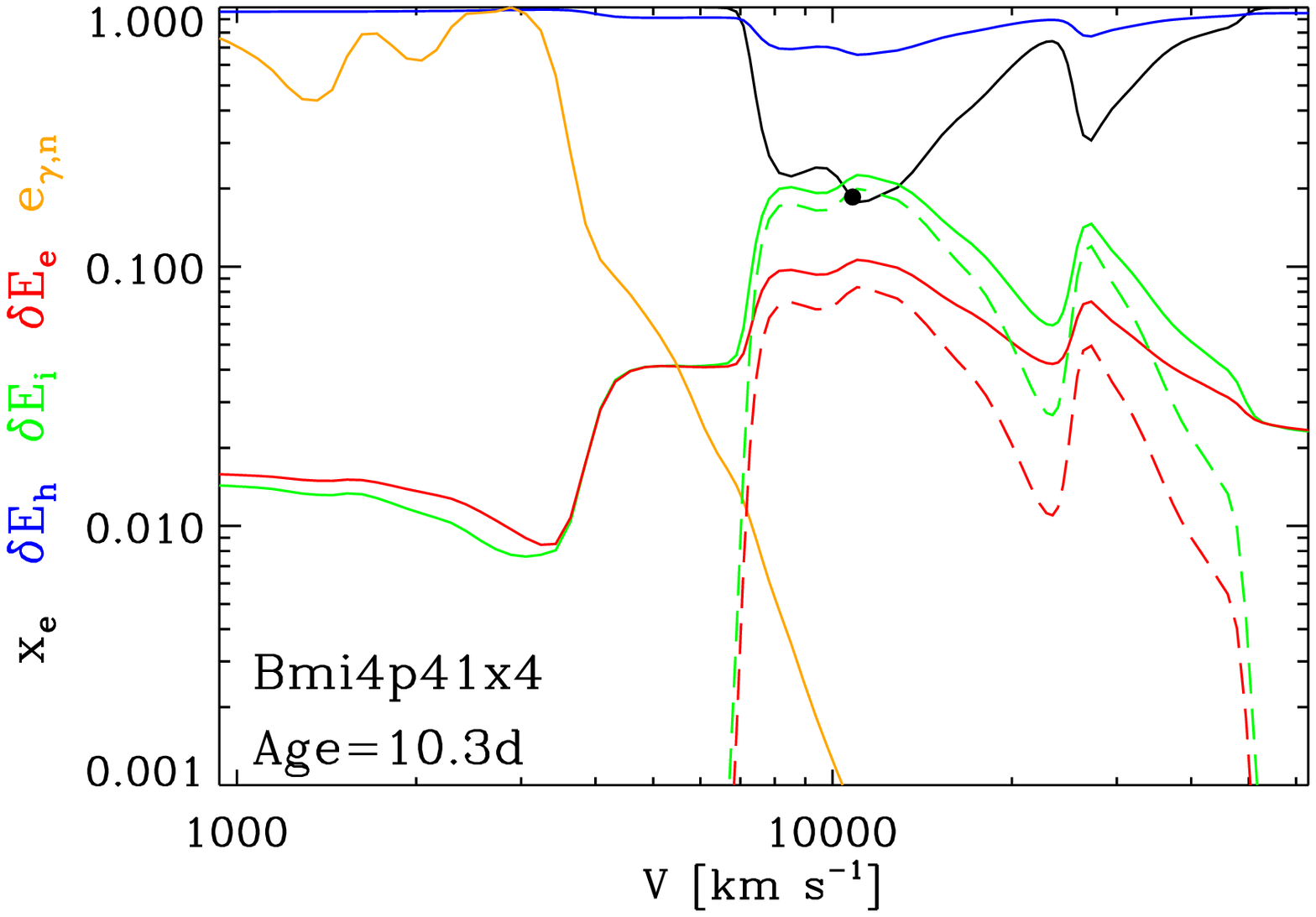,width=8.5cm}
\epsfig{file=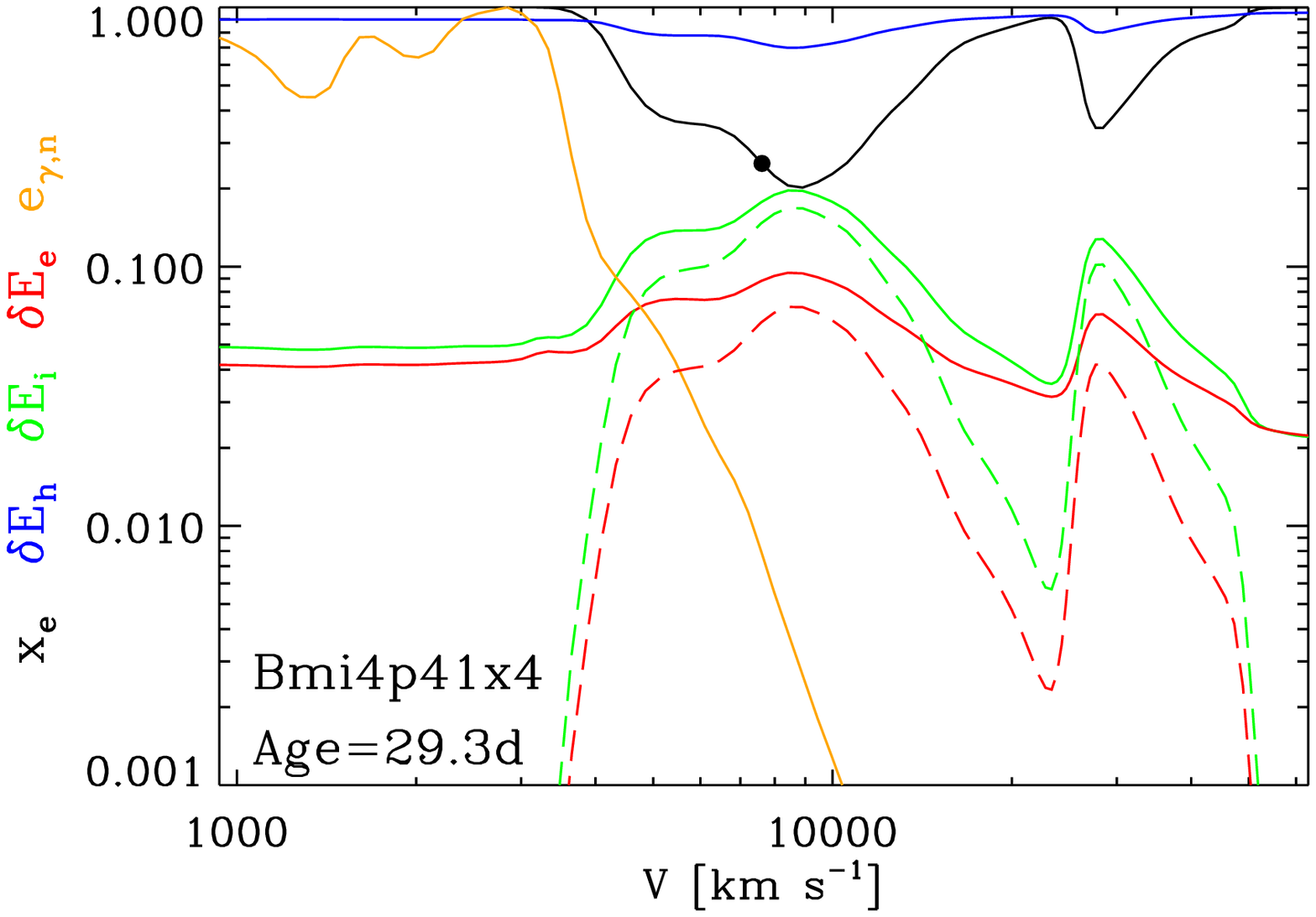,width=8.5cm}
\epsfig{file=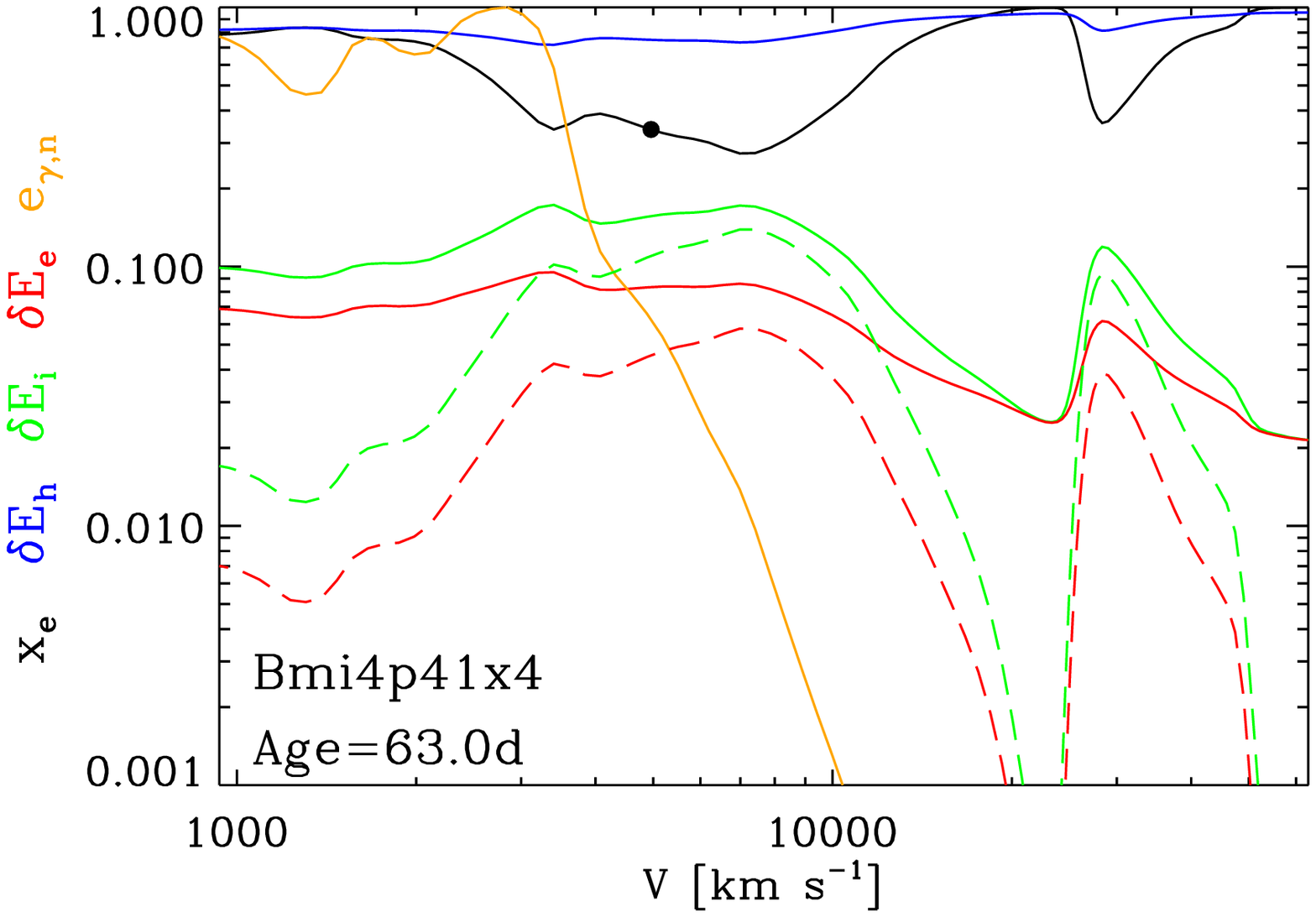,width=8.5cm}
\caption{Illustration of the fractional energies going into heat $\delta E_h$ (blue), ionization $\delta E_i$ (green), and
excitation $\delta E_e$ (red) as a function of ejecta velocity and shown at 10.3 (top), 29.3 (middle), and  63.0\,d (bottom)
after explosion in model Bmf4p41x4.
The fractional energies shown with solid lines denote the total contributions in each energy channel, while the corresponding
dashed lines give the ionization and excitation associated with He\one\ only.
We also show the ejecta ionization fraction $x_e$ (black), which we take here as the ratio of the total ion population to the combined total
atom and ion populations, and the normalised decay-energy deposition profile at the time ($e_{\gamma,{\rm n}}$,
in units of erg\,cm$^{-3}$\,s$^{-1}$, and normalized to its maximum value; orange).
In each plot, the black dot represents the location where the inward-integrated Rosseland-mean optical depth is 2/3.
\label{fig_nonth}}
\end{figure}

\begin{figure}
\epsfig{file=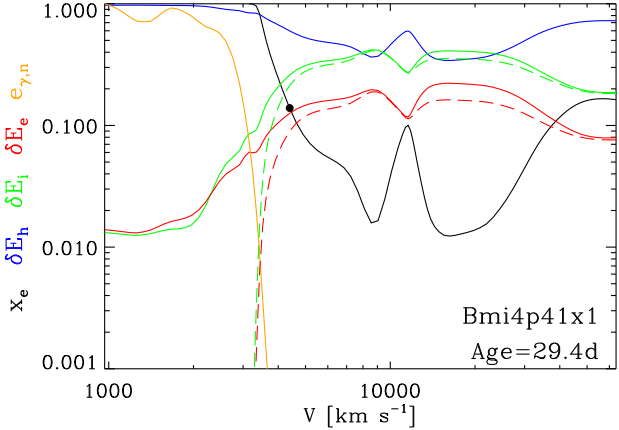,width=8.5cm}
\caption{Same as Fig.~\ref{fig_nonth}, but now for the very weakly mixed model Bmf4p41x1 at 29.4\,d after explosion.
The relative fractions of the energy deposited by radioactive decay are comparable to those obtained for model
Bmf4p41x4 at the same time (middle panel of Fig.~\ref{fig_nonth}), although the lower ejecta ionization enhances somewhat
the contributions to ionization and excitation (which primarily go to He\one). However, the magnitude of these non-thermal
processes is completely dwarfed at and above the photosphere by the very steeply declining energy deposition profile
(orange). In fact, most of the decay energy is deposited in regions that are ionized, located below the photosphere
(black dot), so that the bulk of the decay energy goes into heat in this unmixed model.
\label{fig_nonth_unmixed}}
\end{figure}

\section{Thermal and non-thermal effects associated with \gray\ energy deposition}
\label{sect_nonth}

 Before presenting \LCS\ and synthetic spectra obtained for all model sequences,
 we first describe the effects associated with \gray\ energy deposition
 in two representative models -- one with strong mixing (Bmf4p41x4) and one with  weak mixing (Bmf4p41x1).
 These two models are characterised by the same total mass, the same cumulative
 yields (i.e., same total mass of individual species), the same density structure, and the same explosion energy (or expansion rate).
 Hence both models have, at a given post-explosion time, the same mass-density structure (versus radius or velocity).
 The differences between them arise directly from differences in chemical stratification and in particular how the
 different \isoni\ distributions influence the heating/ionization/excitation of the gas. Further, mixing
 impacts the opacity and emissivity of the gas through changes in the chemical abundances.

  Using the highly mixed model Bmf4p41x4, we show in Fig.~\ref{fig_nonth} the ionization fraction ($x_e$; black), the fractional
  energies going into heat $\delta E_h$ (blue), ionization $\delta E_i$ (green), and excitation $\delta E_e$ (red), and the normalised
  energy deposition rate ($e_{\gamma,{\rm n}}$; orange)  versus ejecta velocity at 10.3 (top), 29.3 (middle), and  63.0\,d (bottom)
  after explosion. To gauge the potential effects of non-thermal electrons on the continuum and line formation regions, which are located
  at and above the optical photosphere, we show this location in each panel with a black dot. We find that at all three epochs,
  most of the energy originally in non-thermal electrons gets degraded into heat.
  Hence, even if injected non-thermally, most of this energy ends up thermal here and directly influences the \LC.
  Near the photosphere, 10\% of the total deposited energy goes into ionization and about half that goes into excitation.
    Because helium is the most abundant species throughout most of the ejecta, it is He\one\ that benefits the most from
  this non-thermal ionization and excitation (dashed curves).

  In a complicated fully-coupled non-LTE situation, there is no strict separation between these various heat, ionization,
  and excitation contributions.
  For example, heat can indirectly cause photo-ionization from a thermal blackbody and collisional ionization with thermal electrons
  at large optical depth where the material is close to LTE. Furthermore, over-excitation of upper levels in an atom/ion
  will facilitate photo-ionization and thus indirectly influence the ionization state of the gas. These complications justify a
  full non-LTE treatment so that all important rates are allowed for in the determination of the level populations, the electron
  density, and the gas temperature. This treatment must also allow for time-dependent terms in the statistical equilibrium
  equations since these are known to influence the ionization state of the gas \citep{DH08_time}, and thus the fate
  of non-thermal electron energy.

  The {\it magnitude} of the effects associated with each channel scales with the actual energy deposited by radioactive decay.
  In the highly mixed model (Fig.~\ref{fig_ejecta_prop}), the decay energy is deposited not only in the inner ejecta
  but also in high-velocity regions at and beyond the photosphere. 
  This extra thermal and ionization energy causes the SN to brighten and peak earlier compared
  to unmixed models (see Section~\ref{sect_results_LC}). As we shall see in Section~\ref{sect_results_spec},
  it also leads to a sizable strengthening of He\one\ lines in this model series.

  In the weakly mixed model Bmf4p41x1 (shown at 29.4\,d only),
  the actual channelling of non-thermal energy into heat, ionization, and excitation, is very
  similar (Fig.~\ref{fig_nonth_unmixed}). The magnitude and distribution of the energy deposited is, however, drastically different.
  It is negligible at, and above, the photosphere, since the bulk of the \isoni\ is confined to the inner ejecta layers located
  below 2800\,\kms. Since models  Bmf4p41x1 and Bmf4p41x4 have the same amount of \isoni\ at a given post-explosion time, the temperature in
  those inner layers is much higher than in the contemporaneous model Bmf4p41x4. As a consequence, the bulk of the decay
  energy is deposited below the photosphere, in deep regions that at this time are ionized.
  Consequently, the bulk of the injected non-thermal energy not only ends up as thermal energy,
  but it also fails to influence non-thermally the line and continuum formation regions.

  Non-thermal ionization and excitation are thus favored when the decay energy is deposited in regions of low ionization, which also
  tend to be external to the photosphere. Mixing is thus critical to propel \isoni\ at large velocities (and in all
  directions; see Section~\ref{sect_gray}), in regions
  that on a short time scale will be above the photosphere. In unmixed ejecta, heating from the \isoni\ at depth
  causes a strong and delayed heat wave that causes the photosphere to be external to the deep-seated \isoni-rich regions
  \citep{dessart_etal_11}. As an aside, the dependency of non-thermal effects to ionization suggests that they
  may not be paramount in SNe Ia ejecta since the large \isoni\ mass causes a large ejecta ionization at all times.

  Interestingly, we find that the strong mixing adopted in models Bmf4p41x4 and Bmf5p09x4 actually quenches
  the impact of non-thermal ionization/excitation on He\one\ in the outer ejecta. Indeed, energy deposited at large 
  velocities causes heating of these
  layers, which over time and through the ionization freeze-out caused by time-dependent effects \citep{DH08_time},
  cause helium to be fully ionized in those outer regions (the ion that gets the largest share of non-thermal ionization
  and excitation energy at large velocities is indeed He\two\ in Fig.~\ref{fig_nonth}).
  In this context, a lower level of mixing with allowance for
  non-local energy deposition could have allowed helium to recombine, and thus a stronger non-thermal excitation
  of He\one\ atoms. In the future, we will need to investigate ways of mimicking as close as possible the
  mixing properties of multi-dimensional explosion simulations.

\begin{figure*}
\epsfig{file=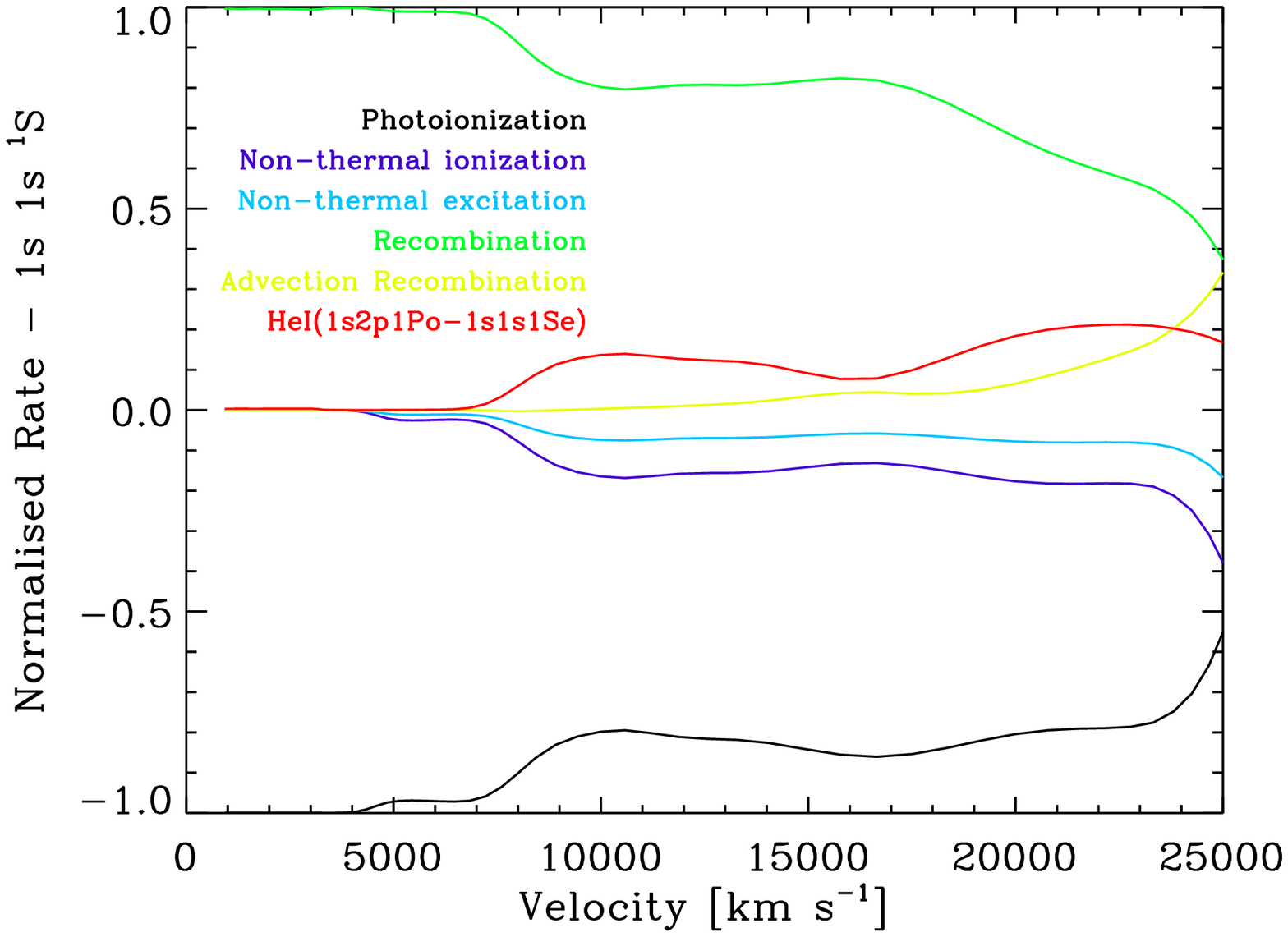,width=7.0cm}
\epsfig{file=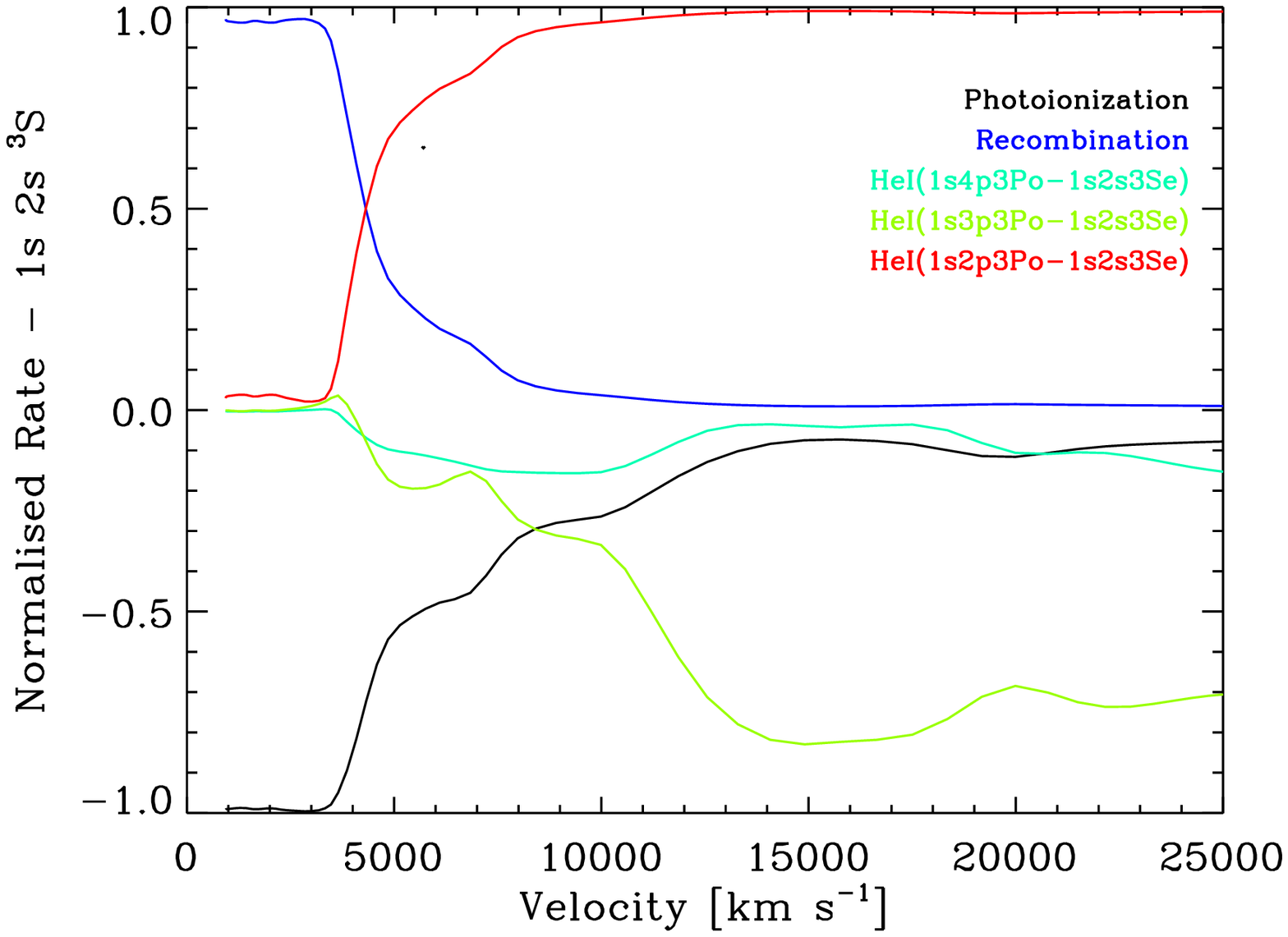,width=7.0cm}
\epsfig{file=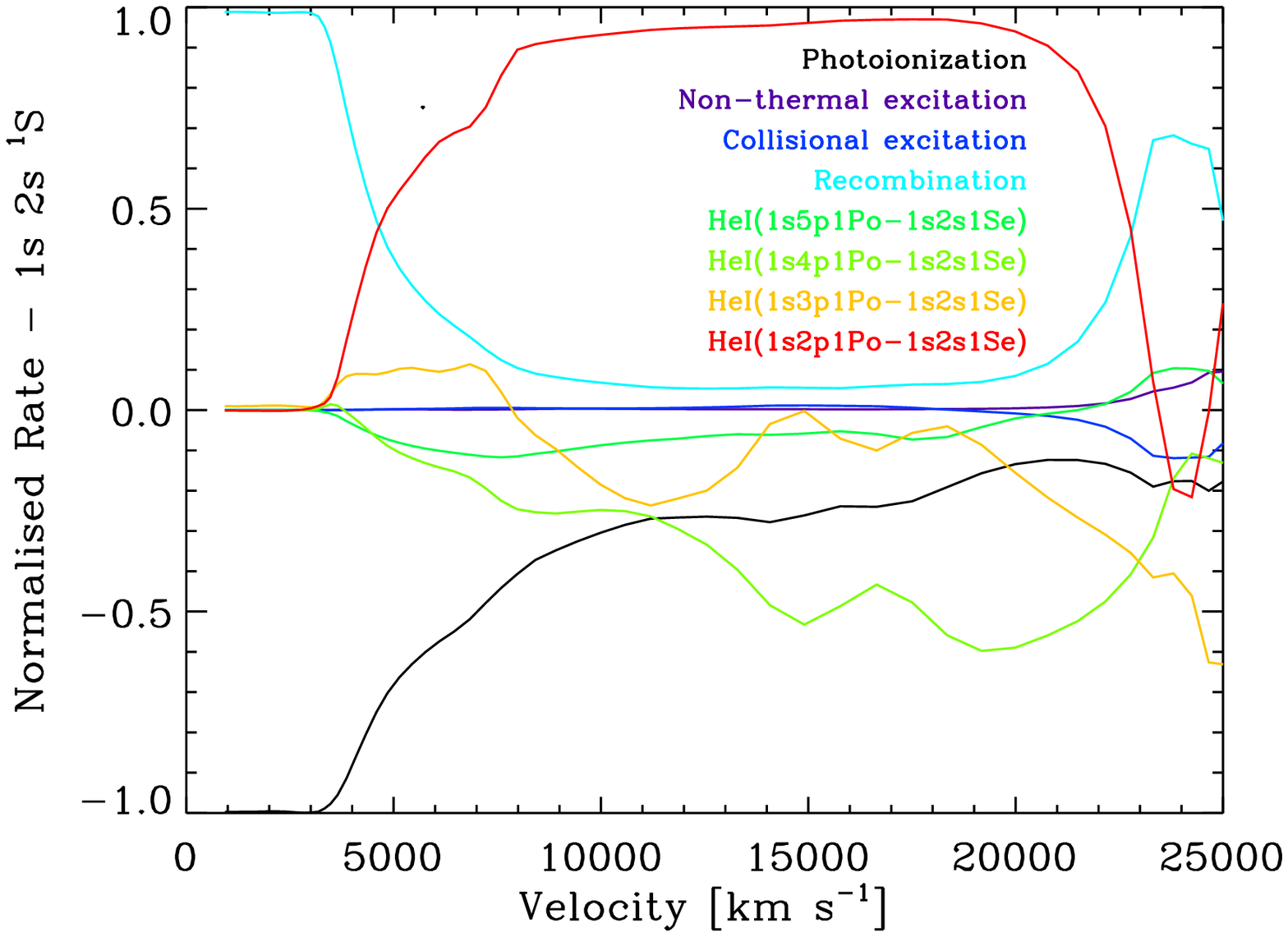,width=7.0cm}
\epsfig{file=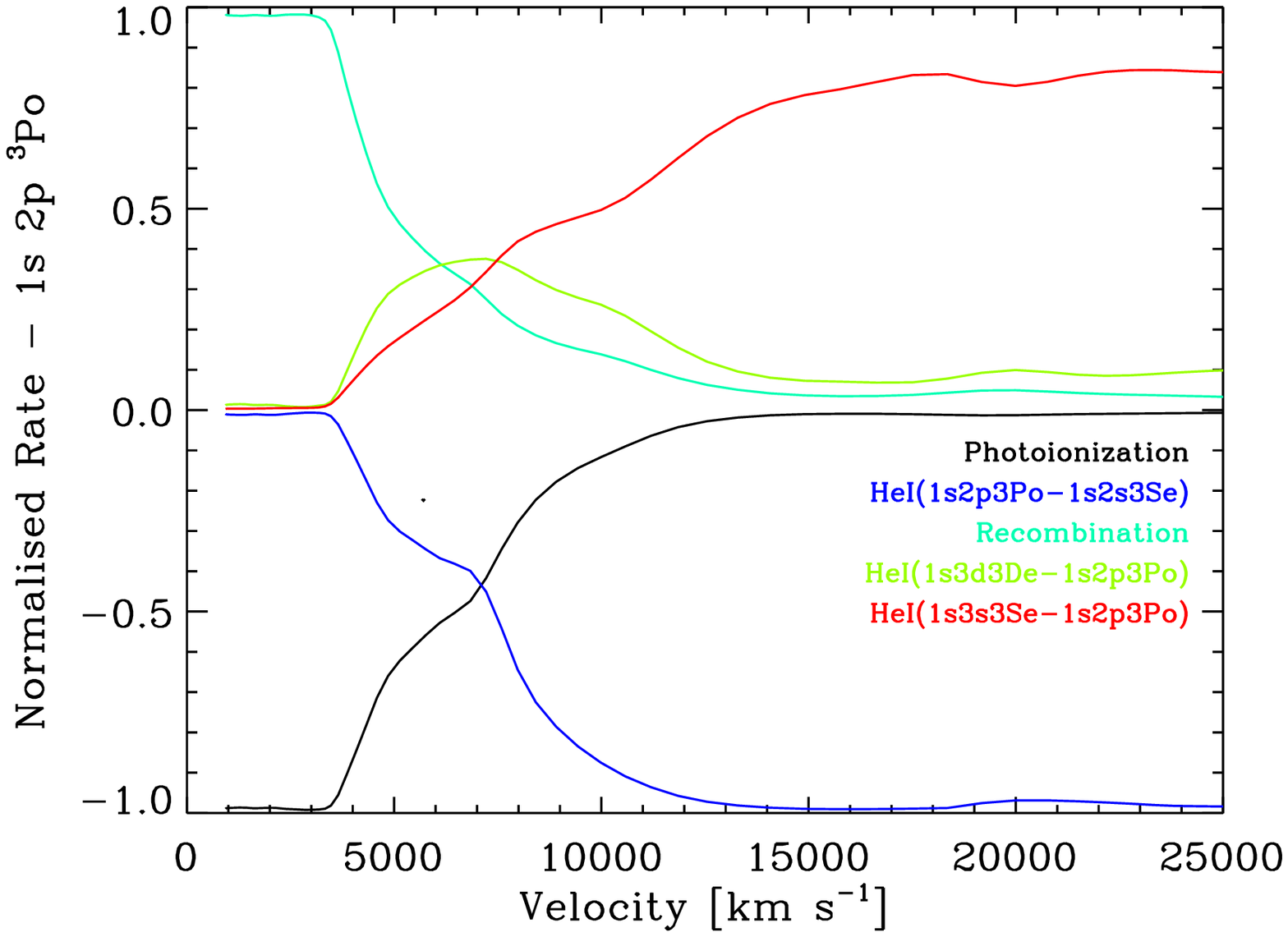,width=7.0cm}
\epsfig{file=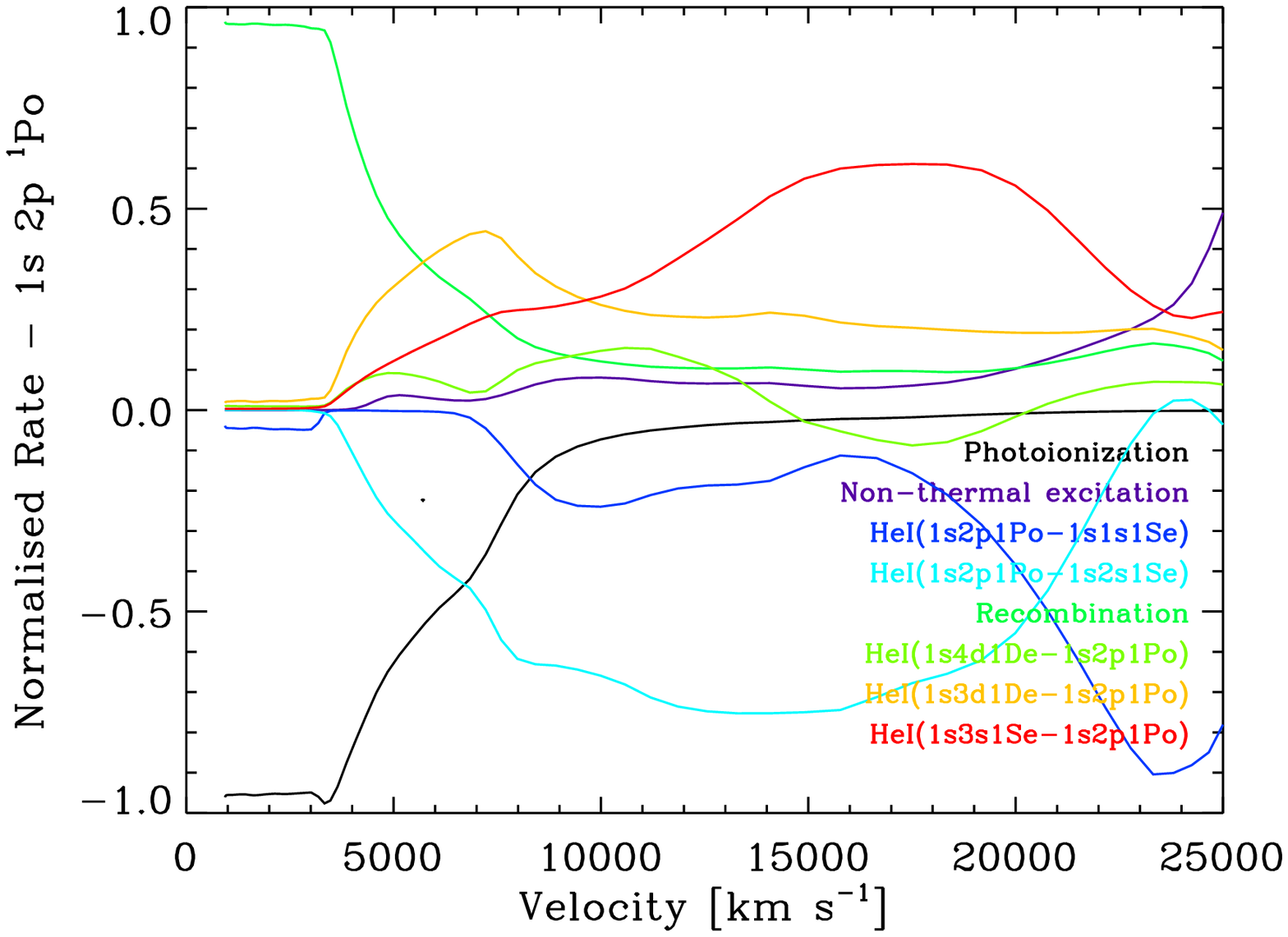,width=7.0cm}
\epsfig{file=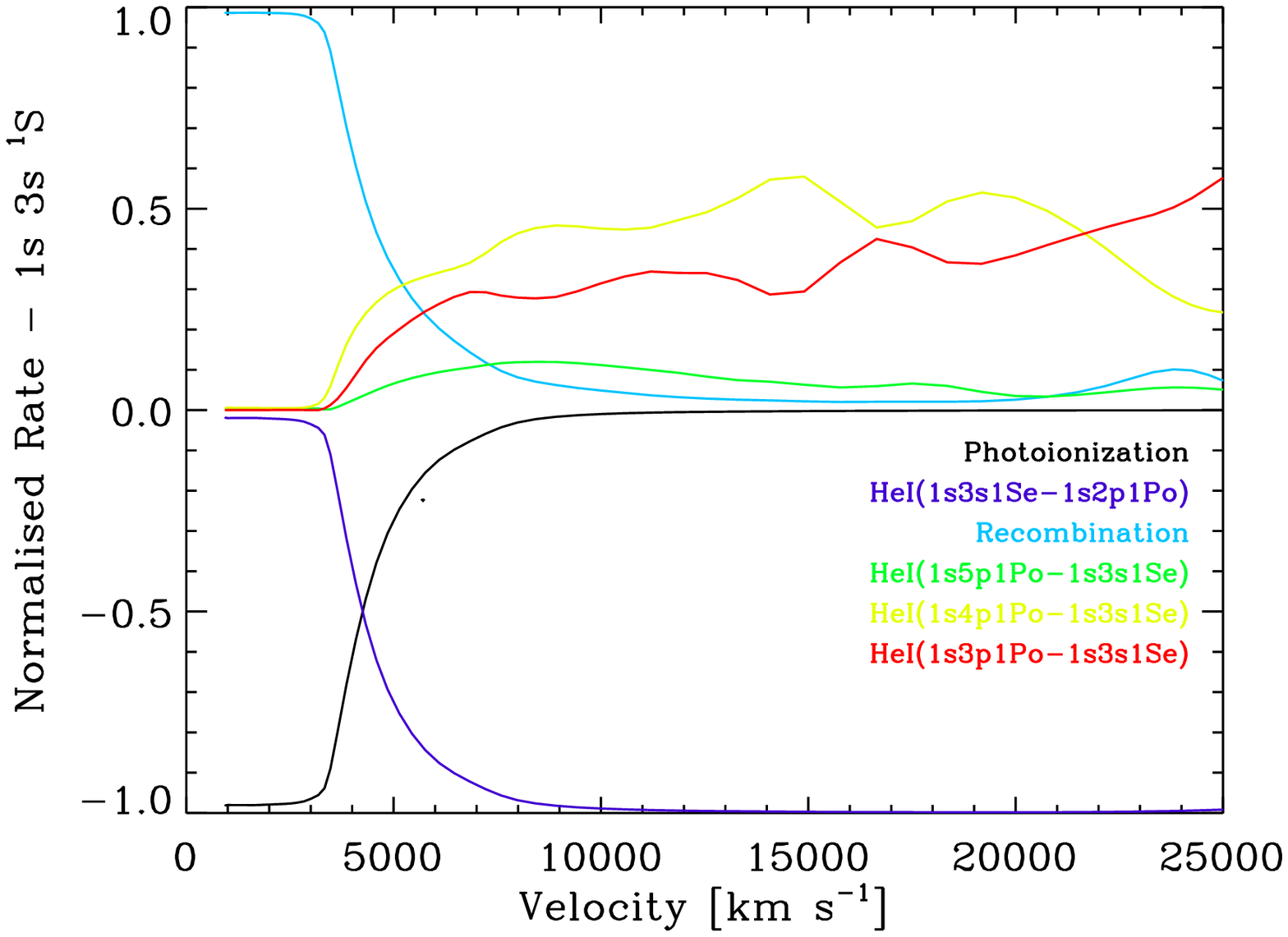,width=7.0cm}
\epsfig{file=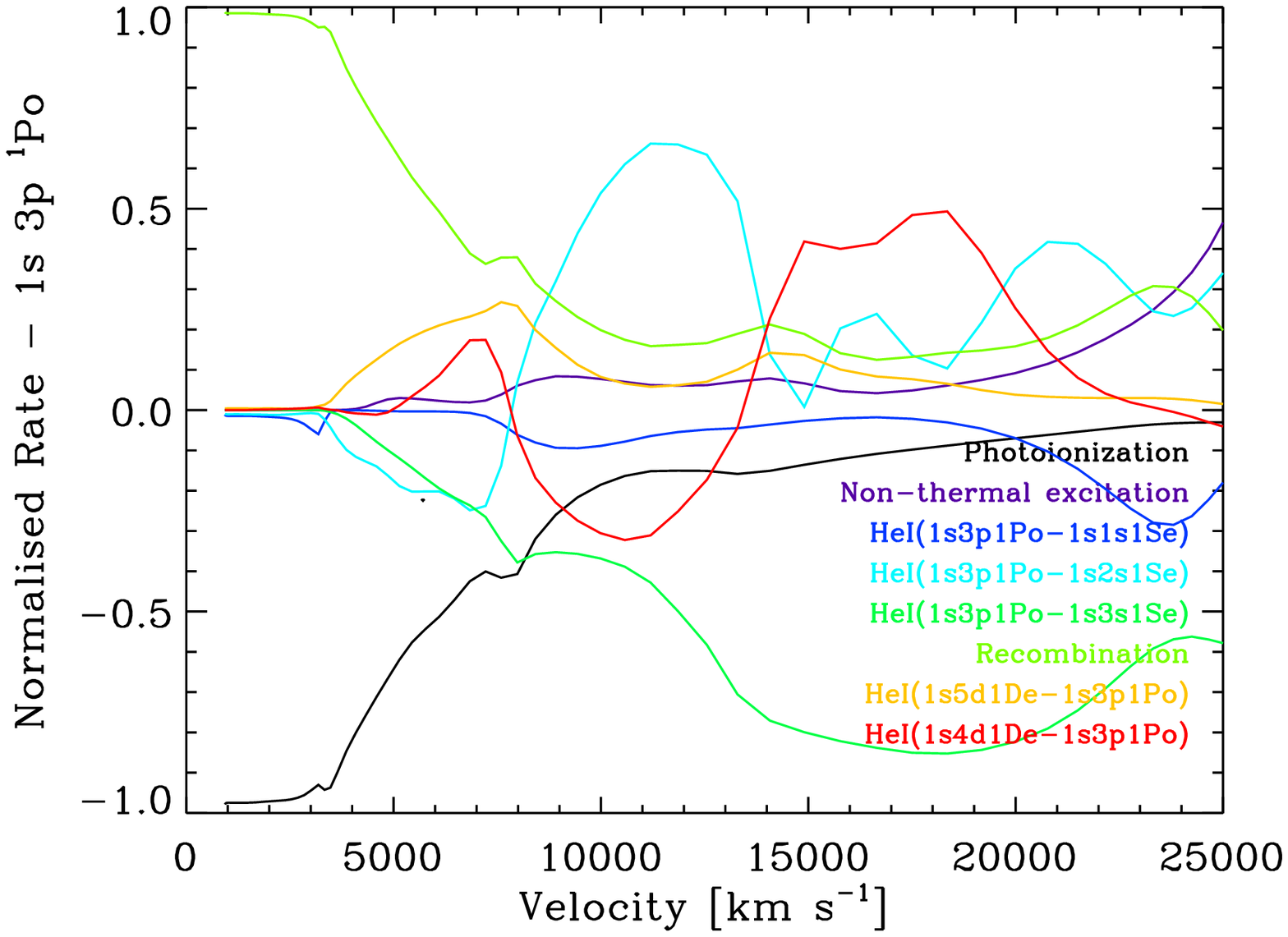,width=7.0cm}
\epsfig{file=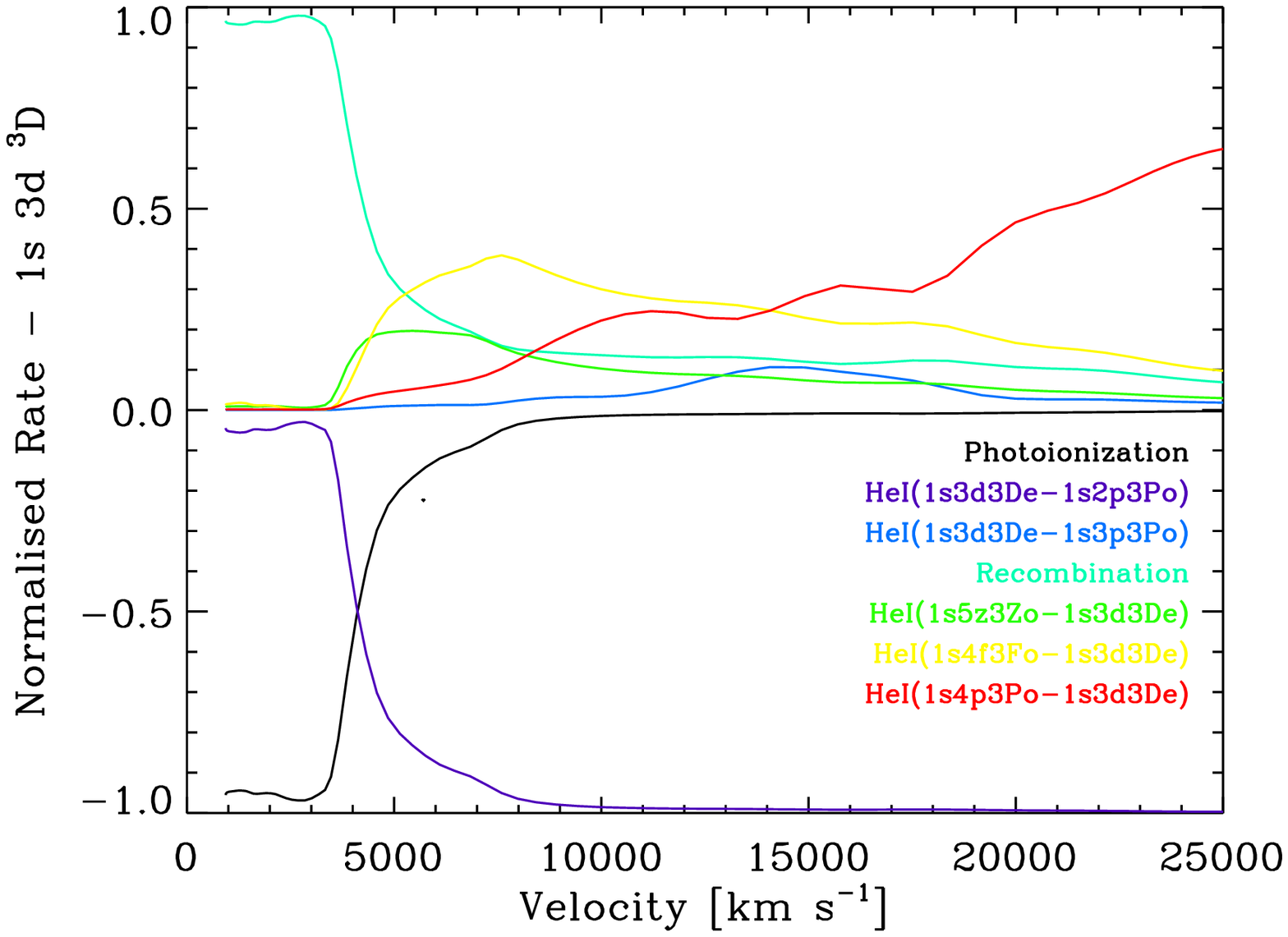,width=7.0cm}
\caption{
Illustration of the dominant fractional rates in (positive) and out (negative) for the first
9 levels of He\one, ordered from left to right and top to bottom
\elevel{1s}{1s}{1}{S}, \elevel{1s}{2s}{3}{S},  \elevel{1s}{2s}{1}{S}, \olevel{1s}{2p}{3}{P}, \olevel{1s}{2p}{1}{P},
\elevel{1s}{3s}{1}{S}, \olevel{1s}{3p}{1}{P}, and \elevel{1s}{3d}{3}{D} (we omit level \elevel{1s}{3s}{3}{S}).
For this illustration, we use model Bmf4p41x4 at light curve peak.
For better visibility, we display the individual rates as a fraction of the total rates for a given level at a given depth.
[See Section~\ref{sect_results_LC} for discussion.]
\label{fig_rates}
}
\end{figure*}

\section{Processes controlling the H\lowercase{e}\one\ level populations}
\label{sect_rates}

 In SN ejecta, departures from LTE are usually very strong at, and above, the photosphere
(see, e.g., \citet{DH11} for an illustration of the departure coefficients of H\one, O\one, and Fe\two\ in a SN II model).
Here, non-thermal effects drive the material even further from LTE.

In this section, we describe in more detail how the He\one\ level populations are controlled by showing the 
dominant rates in and out of the lower nine levels (Fig.~\ref{fig_rates}). To assist with the interpretation, 
we give a succinct summary of  the properties of these levels and the associated bound-bound transitions 
in Table~\ref{tab_hei}.

 Rates controlling the level populations fall into different categories but are fundamentally collisional or radiative.
The former include  thermal and non-thermal (collisional) ionization and excitation, and the latter
photoionization and recombination (bound-free/ free-bound transition), and radiative excitation and de-excitation
(bound-bound).  In each panel of Fig.~\ref{fig_rates} we display a given rate
if it exceeds 5\% of the total rate at any given depth in the model (see label and color coding). The rate associated
with a line transition at frequency $\nu_{ul}$ corresponds to the quantity $n_u A_{ul} (1 - J_{\nu_{ul}}/S_{ul})$ where $n_u$
is the population of the upper level, $A_{ul}$ the Einstein coefficient, $J_{\nu_{ul}}$ the mean intensity at the frequency
$\nu_{ul}$, and $S_{ul}$ is the line source function. This rate may be positive (implying a
net decay of electrons from the upper level) or negative (implying a pumping of electrons from the lower state
into the upper state)  depending on how $J_{\nu_{ul}}$ compares to $S_{ul}$.

Below $\sim$\,4000\,\kms\ recombination and photoionization rates appear to dominate. However at these depth
we are almost in LTE and many processes contribute to setting the level populations. The line rates are also large,
but since we plot the net rates they have been analytically cancelled. Above  $\sim$4000\,\kms\ individual processes
are (generally) no longer in detailed balance, and the level populations are controlled by a variety of processes.

 In our treatment of He\one\, non-thermal ionization is only allowed from the ground state of He\one\ and thus 
 only directly affects the population of level  1s$^2$\,$^1$S (top-left panel) and the ground state of He\two. 
 This is a valid approximation since the ground state is the most populated. Unlike recombination and photoionization 
 rates that tend to cancel each other, there is no significant reverse process, and so, despite the modest rate, 
 the impact on He ionization is important.

While many different non-thermal excitation routes are considered, the most important is from the ground state
(1s$^2$\,$^1$S) to the \olevel{1s}{2p}{1}{P}\ level. When the bound-bound transition to the ground state is
sufficiently optically thick, the \olevel{1s}{2p}{1}{P}\ level decays to \elevel{1s}{2s}{1}{S}\ producing  He\one\,20581\,\AA.
Other $^1$P$^{\scriptsize \rm  o}$ levels are also significantly influenced by non-thermal excitations from the ground state --
in the Bethe approximation \citep{Reg62_col} the rate of excitation to these levels is proportional to the oscillator strength.

In contrast, level \olevel{1s}{2p}{3}{P}\  is primarily populated through downward transitions from higher levels and thus the 10830\,\AA\ line
from transition He\one(\olevel{1s}{2p}{3}{P}-\elevel{1s}{2s}{3}{S}) stems primarily from recombination following photoionization and
non-thermal ionization. He\one\ 10830\,\AA\ and 20581\,\AA\ dominate the cascade into levels 2 and 3, i.e., \elevel{1s}{2s}{3}{S}
and \elevel{1s}{2s}{1}{S} (red curves in the 2nd and 3rd panels from top-left).

The distinction between the  \olevel{1s}{2p}{1}{P}\ and the \olevel{1s}{2p}{3}{P} primarily arises because
of the large difference in collision cross-section. The collision strength between levels with permitted transitions scales as $\log E$
(energy of electron) at high energies (i.e., $E\gg$\,ionization energy), while the collision strength of other transitions is almost constant
(transitions to \elevel{1s}{2s}{1}{S}) or declines with $E$ (transitions to \olevel{1s}{2s}{3}{P}).\footnote{We utilized the online
Los Alamos Atomic Physics Codes at http://aphysics2.lanl.gov/tempweb/lanl/, which are based on Hartree-Fock method of
R.~D. Cowan, to compute collision strengths for both H\one\ and He\one.} Two other factors also come into play --- for
He\one\ the rate of non-thermal ionizations is somewhat larger than the rate of non-thermal excitations, and approximately
3/4 of the recombinations (excluding the ground state) are to triplet levels.

\begin{figure}
\epsfig{file=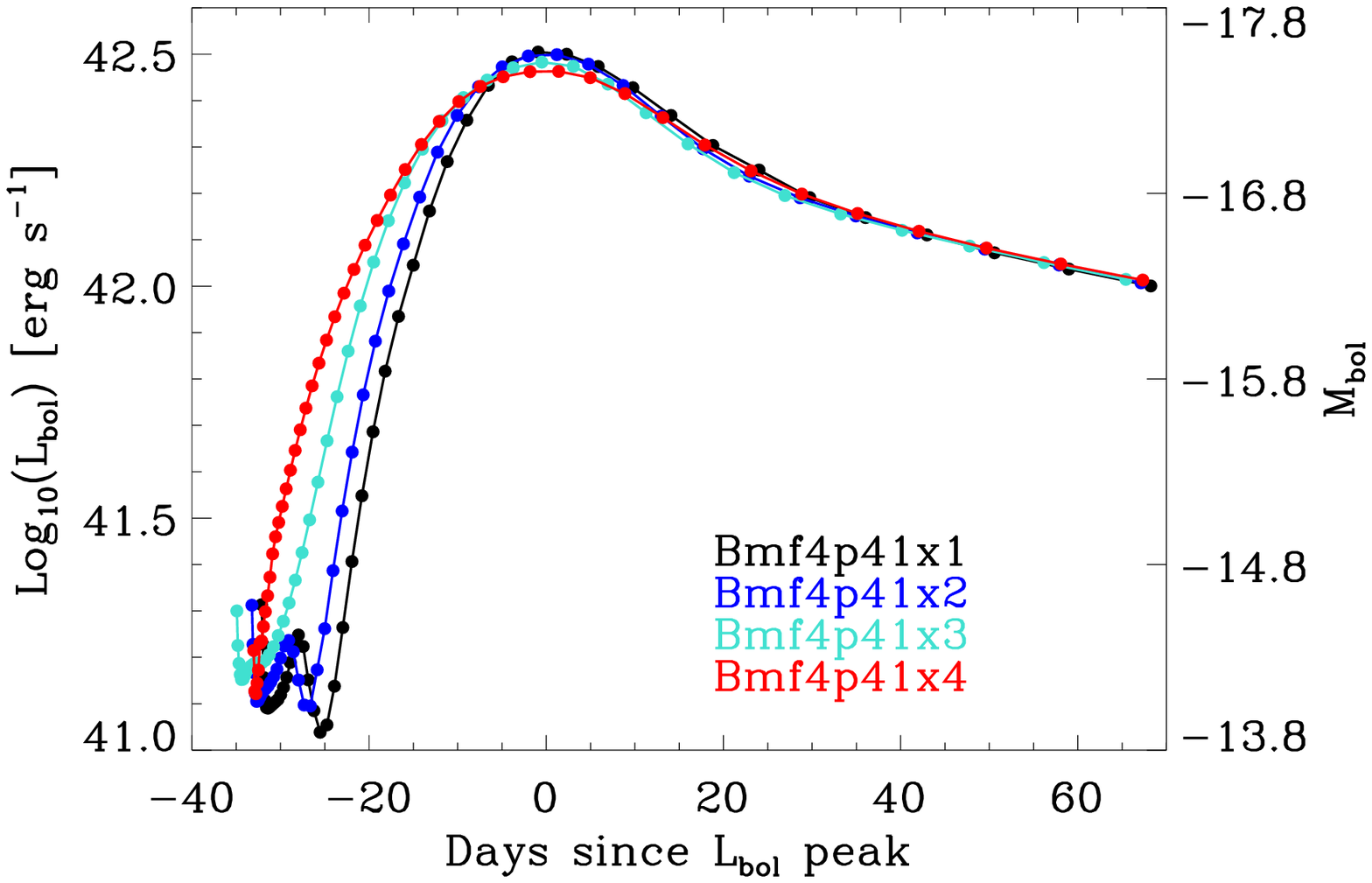,width=8.5cm}
\caption{Bolometric light curve for the Bmf4p41 series shown with respect to the time of peak bolometric luminosity
and assuming full  \gray\ trapping. In practice, \gray\ escape starts a few weeks after the peak in such progenitors
\citep{dessart_etal_11} and is slightly greater for higher mixing magnitudes (at 100\,d, we find that the fraction of \grays\ that
escape increases from $\sim$25\% to $\sim$40\% from Bmf4p41x1 to Bmf4p41x4). [See Section~\ref{sect_results_LC} for discussion.]
\label{fig_lbol}
}
\end{figure}

An important issue to remember when discussing the influence of various processes is that the size of a particular
process does not necessarily give an indication of its importance. An obvious example is bound-free processes
for the ground state of He\one. While the component rates for the process may be larger than other rates, they may be
negligible in controlling the helium ionization structure when the He\one\ ground-state continuum is in detail balance,
and the two rates (photoionization and recombination) cancel with a high degree of accuracy.

In the present case non-thermal excitation and ionizations lead (directly or via recombination and cascades) to
an over population of the n=2 levels (primarily the 2s states) in He\one. The population of these state is set by
the leakage back to the ground state which compensate for the non-thermal excitations and ionizations. When this
occurs photo-ionizations from these states may become the dominant ionization process, even though this process
would be negligible in the absence of non-thermal processes. As noted by \citet{lucy_91} this can amplify the
influence of non-thermal processes.

Non-thermal processes act directly only on a few levels in He\one. However non-thermal ionization affects all levels
indirectly through its influence on the He ionization and hence on recombination. Non-thermal processes are important
throughout the evolution of SNe Ibc, including the nebular phase.

\begin{table}
\caption{Selection of properties of He\one\ lower levels and important transitions in between
them. We present an illustration of the rates controlling the population of the lower levels
of He\one\ in Fig.~\ref{fig_rates} in model Bmf4p41x4 at light curve peak.
We first give the level Id. number, the level name, its statistical weight $g$, its excitation energy $E_i$ (in cm$^{-1}$)
above the ground state. We then list the main transitions between these levels that fall in the optical and near-IR ranges,
giving the oscillator strength and the wavelength for each. \label{tab_hei}}
\begin{tabular}{lccc}
\hline
Id.   & Level         &  $g$   &     $E_i$ (cm$^{-1}$)   \\
\hline
1    &  \elevel{1s}{1s}{1}{S}     &   1      &   0.0           \\
2    &   \elevel{1s}{2s}{3}{S}    &   3      & 159856.1    \\
3    &   \elevel{1s}{2s}{1}{S}    &   1      & 166277.5     \\
4    &  \olevel{1s}{2p}{3}{P}     &   9      & 169087.0     \\
5    &  \olevel{1s}{2p}{1}{P}     &   3      & 171135.0     \\
6    &   \elevel{1s}{3s}{3}{S}    &   3      & 183236.9      \\
7    &   \elevel{1s}{3s}{1}{S}    &   1      & 184864.9   \\
8    &  \olevel{1s}{3p}{3}{P}    &    9      &  185564.7  \\
9    &   \elevel{1s}{3d}{3}{D}   &   15    &  186101.7  \\
10  & \elevel{1s}{3d}{1}{D}     &  5      &  186105.07   \\
11  & \olevel{1s}{3p}{1}{P}      & 3  &  186209.47     \\
\hline
Transition  & $gf$    & Wavelength [\AA] &  \\
\hline
4-2     &   5.399E-01  &     10830.17 &   \\
8-2     &   6.406E-02  &      3888.64 &   \\
5-3     &   3.294E-01  &     20581.28 &   \\
11-3   &   1.558E-01  &      5015.68 &   \\
6-4     &   6.953E-02  &      7065.25 &   \\
9-4     &   6.097E-01  &      5875.66 &   \\
7-5     &   4.703E-02  &      7281.35 &   \\
10-5  &   7.094E-01  &      6678.15 &   \\
11-3  &   1.558E-01  &      5015.68  &  \\
\hline
\end{tabular}
\end{table}

\section{Light curve and photospheric properties}
\label{sect_results_LC}

  In our non-LTE time-dependent radiative-transfer simulations, we compute the emergent spectrum from a few \AA\
to a few hundred microns. The frequency integral of this emergent spectrum at each epoch yields the bolometric
luminosity of each model. We show the  evolution for the bolometric luminosity for the  Bmf4p41x[1--4] model series
in Fig.~\ref{fig_lbol}, using the peak of the \LC\ as the time origin (the Bmf5p09x[1--4] series behaves similarly
and is thus not shown). Full \gray\ trapping is assumed at all times.

Enhanced mixing causes an earlier and more gradual rise to the peak. The post-breakout plateau that
characterises the unmixed or weakly mixed models, as described in \citet{dessart_etal_11} vanishes
for the highest levels of mixing (this is similar to what was found by, e.g., \citealt{shigeyama_etal_90b}, for 
SN\,1987A).
 %
 Enhanced mixing causes an earlier peak (merely a few days in this series
but as much as $\sim$15\,d in the Bmf5p09x4 model), and an earlier transition to \gray\ escape,
making the nebular luminosity  drop earlier and faster (not shown here, but see \citealt{dessart_etal_11}).
This effect is however weak because, irrespective of mixing, the bulk of \gray\ energy deposition reflects the \isoni\
distribution that is coupled with the density, and which steeply increases towards the inner ejecta.
 As we discussed earlier in the paper, some numerical diffusion affects somewhat the total mass of
 unstable isotopes, in particular in unmixed models. Since the main focus of this paper is on the non-thermal effects
 on SNe Ib/c spectra, we defer to a forthcoming study a more quantitative description of mixing on SNe Ib/c \LCS\
 (Dessart et al, in preparation).

\begin{figure*}
\epsfig{file=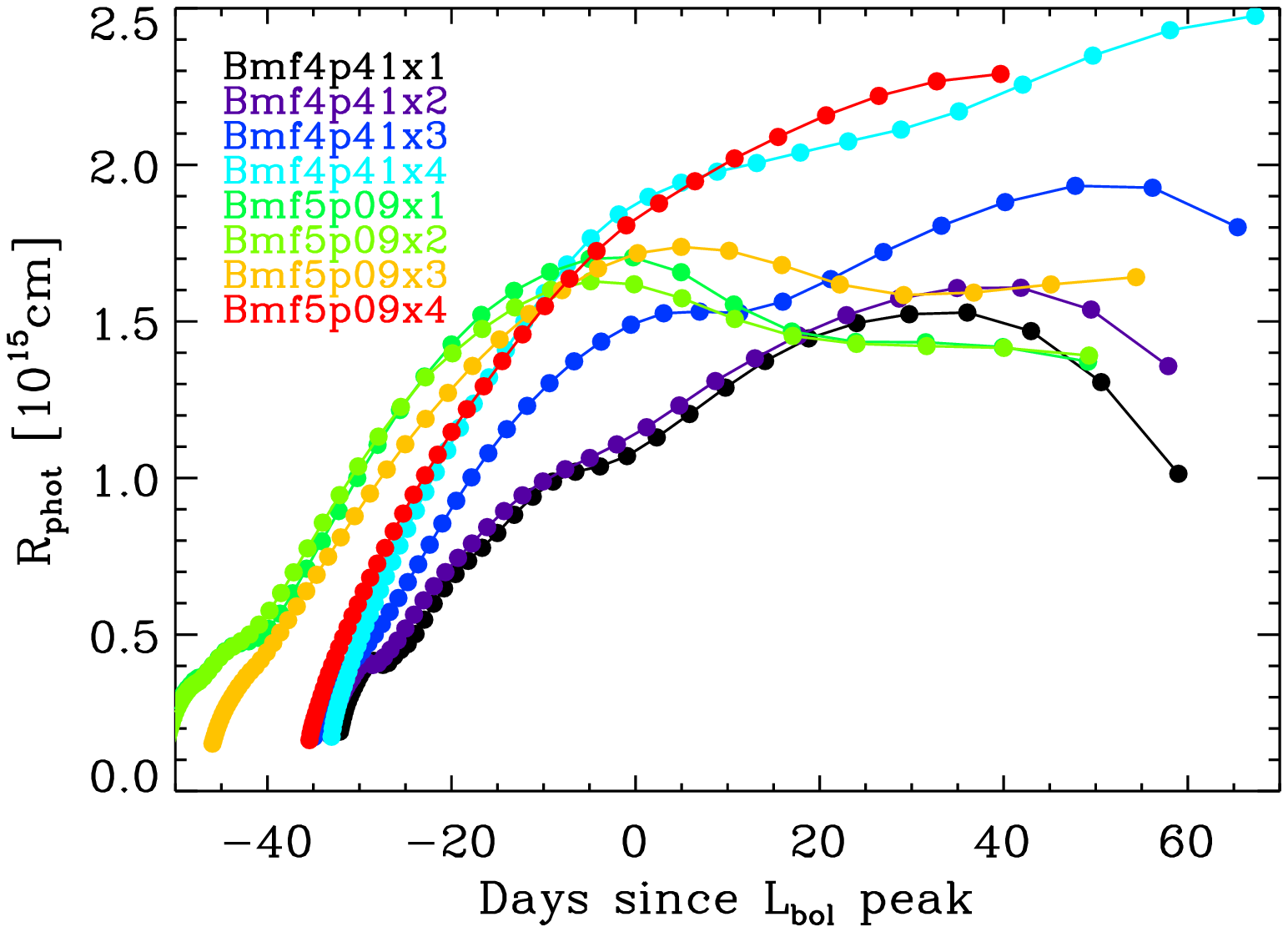,width=8.5cm}
\epsfig{file=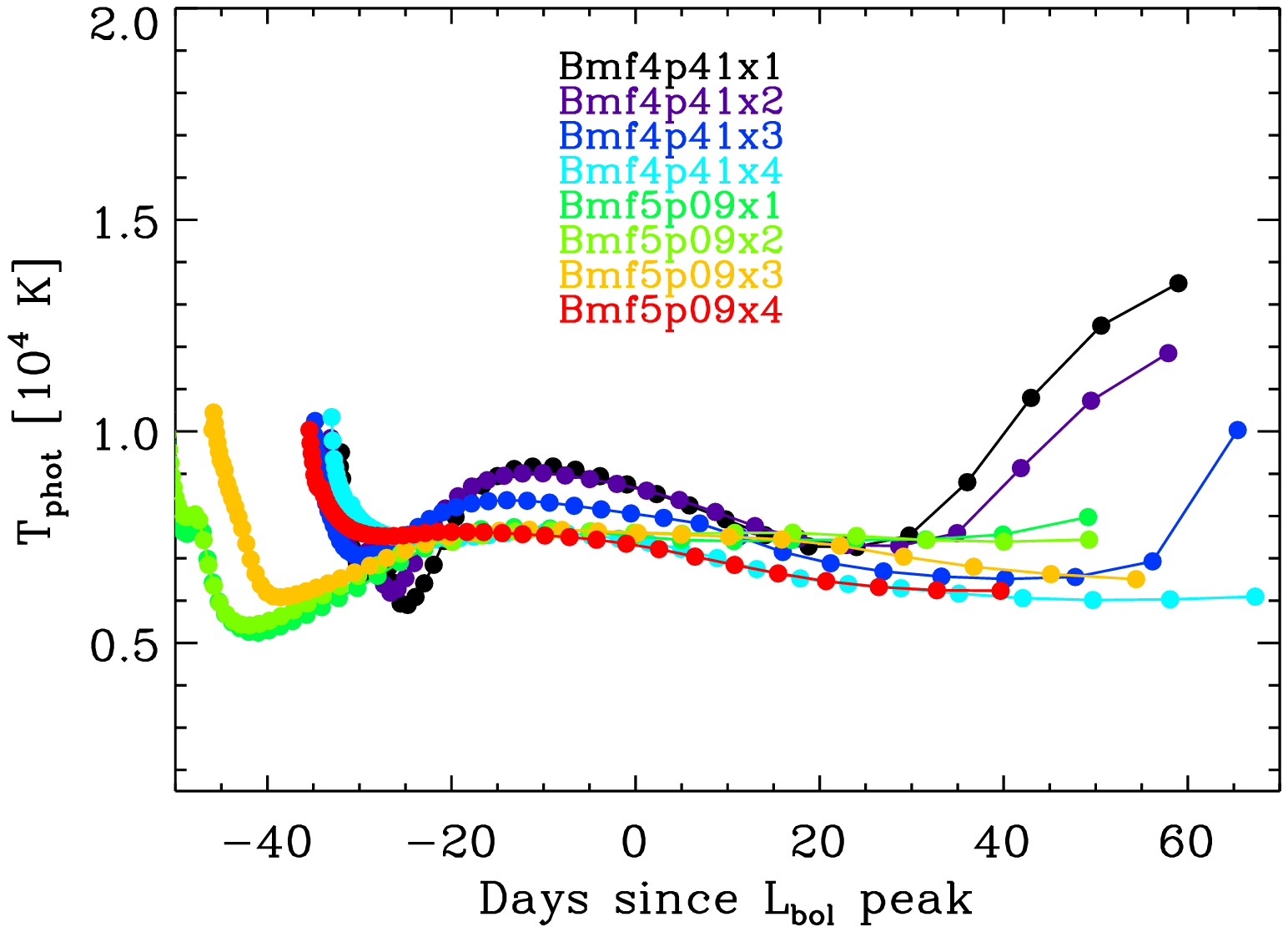,width=8.5cm}
\epsfig{file=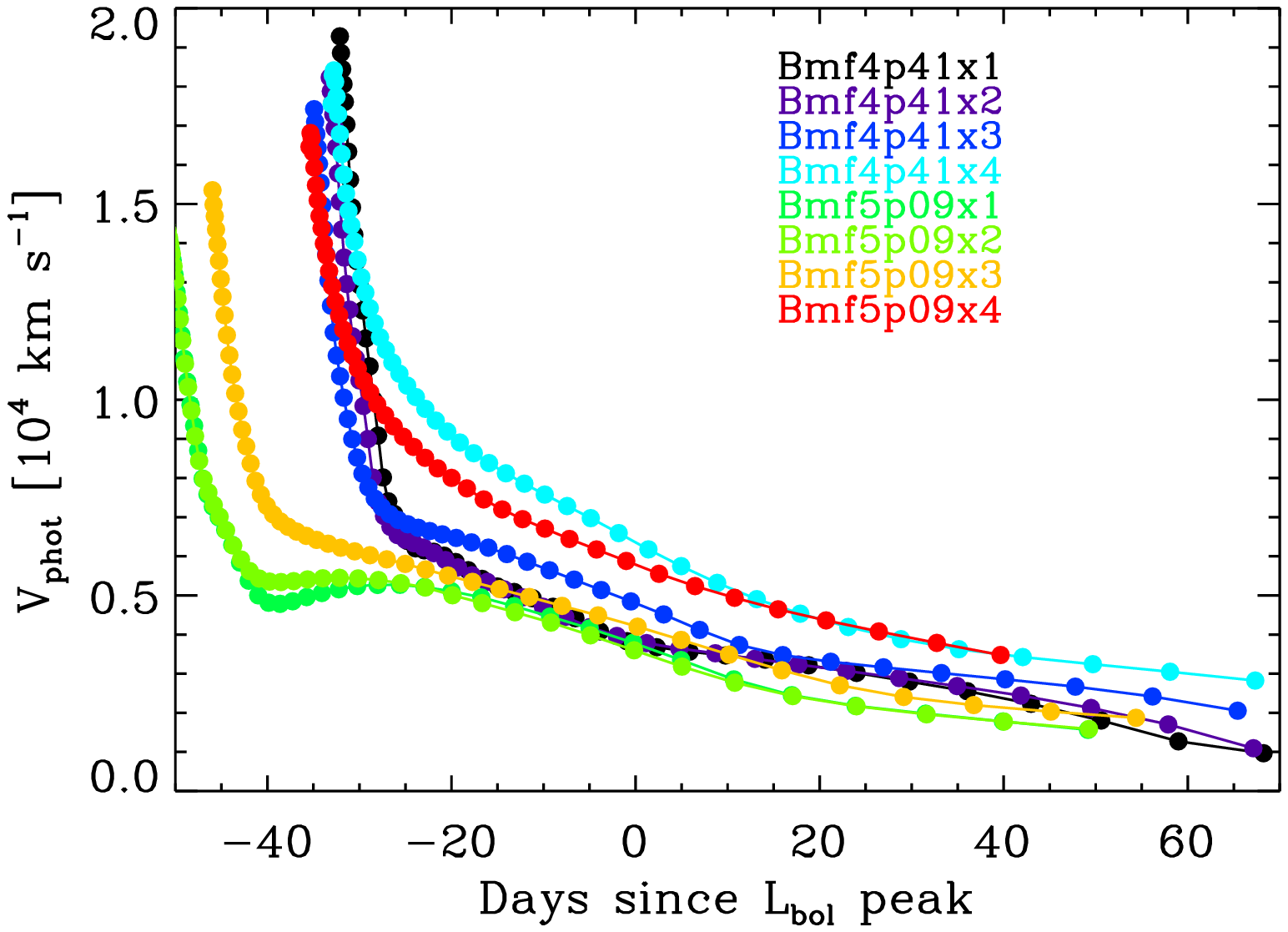,width=8.5cm}
\epsfig{file=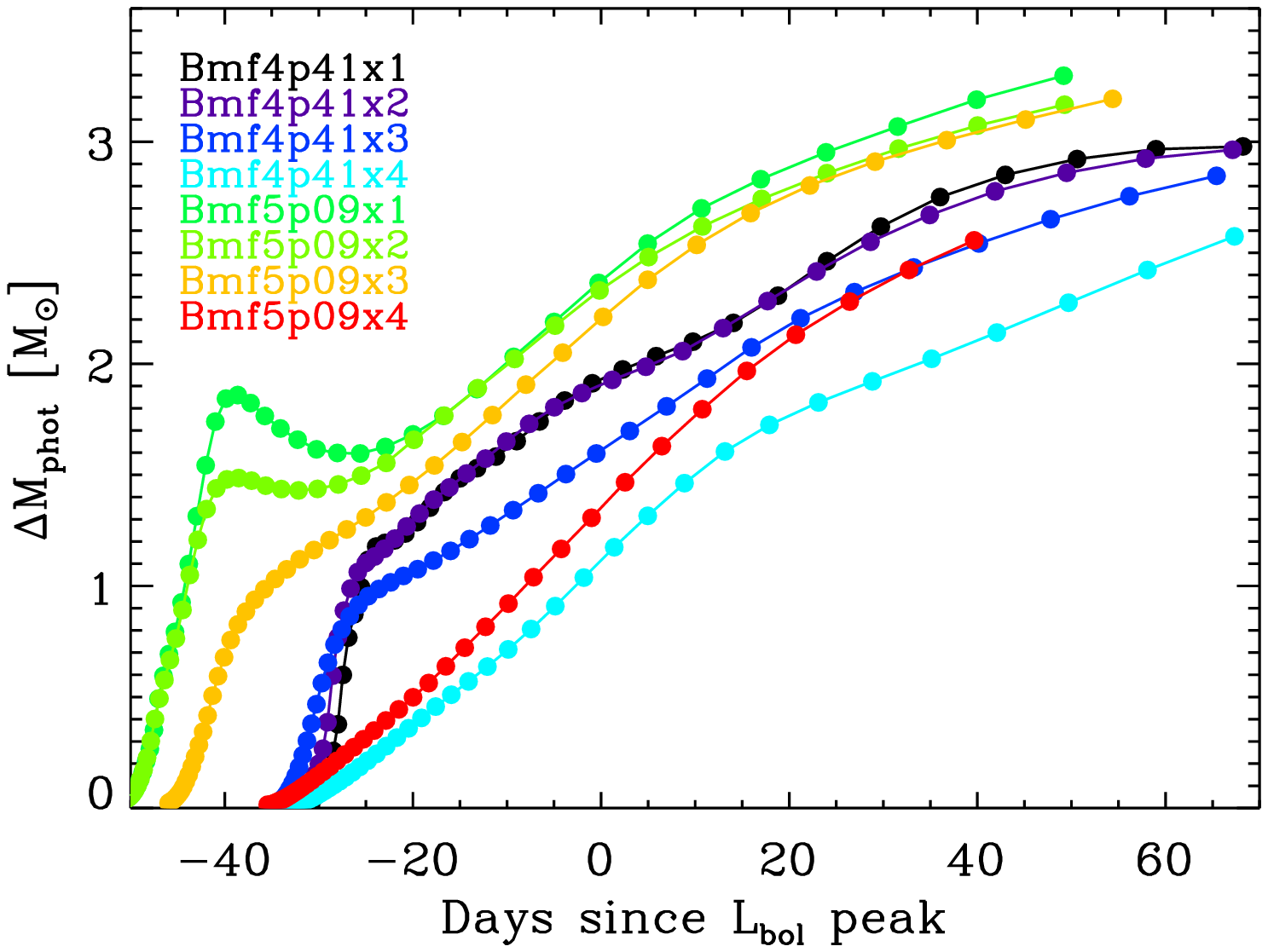,width=8.5cm}
\caption{Evolution with respect to days since the \LC\ peak of the
radius (top-left), the temperature (top-right), the velocity (bottom-left) and the overlying mass
at the electron-scattering photosphere for both model sequences Bmf4p41xn and Bmf5p09xn. Notice
the non-monotonic photospheric velocity in the weakly unmixed Bmf5p09xn models,
a feature that disappears as mixing is enhanced.
\label{fig_phot_prop}
}
\end{figure*}

  The \LC\ behavior reflects the evolution of the photospheric properties (Fig.~\ref{fig_phot_prop})
  and nearly exclusively the radius and the temperature. For both model series Bmf4p41 and Bmf5p09,
  mixing causes the same effect on the photosphere behavior and primarily reflects the induced change
  on the \isoni\ distribution. With enhanced mixing, the heating of the outer ejecta layers is increased
  and the diffusion time of that energy to the outer layers is shorter. Consequently, the photosphere
  migrates outward in radius earlier with enhanced mixing. The extra heat produces a higher temperature
  initially, which thus causes the luminosity to rise earlier, even avoiding a post-breakout plateau for the highest
  level of mixing here. The reduced mass fraction of \isoni\ at depth causes a smaller temperature rise associated with
  the heat wave. With increased mixing, the non-monotonicity of the photosphere trajectory in mass/velocity is reduced
  and even disappears for the highest level of mixing. In this case, the temperature, the velocity and the mass above the
  photosphere continuously and gradually decrease with time. The extra heat generated from the mixed radioactive material
  keeps the photosphere further out in the ejecta, at larger velocities. Without mixing, the photosphere recedes
  in mass very fast and abruptly changes its properties when it suddenly feels the heat wave from decay.

  These effects are all fundamentally of non-thermal origin, but as  described in the previous section,
  the energy from non-thermal electrons is always channeled primarily in the form of heat.
  In the low-ionized conditions above the photosphere, it represents $\sim$70\% of the local energy deposition,
  reaches 95-100\% far above the photosphere where the conditions are ionized
  (which results primarily from ionization freeze-out associated with time-dependent effects; \citealt{DH08_time}).
  Below the photosphere, the heat channel represents $\sim$95\% of the total energy deposited locally, but decreases
  with time at and beyond the peak of the LC, i.e., $\sim$92\% at 29.3\,d and $\sim$82\% at 63.0\,d after explosion.
  Since the energy deposition occurs primarily at depth, the bulk of the non-thermal energy is deposited as heat.
  Decay energy thus influences primarily the \LC, allowing such Type I CCSNe to be bright a few weeks after explosion.



\section{Influence of mixing on synthetic spectra}
\label{sect_results_spec}

\begin{figure*}
 \epsfig{file=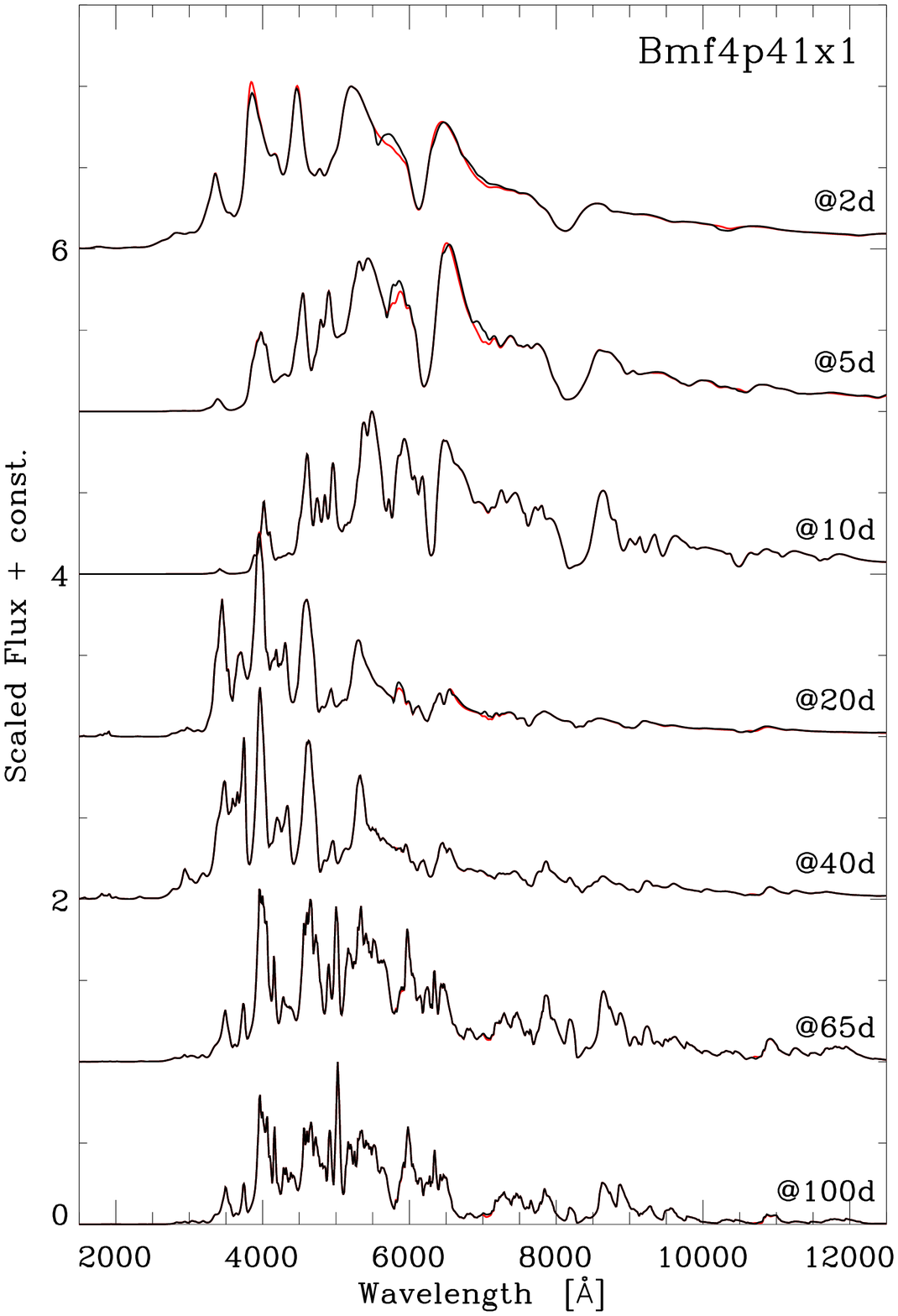,width=5.85cm}
 \epsfig{file=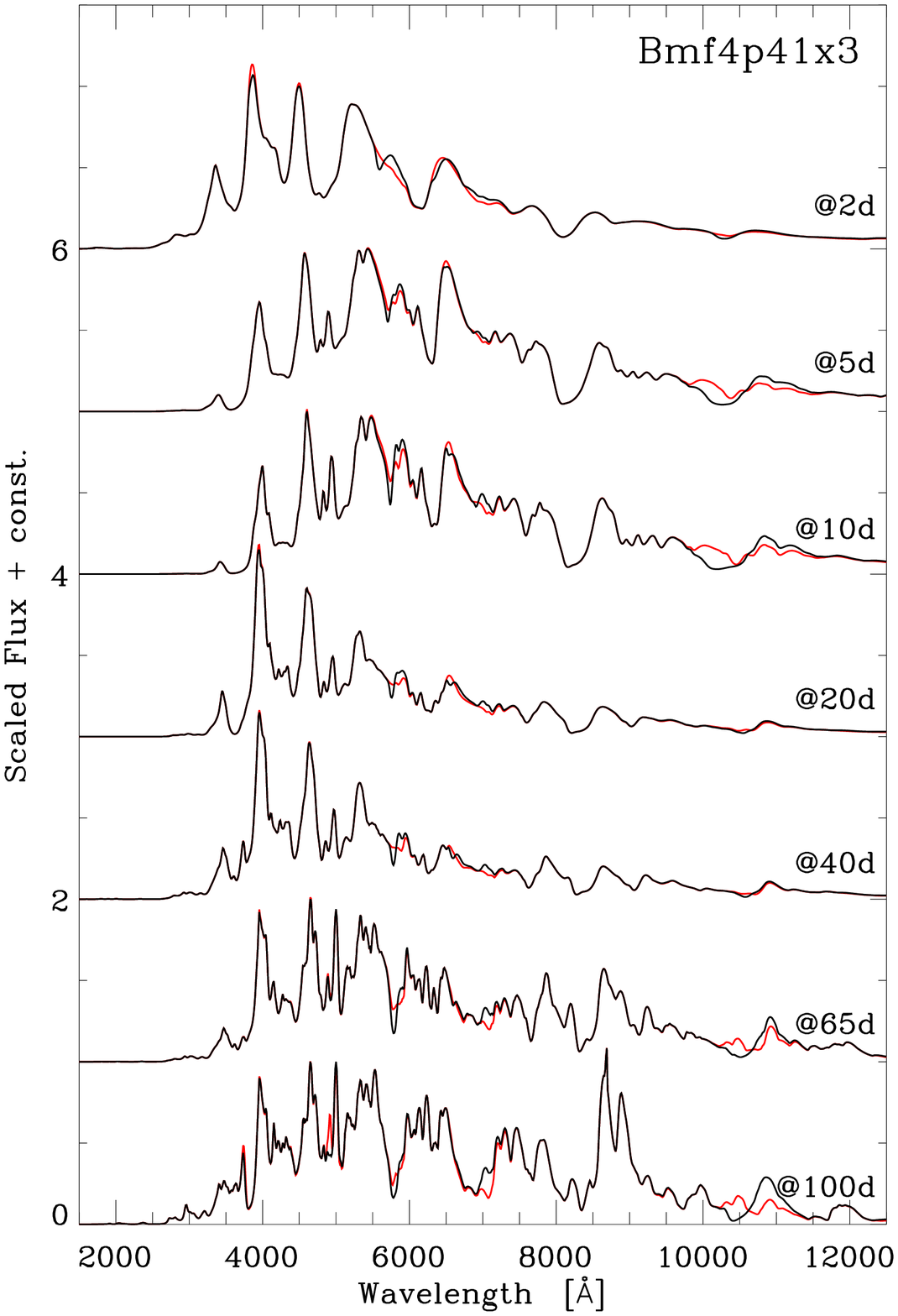,width=5.85cm}
 \epsfig{file=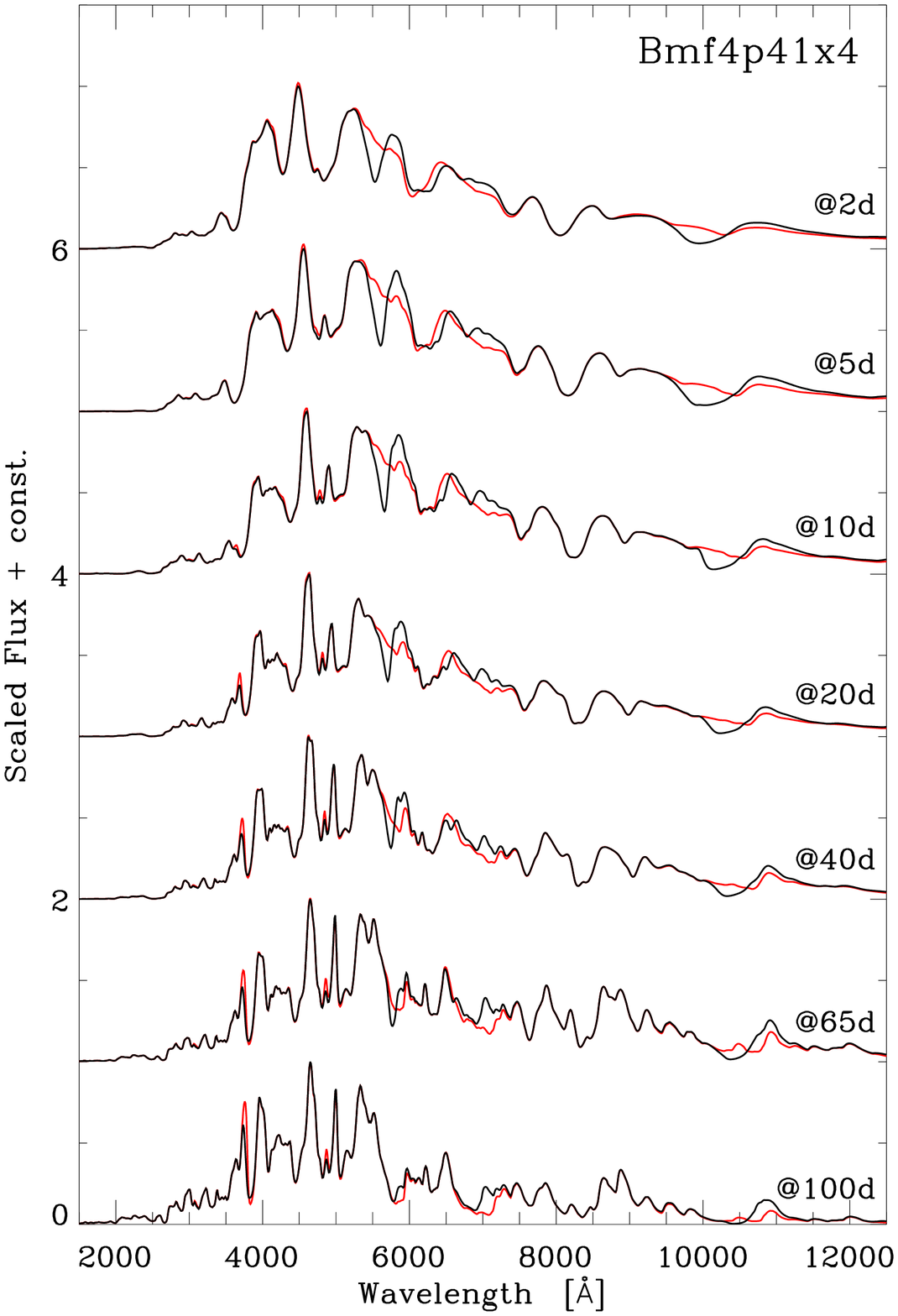,width=5.85cm}
 \caption{{\it Left column:}
 Montage of synthetic spectra (black) for the models Bmf4p41x1 (left; model Bmf4p41x2 has a
 similar evolution and is thus not shown), Bmf4p41x3 (middle),
 and Bmf4p41x4 (right) from 2 until 100\,d after explosion.
For illustration, we overlay the synthetic spectrum that results when excluding He\one\ bound-bound transitions
in the formal solution of the radiative-transfer equation (red).
 The left panel correspond to a SN IIc evolution, owing to the strong H$\alpha$ line initially,
 and the transition to a spectrum without lines of H\one, He\one, nor Si\two\ in the optical after about 15\,d.
 The middle panel  corresponds to a SN IIb evolution, with presence of H$\alpha$ initially and He\one\ lines,
 albeit anomalously narrow, at later times. The right panel corresponds to a rather standard SN Ib evolution, with
 broad He\one\ lines at all times.
 \label{fig_4p4_seq}
 }
 \end{figure*}

\begin{figure*}
\epsfig{file=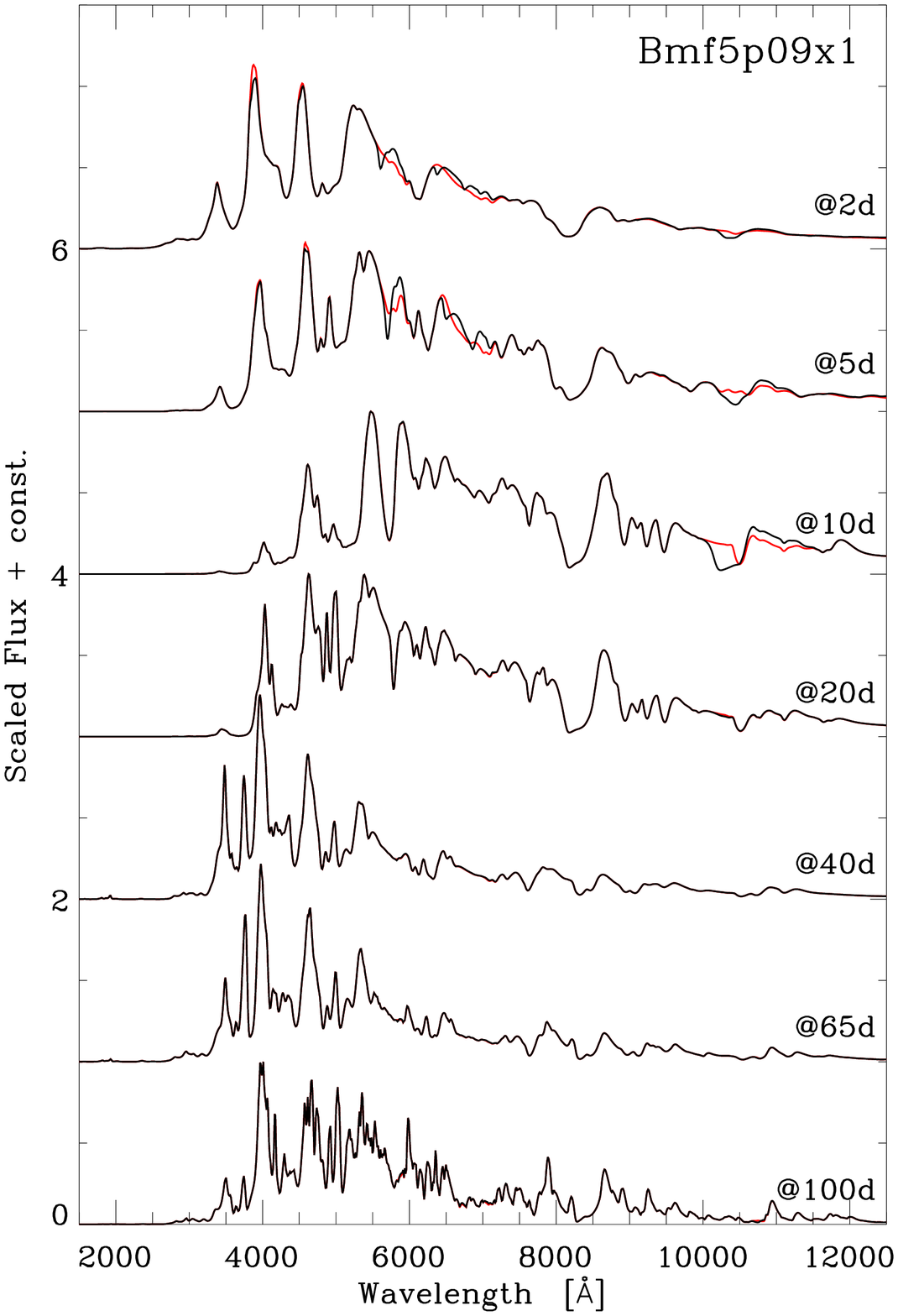,width=5.85cm}
\epsfig{file=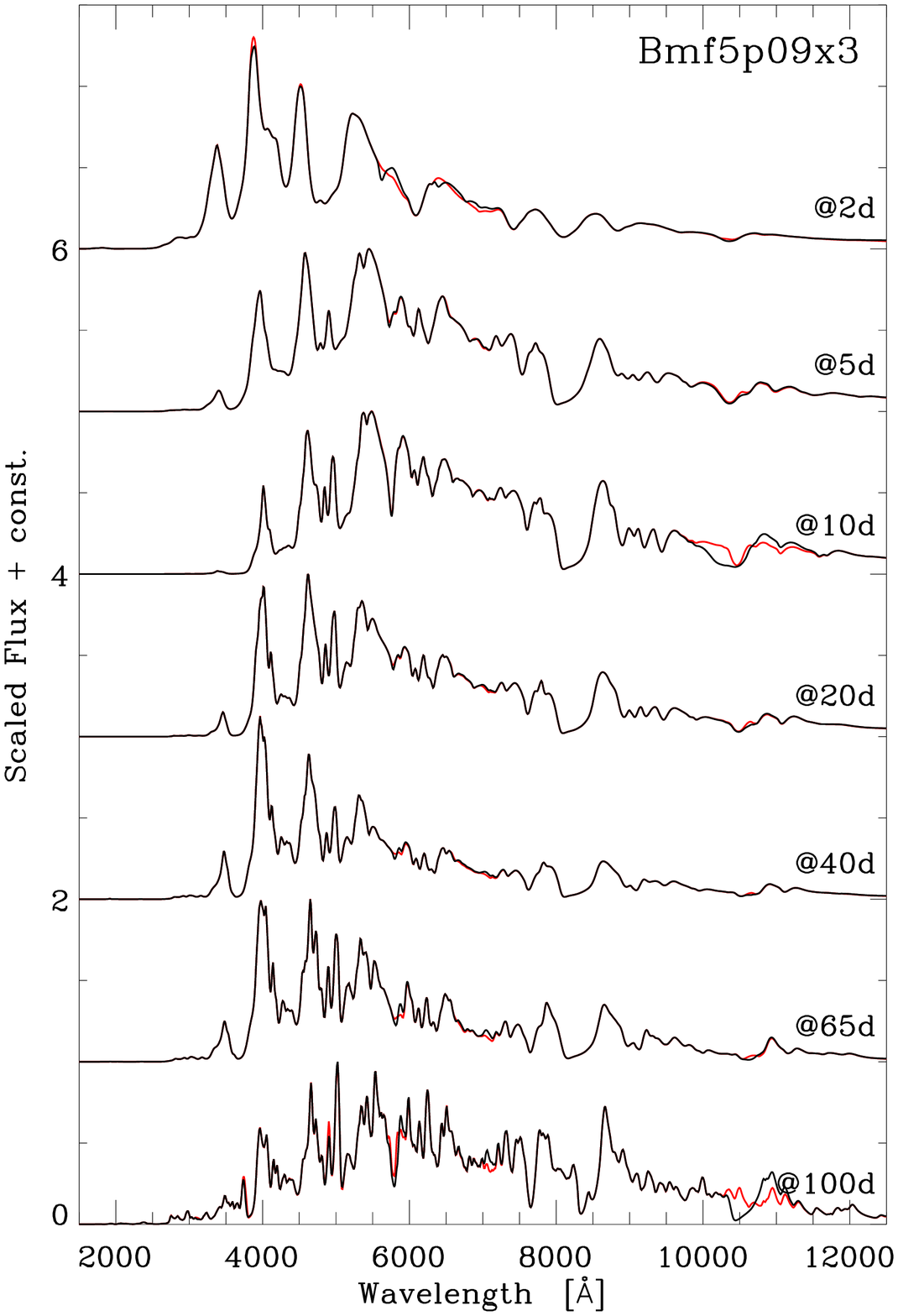,width=5.85cm}
\epsfig{file=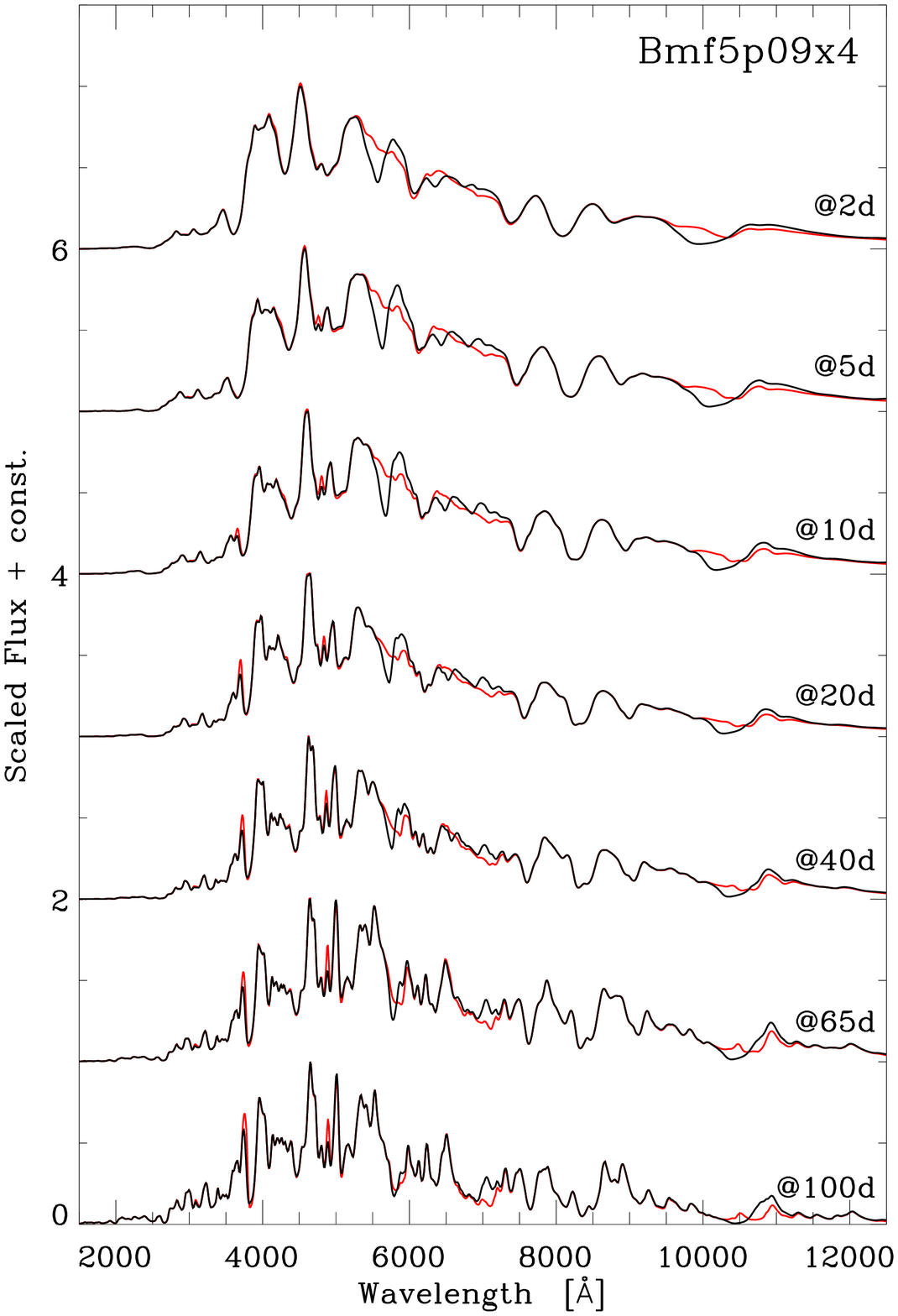,width=5.85cm}
\caption{Same as Fig.~\ref{fig_4p4_seq}, but now for the Bmf5p09x1--4 model set. Owing to the larger
ejecta mass, only the model with the largest mixing (right) produces a genuine SN Ib, while
the other two series correspond to a SN Ic evolution.
\label{fig_5p1_seq}
}
\end{figure*}

  Variations in mixing cause changes in composition (Fig.~\ref{fig_ejecta_prop}), \LC\ evolution
(Fig.~\ref{fig_lbol}), and in non-thermal processes (Figs.~\ref{fig_nonth}--\ref{fig_nonth_unmixed}).
In Figs.~\ref{fig_4p4_seq} and \ref{fig_5p1_seq}, we show the spectral evolution over the first 100\,d for models
Bmf4p41x1--4 and Bmf5p09x1--4, respectively (black curves). We overplot in each panel the
corresponding spectra when He\one\ lines are excluded from the formal solution of the
transfer equation. All simulations are done with large, but not huge, model atoms (Section~\ref{sect_model};
note that these simulations represent nonetheless the most CPU-intensive non-LTE line blanketed simulations ever
performed for SNe Ib/c), so the spectral colors and morphology are not fully converged. For the present
qualitative presentation, this is sufficient but we will explore the dependency of our results on the adopted model
atom in a forthcoming paper. We obtain three different types of spectral evolution, corresponding to different SN classification:

\begin{itemize}
\item The weakly and moderately mixed models (i.e., Bmf4p41x1--2 and Bmf5p09x1--3) have
a similar spectral appearance as the corresponding unmixed models presented in \citet{dessart_etal_11}.
Despite variations in mixing, which impact both \isoni\ and all other species, the differences in this set are
subtle. Most important, none of these models show optical He\one\ lines after about 10 days past explosion,
and when such He\one\ lines  are present early-on, they arise exclusively from non-LTE effects \citep{dessart_etal_11}.
In the optical, model Bmf4p41x1 (and Bmf4p41x2, not shown) shows strong H$\alpha$ initially,
which quickly vanishes to leave a spectrum without H\one, He\one\ lines, nor the strong Si\two\ absorption at 6300\AA.
In a strict sense, this spectral evolution would correspond to a SN IIc, a SN subtype unobserved today.
All other models in this set would be of Type Ic.
We note that the presence of He\one\ and H$\alpha$ in these models at early times is 
influenced by the ionization freeze-out in the outer ejecta, a time-dependent effect
\citep{DH08_time} that dominates in the absence of non-thermal excitation/ionization \citep{dessart_etal_11}.

\item In the moderately mixed  model Bmf4p41x3 (middle panel), He\one\,5875\AA\ is visible but
it is extremely weak and very narrow, disqualified for a standard SN Ib spectrum. Nonetheless, we
can clearly see that the increasing mixing starts to have a visible effect on the spectra. 
Non-local energy deposition at late times compensates for the weak mixing and causes the non-thermal
excitation/ionization of He\one\,10830\,\AA; this effect is however too weak to produce unambiguous He\one\ optical
lines. Because of the presence of H$\alpha$ at early times, this spectral evolution corresponds to a Type IIb SN,
although of a peculiar kind due to the weakness of the He\one\ lines.

\item For the heavily mixed models Bmf4p41x4 and Bmf5p09x4, the spectral appearance is markedly different
from all other models discussed above. In both models, and despite the differences in ejecta mass,
the spectral appearance is the same and very
typical of SNe Ib in the sense that He\one\ lines are now unambiguously seen at all times.
Because of the mixing procedure, which affects the entire ejecta, model Bmf4p41x4
does not show H\one\ lines initially (the surface hydrogen mass fraction is too low). If the surface hydrogen
mass fraction had been kept as high as in model Bmf4p41x1, model Bmf4p41x4 would be a Type IIb SN,
but given the absence of H\one\ lines, model Bmf4p41x4 (just like model Bmf5p09x4) corresponds to a SN Ib.
\end{itemize}

So, while all these He-rich ejecta show He\one\ lines at early post-explosion times, as discussed
in detail by \citet{dessart_etal_11}, only the moderately and heavily mixed models show optical He\one\ lines
after about 10\,d, throughout the peak of the \LC\ and beyond. This also applies to the He\one\ lines at
10830\,\AA\ and 20581\,\AA.

These models suggests that there is a continuum of He\one\ line strengths as we increase the mixing efficiency (this
would have been more obvious if we had had intermediate models between Bmf4p41x3 and Bmf4p41x4, for example).
As mixing increases, non-thermal excitation and ionization become more efficient at and above the photosphere
(Section~\ref{sect_nonth}). The enhanced ionization populates excited levels of He\one\, producing
He\one\ lines that are otherwise vanishingly weak.
In these SN Ib models the He\one\ lines dominate over weaker overlapping lines in the
optical. In contrast, the He\one\ lines are weak  in our SNe Ic models, and one sees instead a complex mixture of weak lines
associated with Si\two, Fe\two, Na\one, C\two, and O\one. In our simulations, these weaker lines tend to be always present
and thus, what yields here the Ib or Ic classification depends on how strong the He\one\ lines are, which in our set of helium-rich
ejecta models is more related to the abundance and spatial distribution of \isoni\ rather than the He abundance
(line identifications for observed SNe Ib and Ic are given in the next section).

\begin{figure*}
\epsfig{file=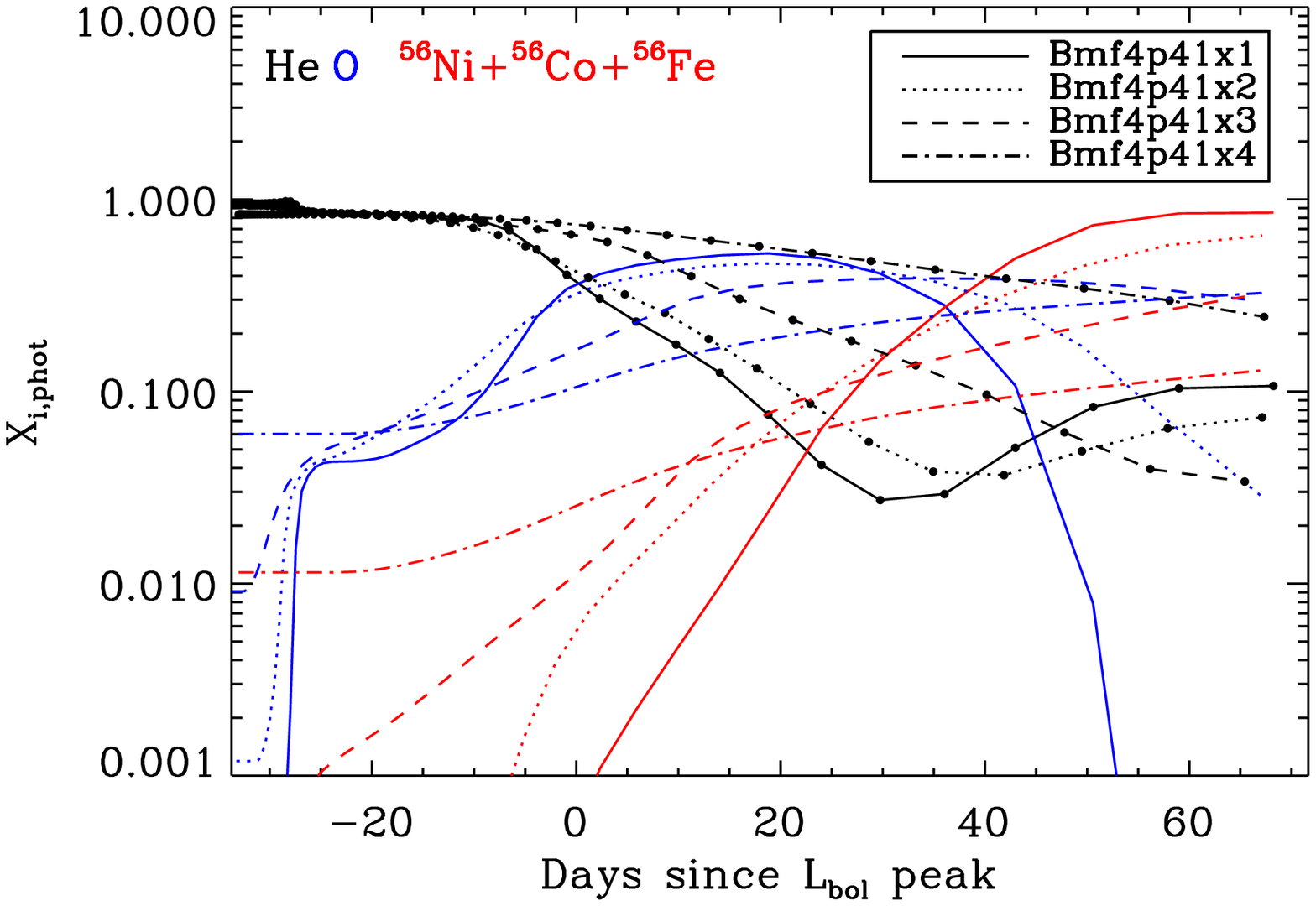,width=15cm}
\epsfig{file=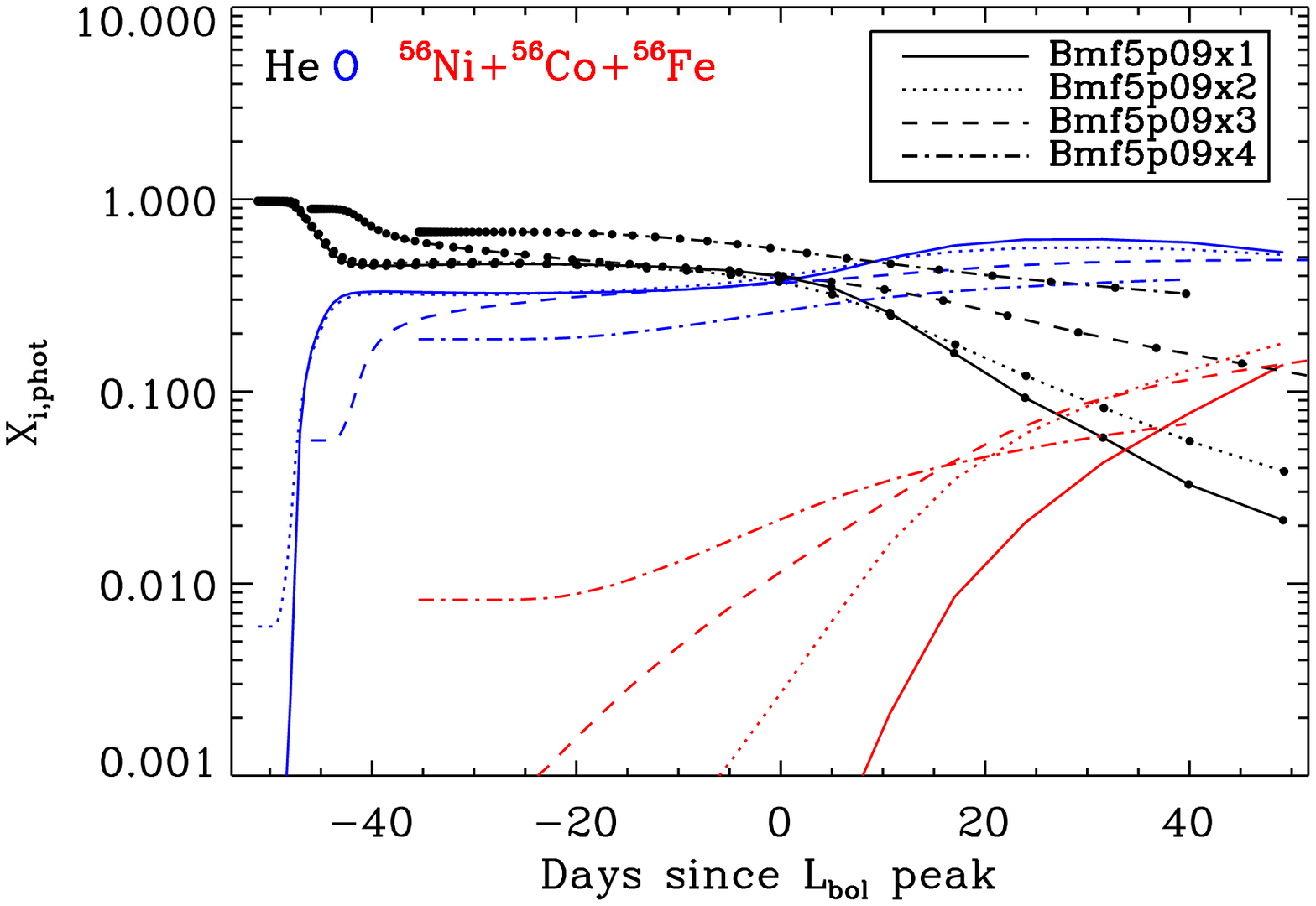,width=15cm}
\caption{Evolution of the helium (black), oxygen (blue), and \isoni\ (including its decay products; red)
mass fractions at the photosphere for the simulations
associated with the Bmf4p41x1--4 (top) and Bmf5p09x1--4 (bottom) model sets and shown with
respect to the time of peak bolometric luminosity. All ``standard'' SNe Ib show strong He\one\ lines at
maximum brightness. Only models Bmf4p41x4 and Bmf5p09x4 have this property, which seems
to require a \isoni\ mass fraction on the order of 0.01 in the helium-rich regions. \label{fig_xfrac_phot}
}
\end{figure*}

To illustrate this feature, we show in Fig.~\ref{fig_xfrac_phot} the variation at the photosphere of the mass fraction of
helium and of unstable isotopes associated with \isoni\ decay in all models (the time origin is the time of \LC\ peak).
Varying the mixing efficiency leads at the light curve peak to variations of a few tens of percent for the helium
mass fraction, but it causes orders of magnitude variations in the \isoni\ mass fraction. We find that
He\one\ lines only appear strongly if the original \isoni\ mass fraction in the corresponding layers is greater that $\sim$0.01.
We expect some sensitivity to the helium mass fraction too but not as dramatic. In the heavily mixed models,
a helium mass fraction of 0.2-0.3 deep in the ejecta is still enough to produce He\one\ lines.

This \isoni\ mass-fraction threshold makes the production of SNe Ib a challenge -- all ejecta without adequate
\isoni\ mixing will produce SNe Ic.
This also suggests a maximum mass for a SN Ib ejecta of no more than a few \msun. Indeed,
the typical SN Ib/Ic \isoni\ mass is 0.1\,\msun. For a fully mixed 10\,\msun\ ejecta, this would produce
a 0.01 \isoni\ mass fraction. Since even in our fully mixed models the \isoni\ mass fraction drops by a factor 5-10
between the inner and outer regions, the SN Ib ejecta mass cannot be more than $\sim$3\,\msun\ for the
successful production of He\one\ lines.
Because \isoni\ needs to be quite abundant in those helium-rich
layers, SNe Ib have to be fundamentally associated with lower mass ejecta. As we discussed recently,
they cannot be associated with high mass ejecta because of the short rise-time to \LC\ peak, the narrow \LC\ peak,
and the fast \LC\ decline at nebular epochs \citep{EW88,dessart_etal_11}.

In the massive-star community, high mass progenitors leading to WR stars are often considered as potential
candidates for SNe Ib/c \citep{HMM04_evol_rot,georgy_etal_09}, but the corresponding stellar-evolution simulations yield final masses that are typically
too large to accommodate
the observed SNe Ib/c light curves. They are also more generally characterized by an envelope with a large oxygen-rich buffer between
the core and any remaining outer helium-rich  shell, potentially compromising the efficiency of mixing, and
hence of non-thermal ionization/excitation. An alternative is fast rotation, which can allow lower-mass massive stars to die
as a moderate-mass WR star with some outer helium-rich shell
and thus one cannot completely exclude this route. To resolve this issue requires that all stellar-evolutionary models for such objects
be evolved until iron-core collapse to permit the modeling of the SN \LCS\ and spectra and a direct confrontation to SN observations.

In helium-rich ejecta, the absence of optical He\one\ lines may simply come from a reduction in \isoni\ mass
fraction in the helium-rich layers, either because of moderate or weak mixing or because of an asymmetric \isoni\ distribution
that fails to thoroughly mix with the helium-rich ejecta layers.
At the very least, helium deficiency is not a prerequisite for the production of a SN Ic, although
it is a sufficient condition.
Additional explanations may be that some SN Ic ejecta are more massive, so that, even if fully mixed, the larger mass
will cause greater dilution and weaker non-thermal excitation. Non-thermal processes are not selective -- atoms/ions
share the available non-thermal ionization and excitation energy largely in proportion to their relative abundances.

\begin{figure*}
\epsfig{file=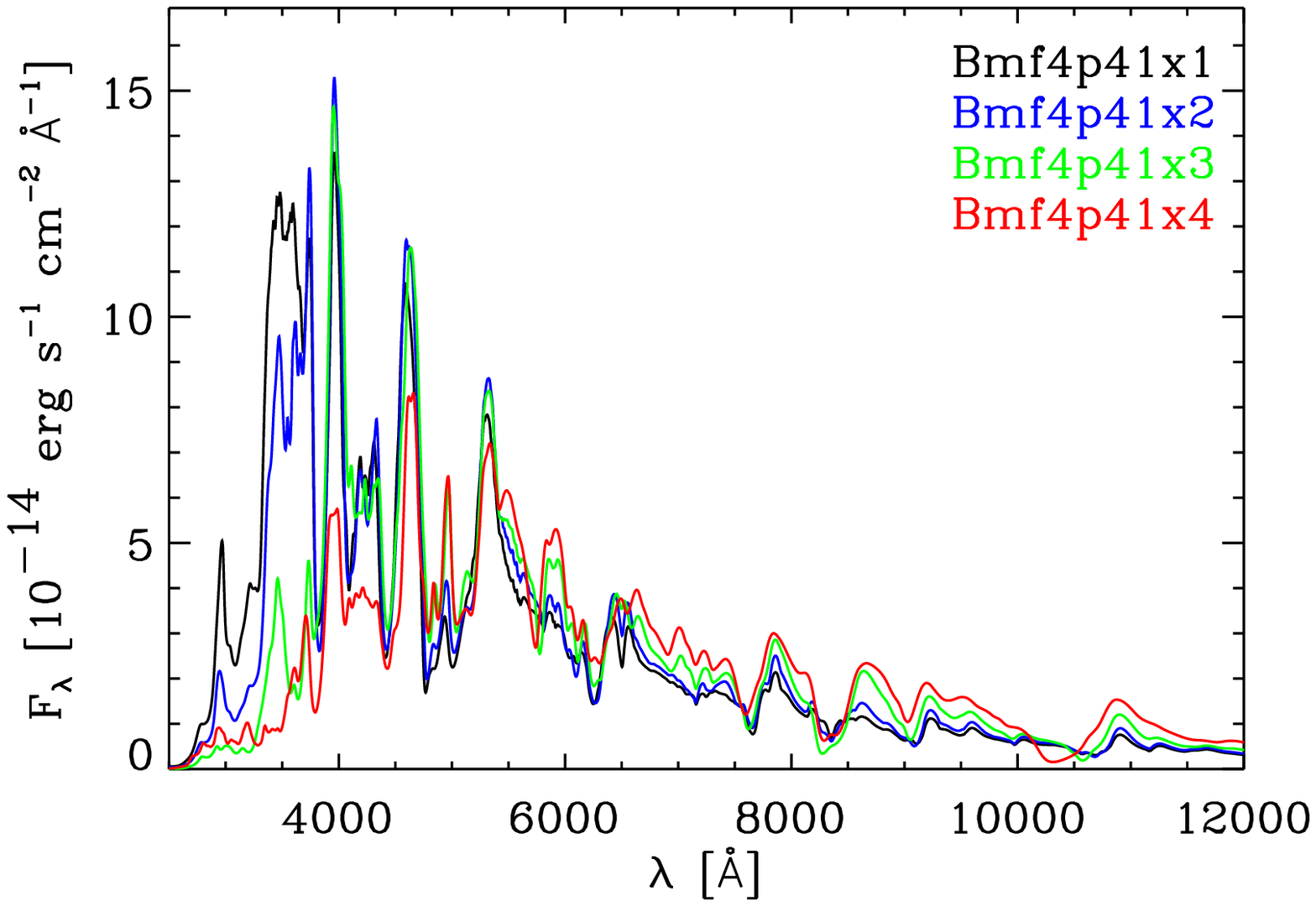,width=16cm}
\caption{Spectral comparison from UV to near-IR and at light-curve peak for models Bmf4p41x1--4.
A distance of 10\,Mpc is adopted.
With increasing mixing, colors are redder (the photosphere is cooler and more strongly affected by line blanketing in the UV), 
line profiles are broader (the strongest lines are optically thick to larger distances/velocities), and the photospheric velocity 
is larger (the continuum also forms further out).
\label{fig_comp_spectra_at_peak}
}
\end{figure*}

To close the discussion on our synthetic spectra, we present evidence that mixing must
occur at some level in all SNe Ib/c. Relative to unmixed models,
mixing causes the colors at peak to be redder and more compatible with the observations
(Fig.~\ref{fig_comp_spectra_at_peak}; e.g., \citealt{modjaz_etal_09}). This issue requires detailed spectral
modelling since an uncertain reddening can compromise the inferred colors of observed SNe Ib/c
and synthetic spectral colors are sensitive to the completeness of the model atoms.
In any case, mixing is likely a generic property of Type I CCSN ejecta,
and it is probably variations in the resulting mass fractions for both helium and \isoni\ that control the classification as Ib or Ic.

\section{Comparison to observations}
\label{sect_comp}

The goal of this paper is not to do a comprehensive comparison of our model results to observations.
Rather, we are primarily interested in understanding the key physics and ejecta properties that influence
the \LCS\ and spectral evolution of Type Ib and Ic SNe.
Although we use only two initial ejecta models, it is interesting to check how well our models fare
against some well observed SNe Ib/c.

  In Fig.~\ref{fig_comp_obs} (left panel), we show a comparison of the heavily mixed model Bmf5p1x4 at 10.3\,d after
explosion with the observations of the Type Ib SN2008D \citep{modjaz_etal_09} on the 2nd of February 2008, which corresponds
to 23.88\,d after the X-ray detection
(we choose a time when the model matches both the SN2008D spectral color and the line widths approximately). Besides a good fit to the He\one\ lines
we reproduce well the effects of line blanketing in the blue part of the optical, as well as the features due to intermediate
mass elements (IMEs) in the red part of the spectrum. Despite ubiquitous line overlap, He\one\ lines are strong enough
to be unambiguously identified. For example, He\one\,5875\,\AA\ remains stronger than the overlapping lines due to Fe\two,
Si\two, and to a weaker extent C\two.
In the near-infrared, He\one\,10830\,\AA\ overlaps with two weak C\one\ and O\one\ lines, and a strong Mg\two\ line,
but the broader component is clearly due to helium. This is in agreement with the chemical stratification of the ejecta,
with a larger abundance of IMEs at smaller velocities in such a helium-rich progenitor star.

  In the right panel of Fig.~\ref{fig_comp_obs}, we show a similar comparison, but now to the Type Ic SN 2007gr
  at $\sim$25\,d after discovery \citep{valenti_etal_08},
  using the moderately mixed model Bmf5p1x3 at 15.1\,d after explosion.
  Unlike the spectrum of SN2008D discussed above, the He\one\ lines are very weak and dominated
  by overlapping components. For example, He\one\,5875\,\AA\ is now weak so that this spectral region is composed of
  a number of overlapping lines due to Na\one, Fe\two, Si\two, and O\one.
  In the near-IR, the He\one\,10830\,\AA\ line is now weaker/narrower and suffers greater contamination from C\one\ and Mg\two\
  lines. While SN 2007gr is classified a Type Ic, a model with weak helium lines (but truly helium-rich), provides a satisfactory
  match to observations. {\it This shows that the evidence behind the helium deficiency of SNe Ic is weak, if not inexistent.}

  A property of SNe Ib that is interesting to examine is the width of the He\one\,5875\,\AA\ line.
  For example, for the standard SN Ib 2008D, the blueshift with respect to line center of the maximum absorption
  in this line corresponds to a velocity of about 10000\,\kms\ at the \LC\ peak. In Fig.~\ref{fig_vline}, we show
  the corresponding velocity versus time since \LC\ peak for the moderately and heavily mixed models
  (we overlay the trajectory of the photospheric velocity for each model as a solid line).
  In the heavily mixed models, we obtain velocities at maximum absorption in He\one\,5875\,\AA\ of
  about 7000-8000\,\kms, which is close but somewhat short of the observed standard.
  Interestingly, the moderately mixed model Bmf4p41x3, which has the same mass and kinetic energy
  as model Bmf4p41x4, has a maximum-absorption velocity in He\one\,5875\AA\ about 50\% less than that of Bmf4p41x4.
  This might mean that this model is too massive for its energy, or its energy is too low for its mass,
  a combination of both, or that the mixing is still too weak.
  In any case, although not surprising in light of the present discussion, this highlights the difficulty of determining
  the kinetic energy of SN Ib ejecta, even with the use of spectroscopic information. This inference
  requires the treatment of non-thermal effects and detailed radiative-transfer modelling (including
  time-dependence since it does also modify the ejecta ionization structure; \citealt{DH08_time}).
  Such a physically-consistent treatment is generally not performed, which suggests
  the inferred characteristics of SNe Ib/c published in the literature today are not converged.

\section{Discussion and conclusions}
\label{sect_conc}

  We have presented results from non-LTE time-dependent radiative-transfer simulations based
on 1.2\,B piston-driven explosions produced from two of the binary-star evolution models of
\citet{yoon_etal_10}. This work departs from the previous study of \citet{dessart_etal_11} by the
explicit treatment of non-thermal processes which are associated with the high-energy electrons arising
from Compton-scattering of \grays\ emitted by decaying \isoni\ and \isoco\ nuclei. Our approach
solves for the relative contribution of these non-thermal electrons to heating, ionizing, and exciting
the ejecta gas, and allows us to compute accurate spectra and \LCS\ through the entire photospheric
phase.

This work primarily focused on exploring the effects of non-thermal processes  on the {\it photospheric phase}
of SNe IIb/Ib/Ic. Rather than using a large grid of models
with very diverse properties, we choose for clarity to focus on only two ejecta models, but arising from
binary star evolution. For each of these models we create four additional models that are characterized exclusively
by different levels of mixing. This mixing is performed through a ``boxcar'' algorithm which affects all
species indiscriminately. We arrive at the following conclusions:

\begin{figure*}
\epsfig{file=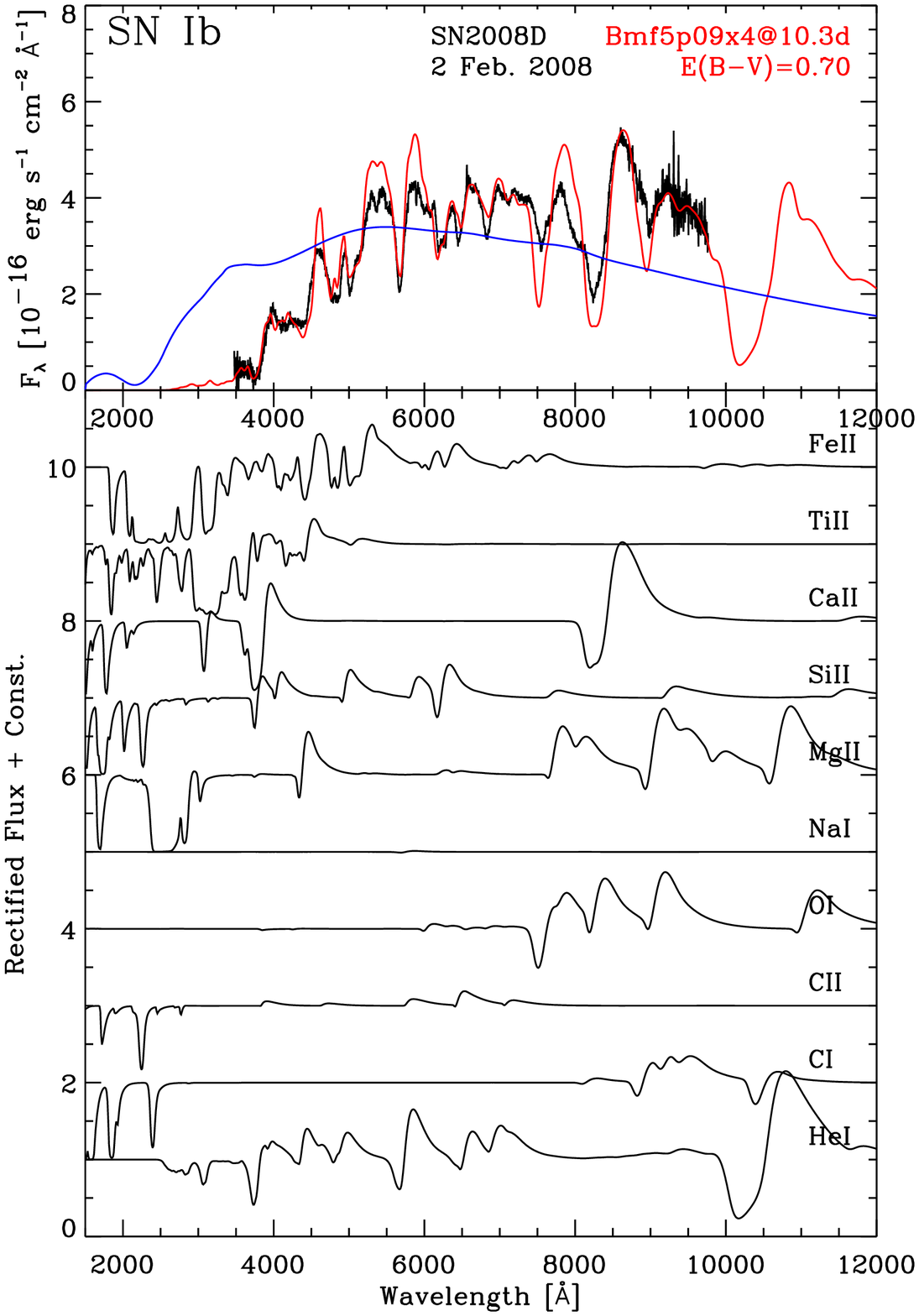,width=8.5cm}
\epsfig{file=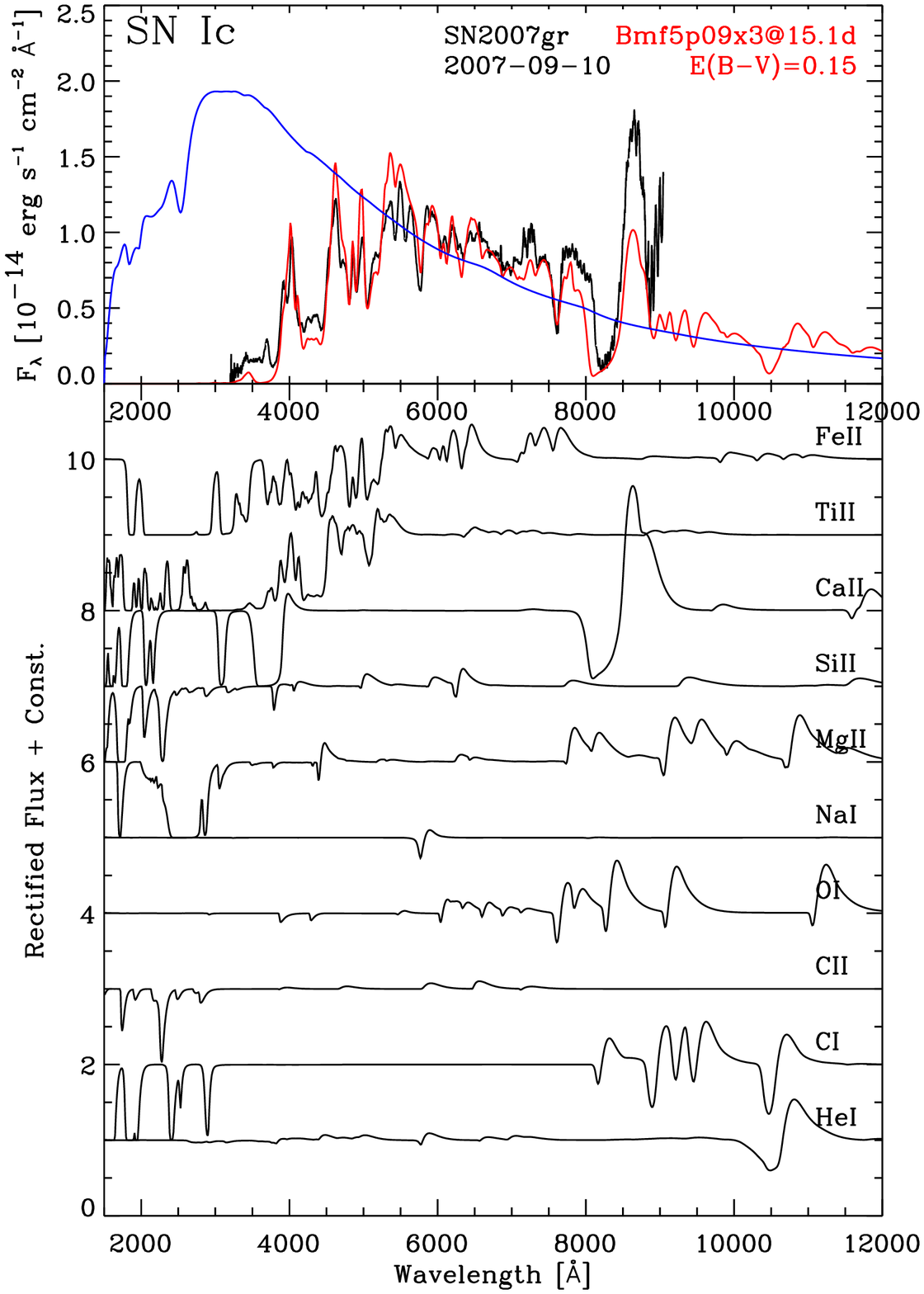,width=8.5cm}
\vspace{-1cm}
\caption{
{\it Left:} Comparison of model Bmf5p1x4 at 10.3\,d after explosion (we show the total synthetic flux in red and the synthetic continuum
flux in blue), reddened by $E(B-V)=0.7$, with the observations of the Type Ib SN2008D on the 2nd of February 2008,
23.88\,d after the X-ray detection \citep{modjaz_etal_09}.
{\it Right:} Similar as the left figure, but now showing a comparison of model Bmf5p1x3 at 15.1\,d after explosion, reddened by $E(B-V)=0.15$,
with the observations of the Type Ic SN2007gr on the 10th of September 2007, $\sim$25\,d after discovery \citep{valenti_etal_08}.
Here, we illustrate the correspondence between these observations and our models when both are characterized by the same color
and line widths.
In each panel, the lower part shows the relative contribution to the total spectrum of bound-bound transitions associated 
with individual species. Notice in particular the distinct He\one\ contribution in the strongly mixed (left) and the moderately mixed model (right).
\label{fig_comp_obs}}
\end{figure*}

\begin{enumerate}

\item The most important consequences of mixing are associated with the change in \isoni\ distribution.
We identify both {\it thermal} and {\it non-thermal} effects.
\item Under ionized conditions, and thus generally at
optical-depths in excess of unity, the non-thermal electrons dump most of
their energy in the form of heat, thus primarily modulating the ejecta temperature, its ionization state,
its optical depth, and thus the \LC\ evolution. With enhanced mixing, the \LC\
starts its post-breakout re-brightening phase earlier, follows a more gradual rise to peak,
and peaks at an earlier time.  The post-breakout plateau is absent in the most mixed models
we study. In all our models, \grays\ are trapped for up
to $\sim$30\,d after peak and thus ejecta differing only in mixing properties have similar
\LC\ slopes in the first month after peak. However, with enhanced mixing, \gray\ escape starts earlier and causes
the nebular luminosity to decrease faster.
This sensitivity to mixing conditions are exacerbated by the small ejecta masses that characterize
all our binary-star progenitor models.

\item While non-thermal ionization and excitation are weak in optically-thick ionized layers, non-thermal electrons
 in partially-ionized regions dump a significant fraction of their energy through  non-thermal ionization and excitation of the plasma.
 In heavily mixed models, the significant decay energy deposited above the photosphere
affects primarily the He\one\ ionization and excitation since
helium is nearly recombined there and is the dominant species.
In contrast, weakly or moderately mixed models deposit most of the energy at depth, below the photosphere,
and fail to non-thermally excite (or ionize) the outer ejecta region where lines and continuum form.
However, in these models, the exclusive deposition of heat at large optical depths creates highly ionized conditions.
\item Consequently, during the photospheric-phase of SNe Ib/c, non-thermal effects are favored
in highly mixed ejecta. As proposed by \citet{lucy_91}, non-thermal excitation and ionization are key for the production
of He\one\ lines. In our present set of helium-rich models, the primary feature is the presence of optical
and near-IR He\one\ lines characteristic of the Type Ib classification. Although characterized by the
same cumulative yields, moderately mixed models fail to non-thermally excite helium atoms and lead
to a Type Ic classification.

\item In spite of our small model set, we identify a diverse range of spectral evolution and
classification, and obtain both Type Ib and Ic SNe. Furthermore, some models show hydrogen at early
times (produced by non-LTE effects alone; \citealt{dessart_etal_11}),
followed by the presence or absence of He\one\ lines. These would correspond to classifications as
Type IIb and IIc, respectively. The latter has never been identified with confidence (see \citealt{branch_etal_06}
for a possible identification of H$\alpha$ in the Type Ic SN 1994I) but this may just be a bias of observational
campaigns which rarely capture Type Ib/c at a few days past explosions.
In view of the present simulations, a type IIc classification is easily understood as stemming from a progenitor with both hydrogen and helium,
the former being under-abundant but easily seen, and the latter arbitrarily abundant but quickly invisible in the
absence of non-thermal excitation \citep{dessart_etal_11}.
A transition from type Ic to type Ib is more difficult to achieve as it would require a large amount of helium
together with moderate mixing. Non-thermal excitation of helium may then become effective only when the photosphere
has receded to deeper layers where \isoni\ is more abundant. This unique transition has been seen for SN2005bf by
\citet{folatelli_etal_06}, although mixing is unlikely the cause in this special case \citep{maeda_etal_07}.

\item Apart from the cases where He\one\ lines are really strong, line overlap considerably complicates
the identification of spectral features. We identify a slew of lines from He\one, C\one, O\one, Na\one, Mg\two,
and Fe\two\ in the optical and the near-IR. One can, for example, easily mistake He\one\,5875\,\AA\ with
Na\one\,D, Si\two\,6350\,\AA\ with H$\alpha$ etc. Multi-epoch spectroscopic observations from a few days
after explosion until the nebular phase are critical to build a comprehensive and accurate picture of the
SN-ejecta composition.

\item From our explorations, we are led to associate SN Ib ejecta with the explosion of low-mass helium cores
in which efficient small-scale mixing takes place. The scale of relevance for non-thermal excitation/ionization
is the \gray\ mean-free-path, which varies with both depth and time. We favor
progenitors that have only a small oxygen-rich mass buffer between the \isoni\ production site and the outer
helium-rich layers, facilitating the mixing, and preventing too much dilution of \isoni\ into a large mass.
In our set of models, we find that a minimum \isoni\ mass fraction of 0.01 is necessary at large velocities
to trigger significant non-thermal effects early on. The small-scale mixing necessary in this context
may originate from the fragmented structure of the SN shock \citep{nordhaus_etal_10}.

\item It requires special circumstances for our low-mass quasi-hydrogen-less cores to become a Ib ---
failure to meet these requirements will typically lead to a type Ic. This may stem from any combination of properties
that prevent helium atoms from being non-thermally excited.
This difficulty to make SNe Ib might be reflected by the 2:1 occurrence rates of SNe Ic:Ib \citep{smartt_09}.
Inefficient non-thermal excitation of He\one\ may result from 1) helium deficiency or under-abundance
(\citealt{yoon_etal_10} argues for the possibility of even lower mass helium-deficient progenitors
than those presented here); 2) a higher-mass helium core with a sizeable oxygen-rich mass
buffer, which causes a stronger dilution of \isoni\ --- this could naturally occur for single WR stars which are
born from high-mass stars with higher-mass helium cores \citep{georgy_etal_09}; 3) explosions that are not mixed
on small scales so that most of the \isoni\ is contained within a limited solid-angle and ejecta volume. In this context,
a ``jet'' of \isoni-rich material is not expected to be a suitable morphology for non-thermal excitation
of ejecta material because the \grays\ would not be able to penetrate the other latitudes that would contain
the bulk of the helium-rich material, i.e., only a tiny fraction of He\one\ atoms would be excited.

\item Because the combination of thermal and non-thermal processes considerably affect ion/atom
excitation and thus the associated line optical depths, we find that models of the same
mass and kinetic energy but with enhanced mixing display broader line-profile widths at a given post-explosion time.
This sensitivity complicates the reliable inference of SN-Ib/c ejecta masses
and explosion energies, and suggests the importance of accounting for
both time-dependent and non-thermal effects when modelling their spectra and \LCS.

\item Adding to conclusions of earlier studies \citep{EW88,shigeyama_etal_90,dessart_etal_11},
our spectral simulations including non-thermal effects strengthen the notion that the bulk of SNe Ib/c arise from low-mass ejecta
likely resulting from binary-star evolution.
The broad line Ic's with and without GRBs, associated with larger mass ejecta and higher kinetic energy \citep{woosley_etal_99},
may stem from different progenitors and different explosion mechanisms. Our simulations suggest that these may in fact contain some helium,
which is invisible because it is not excited. This result does not conflict with the predictions of
massive star evolution for carbon and oxygen WR stars \citep{georgy_etal_09}. Furthermore, it would not conflict with the
prerequisites of the collapsar model \citep{woosley_93} or those of the proto-magnetar model
\citep{wheeler_etal_00,metzger_etal_11} of long-duration GRBs.

\item We identify a negative feedback on non-thermal effects. As the \isoni\ abundance is increased, the heating
caused by energy deposition increases and forces non-thermal electrons to channel their energy in the form of heat,
thereby quenching the non-thermal ionization and excitation channels. In contrast, non-thermal effects thrive under
neutral conditions, i.e., plasmas that are cool, and thus not too \isoni-rich. We thus anticipate non-thermal ionization
and excitation to be stronger for moderate \isoni\ abundances. They should be more efficient in the cool photospheres
of SNe Ib/c than in the hot ionized photospheres of SNe Ia,  highly \isoni-rich, where non-thermal electrons will tend
to dissipate their energy as heat.

\item Our radiative-transfer modeling is 1D so the emergent specific intensity or flux is independent of viewing angle.
As we independently argue in Sect.~\ref{sect_gray}, multi-dimensional effects are unlikely to turn a SN Ib into a SN Ic,
or a IIb into a Ib, but instead  should introduce slight variations in line width and/or strength with viewing angle,
as illustrated in our 1D models differing in mixing magnitude (Fig~\ref{fig_vline}). 
Recently, \citet{rest_etal_11} have used light echoes to reveal the viewing-angle dependent intensity from the type IIb SN 
associated with Cas A, typically characterized by variations in line widths along different directions. 
However, each recorded light echo supports the IIb classification for the SN. 
Using a similar approach to \citet{DH11b}, we will investigate the magnitude of the asymmetries required to explain 
such observations.
\end{enumerate}

\begin{figure}
\epsfig{file=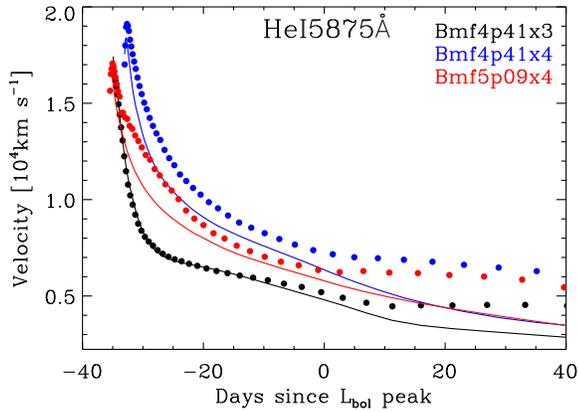,width=8.5cm}
\caption{Time evolution of the velocity at maximum absorption in the strongest optical He\one\ line
at 5875\AA. We only select here the models that show weak (Bmf4p41x3) and strong (Bmf5p09x4 and Bmf5p09x4)
He\one\ lines. We also overlay in each case the evolution of the velocity at the electron-scattering photosphere (solid line).
\label{fig_vline}}
\end{figure}


The next step in our modeling effort is to compute \LCS\ and spectra for a large grid of progenitor
stars, both single and from binary systems, exploded with a range of explosion energy,
and characterized by different levels of mixing. Such a diversity of initial conditions should reflect the
diversity of SNe IIb/Ib/Ic seen in Nature and allow a better assessment of the progenitor and
explosion properties of these CCSNe.

\section*{Acknowledgments}
LD acknowledges financial support from the European Community through an
International Re-integration Grant, under grant number PIRG04-GA-2008-239184,
and from ``Agence Nationale de la Recherche" grant ANR-2011-Blanc-SIMI-5-6-007-01.
DJH acknowledges support from STScI theory grant HST-AR-11756.01.A and
NASA theory grant NNX10AC80G.
This research was supported at UCSC by the DOE SciDAC Program under contract
DE-FC02-06ER41438, the National Science Foundation (AST 0909129) and
the NASA Theory Program (NNX09AK36G).
This work was granted access to the HPC resources of CINES under the allocation 2011--c2011046608 made by
GENCI (Grand Equipement National de Calcul Intensif).


\label{lastpage}

\end{document}